\pdfoutput=1
\documentclass{JINST}

\usepackage{amssymb}
\usepackage{lineno}
\usepackage{graphicx}
\usepackage{amssymb}
\usepackage{amsmath}
\usepackage{epstopdf}
\usepackage{relsize}

\def\cesrta{{C{\smaller[2]ESR}TA}}
\def\cesrta{{C{\smaller[2]ESR}TA}}
\newcommand{\beq} {\begin{equation}}
\newcommand{\eeq} {\end{equation}}

\title{The Conversion of CESR to Operate as the Test Accelerator, CesrTA, Part 2: Vacuum Modifications}

\author{M.G.Billing, Y.Li\
Cornell University, \\
Ithaca, NY, U.S.A.}

\abstract{ Cornell's electron/positron storage ring (CESR) was
modified over a series of accelerator shutdowns beginning in May
2008, which substantially improves its capability for research and
development for particle accelerators.  CESR's energy span from 1.8
to 5.6 GeV with both electrons and positrons makes it 
ideal for the study of a wide spectrum of
accelerator physics issues and instrumentation related to present
light sources and future lepton damping rings. Additionally a number
of these are also relevant for the beam physics of proton
accelerators.  This paper, the second in a series of four,
discusses the modifications of the vacuum
system necessary for the conversion of CESR to the test accelerator,
{\cesrta}, enhanced to study such subjects as low emittance tuning
methods, electron cloud (EC) effects, intra-beam scattering, fast
ion instabilities as well as general improvements to beam
instrumentation.  A separate paper describes the vacuum system
modifications of the superconducting wigglers to accommodate the
diagnostic instrumentation for the study of EC behavior within
wigglers.  While the initial studies of {\cesrta} focussed on
questions related to the International Linear Collider (ILC) damping
ring design, {\cesrta} is a very flexible storage ring, capable of
studying a wide range of accelerator physics and instrumentation
questions.}

\keywords{Accelerator Subsystems and Technologies, Beam-line Instrumentation}

\begin{document}

\section{Overview of CESR Modifications}
\label{sec:cesr_conversion.overview}

The conversion of CESR to permit the execution of the {\cesrta}
program\cite{jinst:billing1} required several extensive modifications.  These included
significant reconfiguring of CESR's accelerator optics by removing
the CLEO high energy physics detector and its interaction region,
moving six superconducting wigglers and reconfiguring the L3
straight section.  There were also major vacuum system modifications
to accommodate the changes in layout of the storage ring guide-field
elements, to add electron cloud diagnostics and to prepare regions
of the storage ring to accept beam pipes for the direct study of the
electron cloud.  A large variety of instrumentation was also
developed to support new electron cloud diagnostics, to increase the
capabilities of the beam stabilizing feedback systems and the beam
position monitoring system, to develop new X-ray beam size
diagnostics and to increase the ability for studying beam
instabilities.  This conversion process for the vacuum system for the 
{\cesrta} program is described in the following sections.

%\input{parameters_reach}
%% done
%---------------------------------------------------------------%
% CesrTA Phase I Report -                                       %
% Chapter - The CESR Conversion to a Damping Ring Configuration %
% Section - Vacuum System Modifications                         %
% Section coordinator:  Yulin Li                                %
% Page estimate: 35pp.                                          %
%---------------------------------------------------------------%
%!TEX root =  cesr_conversion.tex

\section[Vacuum System Modifications]{Vacuum System Modifications}
\label{sec:cesr_conversion.vac_system}

% --- Subsection: Overview
\subsection{Overview}
\label{ssec:cesr_conversion.vac_system.overview}
  CESR's vacuum system is an essential part of its accelerator beam transport system, which is capable of storing total electron and positron beam currents up to 500~mA (or single beam up to 250 mA) at a beam energy of 5.3~GeV.  As shown in Figure~\ref{fig:cesr_conversion:vac_fig1} the CESR vacuum system with a total length of 768.44~m consists primarily of bending chambers in the arcs, two long straight sections, called L0 (18.01~m in length) and L3 (17.94~m in length), and four medium length straights (called $L1$~and~$L5,$ both 8.39~m in length and $L2$~and~$L4,$ both 7.29~m in length).

  With exception of L0 and L3, CESR vacuum beam pipes are made from aluminum extrusions (Type 6063 alloy) with built-in pumping and cooling channels.  The vacuum pumping in the arcs is dominated by the home-made distributed ion pumps (DIPs) inserted into the pumping channel of the extrusion, although conventional ion pumps are installed periodically.  The L0 straight section was the interaction region for the CLEO High-Energy Physics (HEP) experiment, hosting the CLEO detector at the center of L0.  Massive titanium sublimation pumping (TiSP) \cite{JVSTA15:716to722} was implemented in L0 to achieve ultra-high vacuum (UHV) during the high current HEP operations for CLEO.  In the CESR-c/CLEO-c era 12 home-built superconducting wigglers (SCWs) were installed in CESR, with two triplets at $L1$ and $L5$, two doublets and two singlet SCWs in the arcs (see Figure~\ref{fig:cesr_conversion:vac_fig1}).  This complement of SCWs plays an important role in the {\cesrta} program. Over the nearly 29 year of operation prior to the {\cesrta} conversion, the CESR vacuum system performed satisfactorily, with average dynamic pressure in the low $10^{-9}$~Torr with more than 400~mA of stored beams of electrons and positrons.

   At the conclusion of CLEO-c HEP program in March 2008, staged modifications were carried out to convert CESR into the test accelerator {\cesrta}.  The motivation for the vacuum system modification was to support the physics programs of {\cesrta}, including: (1)~Ultra-low emittance lattice design, tuning and associated beam instrumentations, and (2)~Electron cloud studies, including the development and verification of suppression techniques for ECs.  During the design and implementation of the {\cesrta} vacuum system conversion, the following two aspects were of pre-eminent importance:

\begin{figure}
    \centering
    \includegraphics[width=0.75\textwidth]{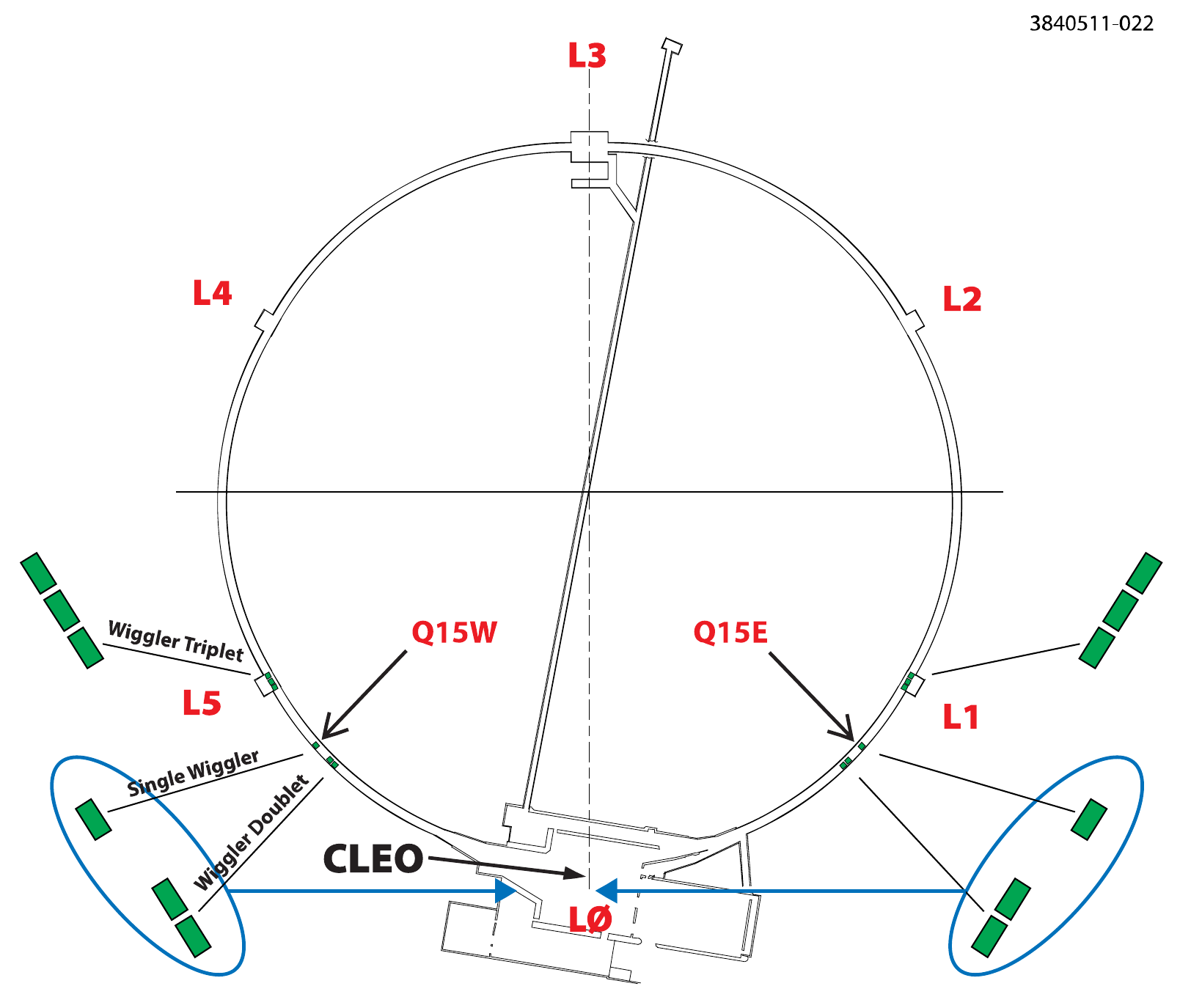}
    \caption[{\cesrta} Vacuum System]{The reconfiguration of the CESR vacuum system provided space in two long experimental regions in L0 and L3, and two flexible short regions near Q15W and Q15E. Hardware for electron cloud studies was installed in these regions. \label{fig:cesr_conversion:vac_fig1}}
\end{figure}

\begin{itemize}
\item To create environments where both local- and collaborator-provided test chambers and equipment can be easily installed and tested.
\item To ensure the continuing successful operations of Cornell High Energy Synchrotron Source (CHESS) at CESR.
\end{itemize}

The physical modification of CESR vacuum system started in May 2008 during a scheduled accelerator shutdown.  The majority of the vacuum system reconfiguration was carried out in two long shutdowns, July 2008 and February 2009.  The following list summarizes the scope of vacuum modifications during scheduled accelerator shutdowns.

\begin{enumerate}
\item May 27 -- June 3, 2008:
    \begin{itemize}
        \item Implemented a Cornell thin-style retarding field analyzer (RFA) in a CESR dipole chamber at B12W location.
        \item Replaced two CESR-c SCWs with RFA-equipped drift chambers at Q14W location.
    \end{itemize}
\item July 7 -- October 23, 2008:
    \begin{itemize}
        \item Replaced four additional CESR-c SC wigglers from Q15W, Q14E/Q15E (see Figure~
\ref{fig:cesr_conversion:vac_fig1}) with new beam pipes equipped with EC diagnostics.
        \item Created Q15W and Q15E short experimental sections.
        \item Completed L0 reconfiguration, including removal of the central part of the CLEO HEP Detector and the IR vacuum
            chambers, installation of 6 SCWs (4 CESR-c and two RFA-equipped \cesrta\ SCWs) and many new
            chambers suited with EC diagnostics, such as RFAs and BPM/TE-wave buttons.  N.b. the RFA-equipped
            SCWs are described in a companion paper.
        \item Partially reconfigured L3 region, including removal of a pair of (high beam impedance)
             electrostatic vertical separators.
    \end{itemize}
\item February 23 -- March 2, 2009:
    \begin{itemize}
        \item Completed L3 central region reconfiguration, including installation of SLAC's EC beam pipes
            and chicane magnets, which had been removed from  PEP-II after their EC 
            studies\cite{NIMA621:22to38}.
        \item Installed a photon-stop vacuum chamber to replace a CLEO-c HEP-era fast-luminosity monitor
            chamber \cite{PAC05:RPPE044} at the Q3W location.  This new chamber enabled the storing up to 100~mA
            positron beam currents in CESR at 5~GeV beam energy when all 6 L0 SCWs are in operation.
    \end{itemize}
\item June 16 -- July 23, 2009:
    \begin{itemize}
        \item Installed first RFA-equipped quadrupole vacuum chamber and in-situ SEY measurement
            stations in the L3 experimental region.
        \item Installed test chambers within the Q15W/E sectors.
        \item Implemented three RFA-equipped SCWs in the L0 region.  N.b. these RFA-equipped
            SCWs are described in a companion paper.
    \end{itemize}
\end{enumerate}

The {\cesrta} vacuum system reconfiguration, as depicted in Figure \ref{fig:cesr_conversion:vac_fig1}, created four dedicated {\cesrta} experimental regions.  The list below highlights the main features of these
experimental sections.

\begin{itemize}
\item L0 EC Region:  This region hosts a string of 6 superconducting wigglers, having 3 of them
    fitted with RFA-equipped beam pipes, and adjacent chambers containing beam position monitor (BPM) buttons,
    used for TE-Wave measurements.  The main function of this experimental section is to investigate
    EC dynamics and suppression techniques in the wiggler field.
\item L3 EC Region:  This region features EC studies in dipole and quadrupole fields and field-free drifts, as well as
    in-situ SEY measurement stations.
\item Q15W \& Q15E Test Sectors:  These two short sectors each contain one CESR normal
    bend and a short straight and are intended for testing vacuum chambers with new EC diagnostics and EC
    suppression coatings.
\end{itemize}

During the reconfiguration and the first 3 years of the {\cesrta} program, over 40 new vacuum chambers and components
were installed in or rotated through CESR vacuum system for the {\cesrta} program.  Almost all of these new
chambers were designed and constructed in-house.  Many EC diagnostics and/or EC suppression features are
successfully incorporated in these new chambers.  The implemented EC diagnostics include RFAs at various types of magnets, shielded pickup (SPU) buttons and beam buttons for TE-wave measurement.  The evaluated suppression techniques include many types of coatings (TiN, amorphous- and diamond-like carbon and non-evaporable getter), grooved walls and a clearing electrode.  Stringent QA/QC procedures were followed in the production of these new UHV components and chambers, including design reviews/approvals, UHV-compatible practices during assembly and installation, and pre-installation vacuum bakeouts.  These efforts have resulted in very high degree of availability of the vacuum system for both the {\cesrta} program and CHESS operations.

Table~\ref{tab:MitigationTests} summarizes the range of vacuum chambers that have been
tested as part of the \cesrta\ program to examine the efficacy of various EC mitigations.
Each of the listed chambers incorporates one or more RFAs, utilized to measure the
beam-induced EC build-up within the chamber.  These studies have spanned each of the key
magnetic environments where mitigations will be required in an ILC positron damping ring (DR.)  In many
cases reference chambers without mitigations were employed to improve our understanding
of the relative performance of the various techniques as well as to provide data to help
characterize the models of the EC development.

\begin{table}[htpb]
   \centering
   \caption{\label{tab:MitigationTests} Vacuum chambers fabricated for testing during the
            initial phase of the \cesrta\ R\&D program studying electron cloud mitigation for
            ILC damping rings.  Mitigation studies have been conducted in drift, dipole,
            quadrupole, and wiggler magnetic field regions. }
   \vspace*{1ex}
   \begin{tabular}{cccccc}
   \hline\hline
                             &            &            &              &               & {\bf Contributing}\\
   {\bf Mitigation}          & {\bf Drift}& {\bf Quad} & {\bf Dipole} & {\bf Wiggler} & {\bf Institutions}\\
   \hline
   Al                        & \checkmark & \checkmark & \checkmark   &               & CU, SLAC \\
   Cu                        & \checkmark &            &              & \checkmark    & CU, KEK, \\
    & & & & & LNBL, SLAC \\
   TiN on Al                 & \checkmark & \checkmark & \checkmark   &               & CU, SLAC \\
   TiN on Cu                 & \checkmark &            &              & \checkmark    & CU, KEK, \\
    & & & & & LNBL, SLAC \\
   Amorphous C on Al         & \checkmark &            &              &               & CERN, CU \\
   Diamond-like C on Al      & \checkmark &            &              &               & CU, KEK \\
   NEG on SS                 & \checkmark &            &              &               & CU \\
   Solenoid Windings         & \checkmark &            &              &               & CU \\
   Fins with TiN on Al          & \checkmark &            &              &               & CU, SLAC \\
   \hline
   {\em Triangular Grooves:} &            &            &              &               &  \\
   On Cu                     &            &            &              & \checkmark    & CU, KEK, \\
    & & & & & LNBL, SLAC \\
   With TiN on Al            &            &            & \checkmark   &               & CU, SLAC \\
   With TiN on Cu            &            &            &              & \checkmark    & CU, KEK, \\
    & & & & & LNBL, SLAC \\
   \hline
   Clearing Electrode        &            &            &              & \checkmark    & CU, KEK, \\
    & & & & & LNBL, SLAC \\
   \hline \hline
   \end{tabular}
\end{table}

% --- Subsection: Electron Cloud Experimental Regions
\subsection{Electron Cloud Experimental Regions}
\label{ssec:cesr_conversion.vac_system.ec_exp_regions}
In this section, we provide details of the {\cesrta} EC experimental regions, described in terms of the vacuum system modification and performance.

\subsubsection{L0 Wiggler Test Region}
\label{sssec:cesr_conversion.vac_system.ec_exp_regions.L0}

\begin{figure}[hbt]
    \centering
    \includegraphics[width=\textwidth]{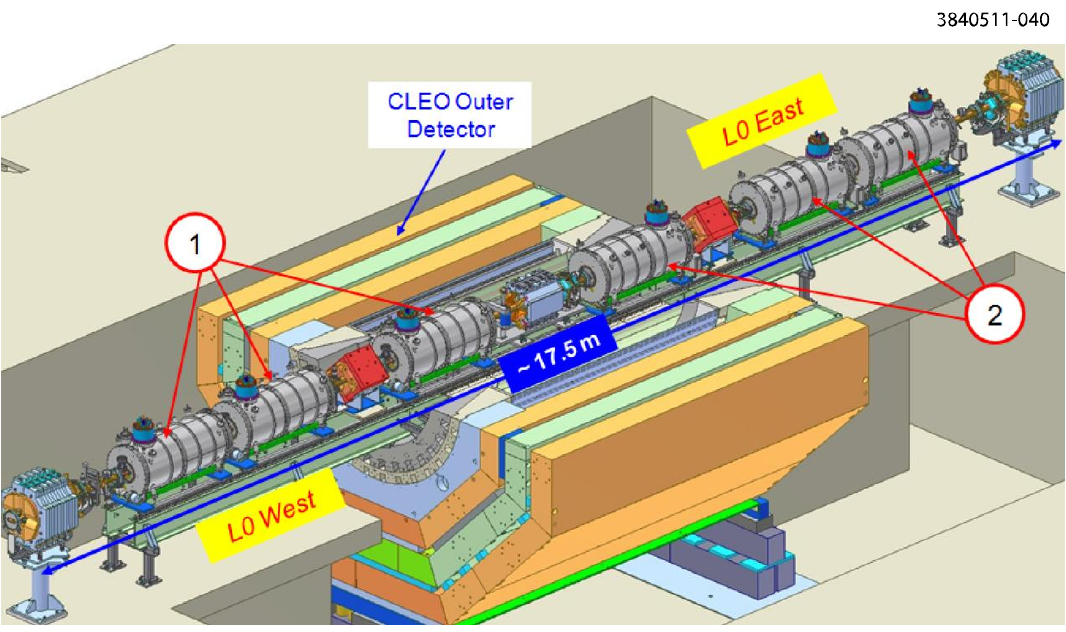}
    \caption[L0 EC Experimental Region Vacuum Layout]{L0 center {\cesrta} EC experimental region, consists of (1) three RFA-equipped SCWs and (2) three CESR-c SCWs.  Many other EC diagnostics, such as RFA in the drifts, BPMs and TE-Wave buttons, are also implemented.  The six SCWs in the region are labelled, from west to east, SCW02WB, SCW02WA, SCW01W, SCW01E, SCW02EA and SCW02EB.  Detailed engineering information of the L0 experimental region can be found in two CLASSE drawings with LEPP Drawing Numbers 6085-012 (Vacuum String + Magnets + Supports) and 6085-045 (Vacuum String only).\label{fig:cesr_conversion:vac_l0_center}}
\end{figure}

The CESR L0 long straight section (18.01~m in length), formerly the CLEO-c HEP Interaction Region (IR), was completely re-configured for the {\cesrta} research programs during two major CESR Shutdowns.  The modified L0 EC experimental region is shown in Figure \ref{fig:cesr_conversion:vac_l0_center}.  During a 4-month shutdown starting on July 7, 2008, the entire L0 straight section was reconfigured.   The modification included the following tasks:

\begin{itemize}
    \item All L0 IR vacuum chambers as well as the central portion of the CLEO detector assembly were removed.
    \item A new rail system to support the accelerator magnetic components was designed and installed through
             the inner-bore of the remaining parts of the CLEO Outer detector.
    \item Six SCWs were moved to the new supporting rails and new
        cryogenic transfer lines were installed. The three SCWs toward the eastern-side of L0 are the original CESR-c
        style, while the
        three SCWs toward the western-side of L0 were fitted with RFA-equipped beam pipes.
    \item Seven new vacuum beam pipes were also constructed and installed in this region in order to bridge the space
        between the SCWs.  Many EC diagnostics were attached to these new chambers, including three insertable,
        segmented RFAs, eight sets of beam buttons as BPMs and TE-wave transmitters/receivers.
    \item A pair of RF-shielded expansion bellows assemblies (aka sliding joints) were included with the
        new chambers to allow thermal expansions during beam operations and to provide flexibility for
        vacuum components exchange at various stages of the {\cesrta} program.
\end{itemize}

%% ??? about last line of caption.

\begin{figure}
    \centering
    \includegraphics[width=\textwidth]{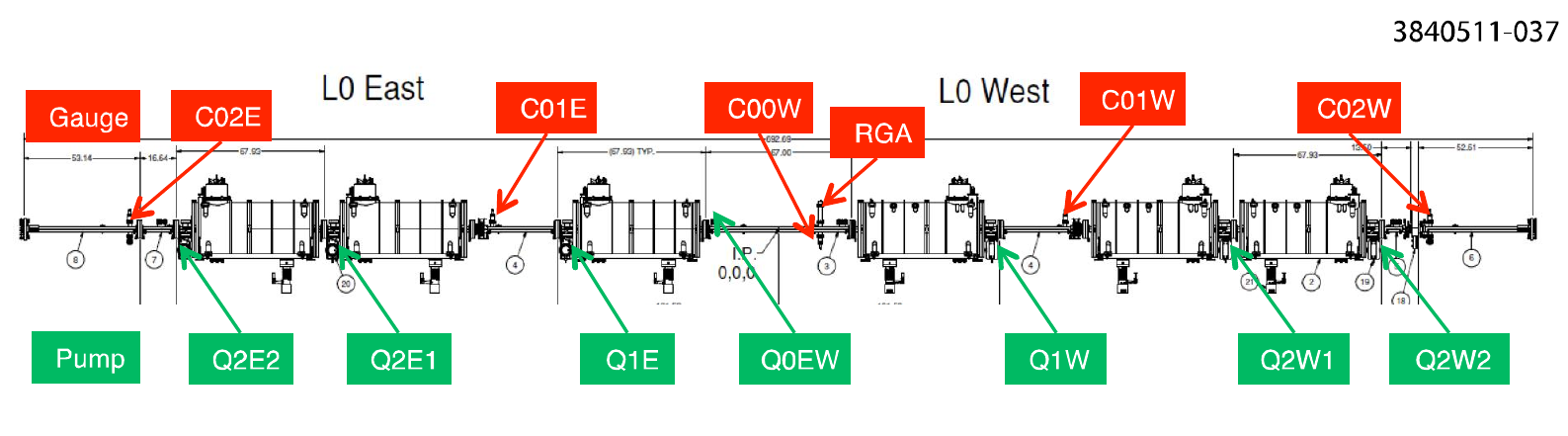}
    \caption[L0 Vacuum Pumping and Gauging]{Vacuum Pumping and Instrumentation in the L0 Center {\cesrta} EC experimental region.  At locations (labeled in {\it GREEN} as Q2E2, Q2E1, Q1E, Q1W, Q2W1 and Q2W2) a combination of a NEG and a small SIP is installed, while at the center of the region (labeled in {\it GREEN} as Q0EW) a large SIP is installed.  Five cold-cathode ion gauges (Model 421, MKS Instruments) and a residual gas analyzer (MicroVision 2 with radiation-resistant extension, MKS Instruments) (labeled in {\it RED}) are also installed in the region to monitor the vacuum system performance.
\label{fig:cesr_conversion:vac_l0_pumping}}
\end{figure}

\begin{figure}
    \centering
\begin{tabular}{cc}
\includegraphics[width=0.45\textwidth]{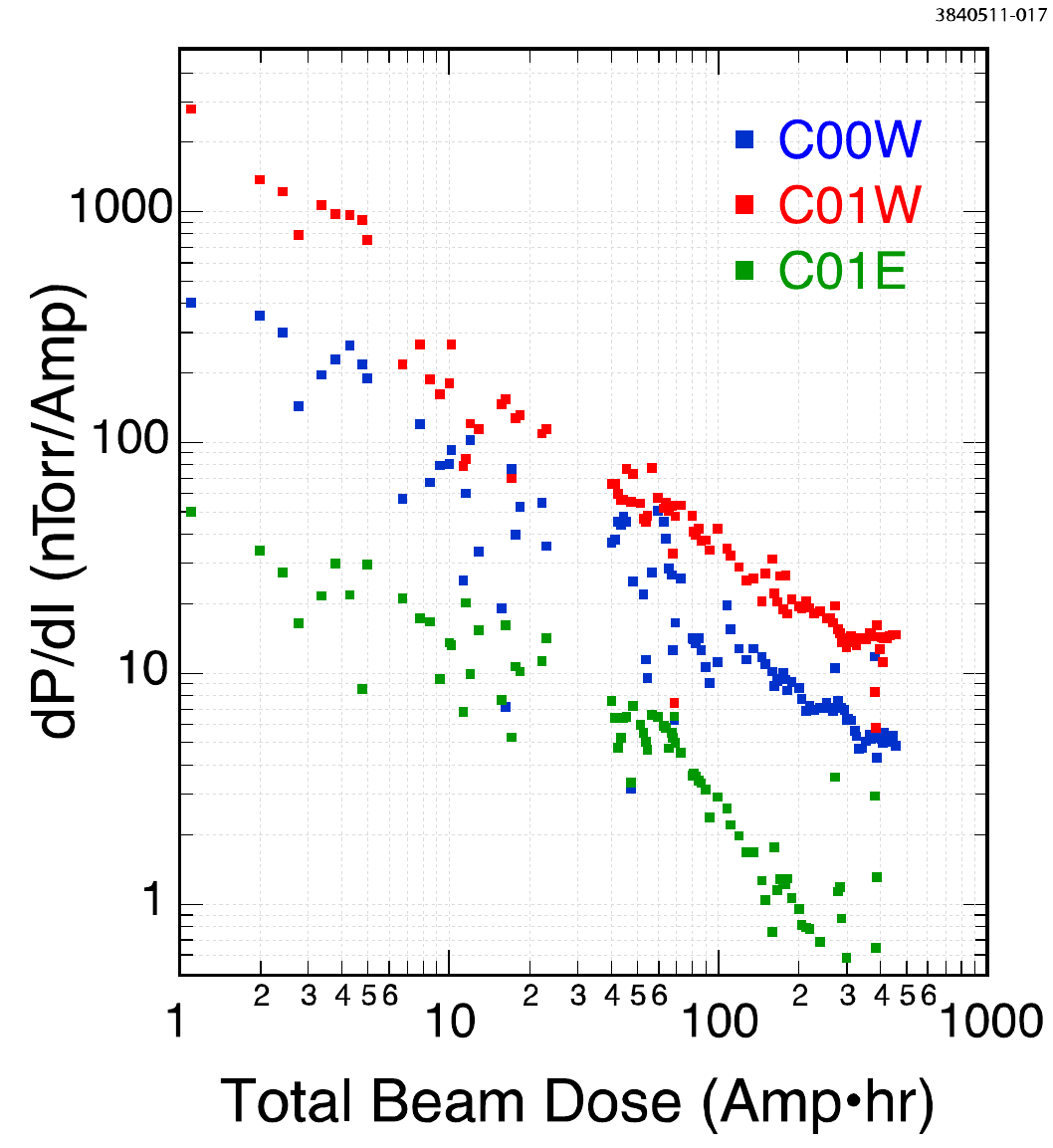} &
\includegraphics[width=0.45\textwidth]{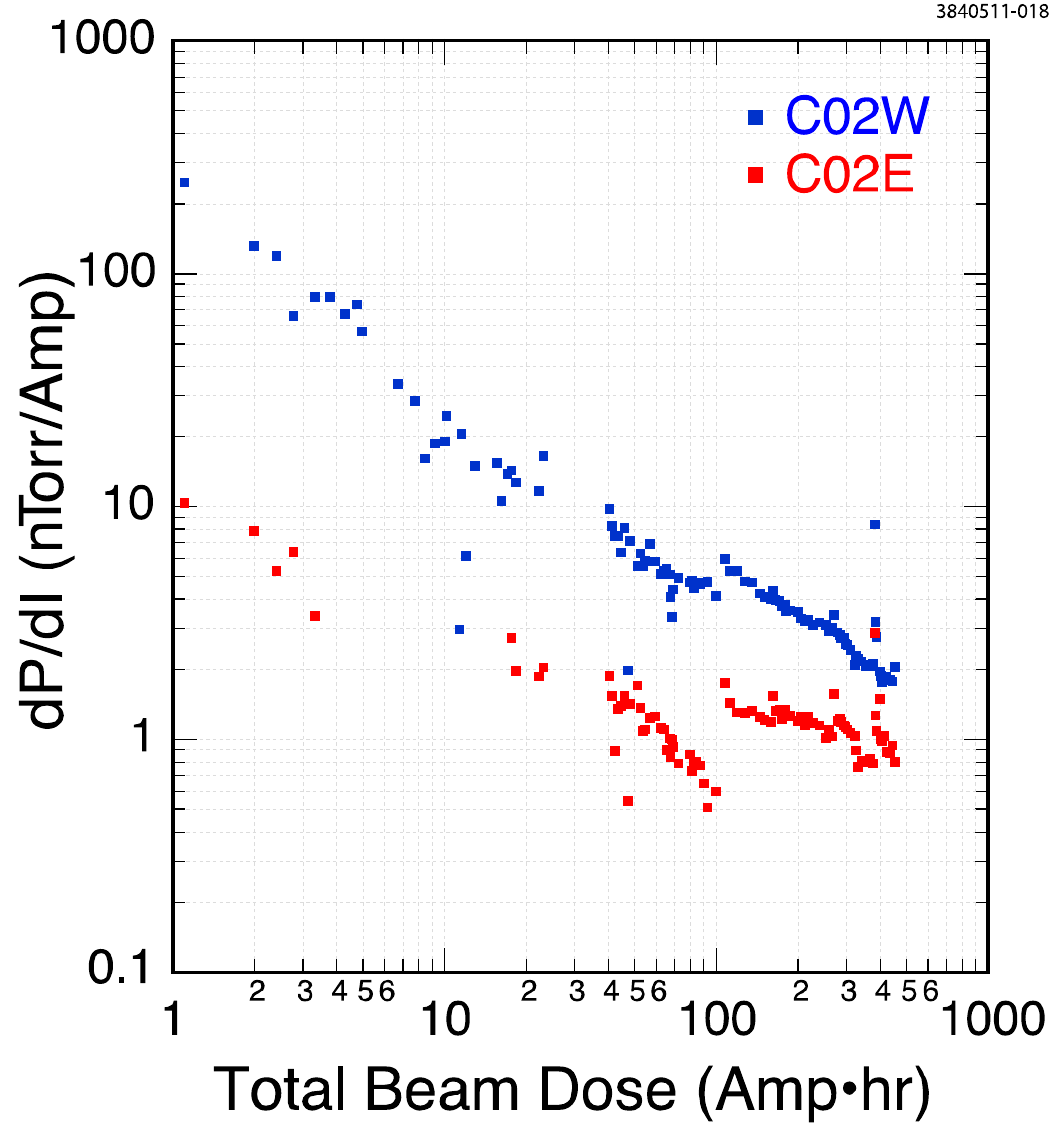}\\
\end{tabular}
    \caption[Beam-Induced Pressure Rises in L0]{Beam induced pressure rise (measured in $10^{-9}$~Torr/A) in the L0 EC experimental region during CHESS operations as a function beam dosage. The locations of the vacuum gauges are shown in Figure~\ref{fig:cesr_conversion:vac_l0_pumping}.  The synchrotron radiation for this region has a critical energy of 1.97~keV for 5~GeV beam energy.
\label{fig:cesr_conversion:l0_dpdi}}
\end{figure}

Most of the new vacuum chambers in L0 region are made of copper extrusions (the same as used for the CESR-c SCW beam pipe) and have cooling channels welded on both sides of the beam pipes.  Achieving ultra-high vacuum (UHV) during both the {\cesrta} and CHESS operations is essential in the vacuum system design for this experimental region.  Figure~\ref{fig:cesr_conversion:vac_l0_pumping} shows installed vacuum pumps and gauges.  The two types of UHV pumps in use are sputter ion pumps (SIPs) and non-evaporable getter pumps (NEGs).  At each SCW location one NEG and one small SIP are installed at a single pumping port.  At the center of this region a large SIP was installed.  The 6~NEGs provide the main vacuum pumping for the region although periodic re-activations of these NEGs are required.  The 7~SIPs provide primary pumping during the initial beam conditioning of the vacuum chambers after a vacuum intervention of the region as well as supplementary pumping of non-gettable gases (such as Ar and CH$_4$).  The vacuum performance of the region is monitored by 5~cold-cathode ion gauges (CCGs) and one residual gas analyzer (RGA). During the initial phase of the {\cesrta} program, the L0 region had been vented with ultra-high purity (UHP) nitrogen three times for installations of RFA-equipped SCWs with different EC suppression techniques.  During the process of SCW installation the existing L0 beampipes were heavily purged with UHP nitrogen.  However, moisture from the ambient air was unavoidably diffusing into the vacuum chambers near the open flange joints, owing to a lack of humidity control in the experimental hall in L0. Although a post-installation in-situ bakeout is not the general practice, for this case the L0 beampipes were heated to $70\,^{\circ}{\rm C}$ with hot water for up to 48~hours after each installation, if the schedule permitted.  As expected, a period of beam conditioning is required to achieve acceptable vacuum level after each vacuum intervention.  Typical beam conditioning behavior is displayed in Figure~\ref{fig:cesr_conversion:l0_dpdi}, where the beam-induced pressure rise ($dP/dI$ in $10^{-9}$~Torr/A) is plotted against accumulated beam dosage (in A~$\cdot$~hr).

\begin{figure}
    \centering
\begin{tabular}{cc}
\includegraphics[width=0.45\textwidth]{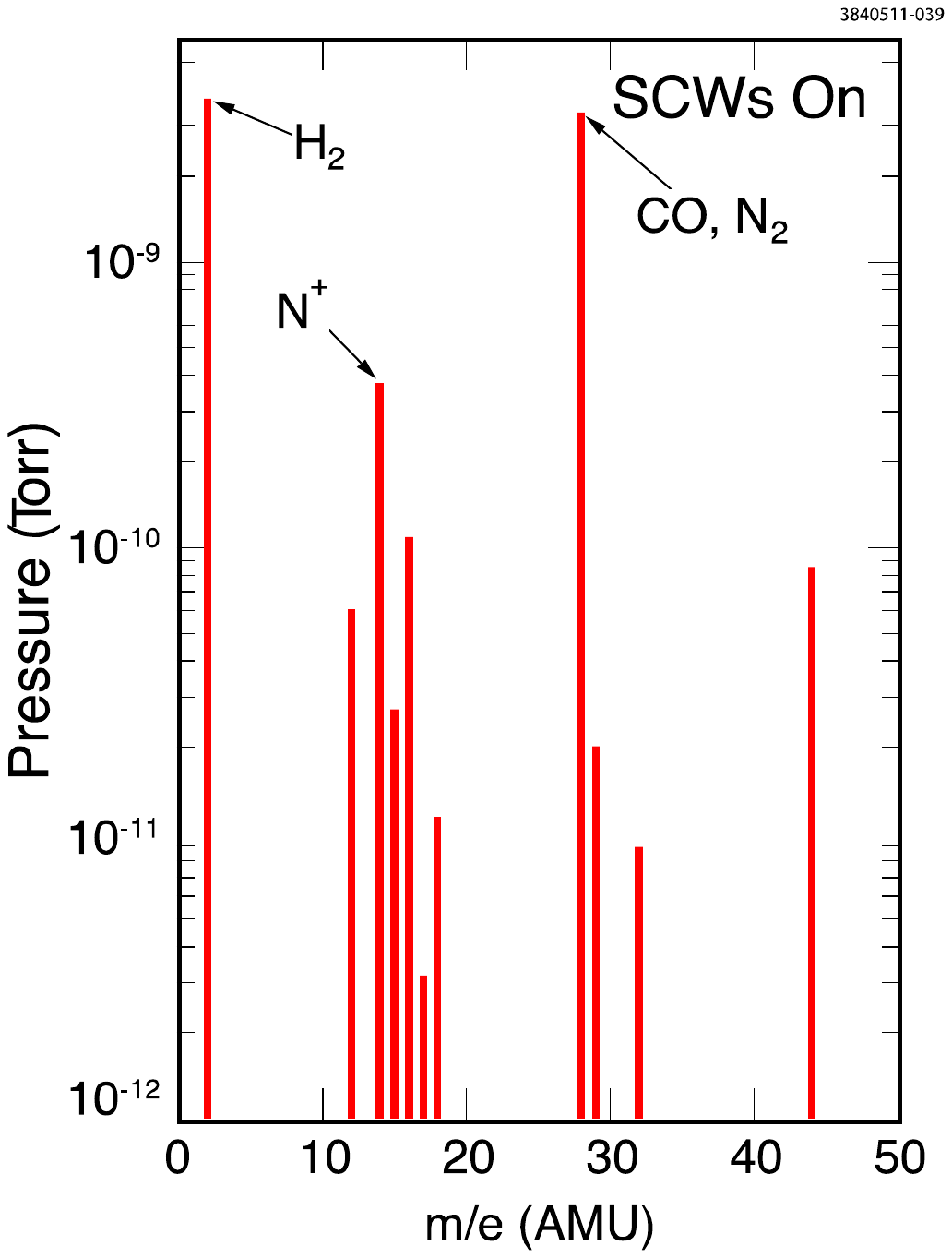} &
\includegraphics[width=0.45\textwidth]{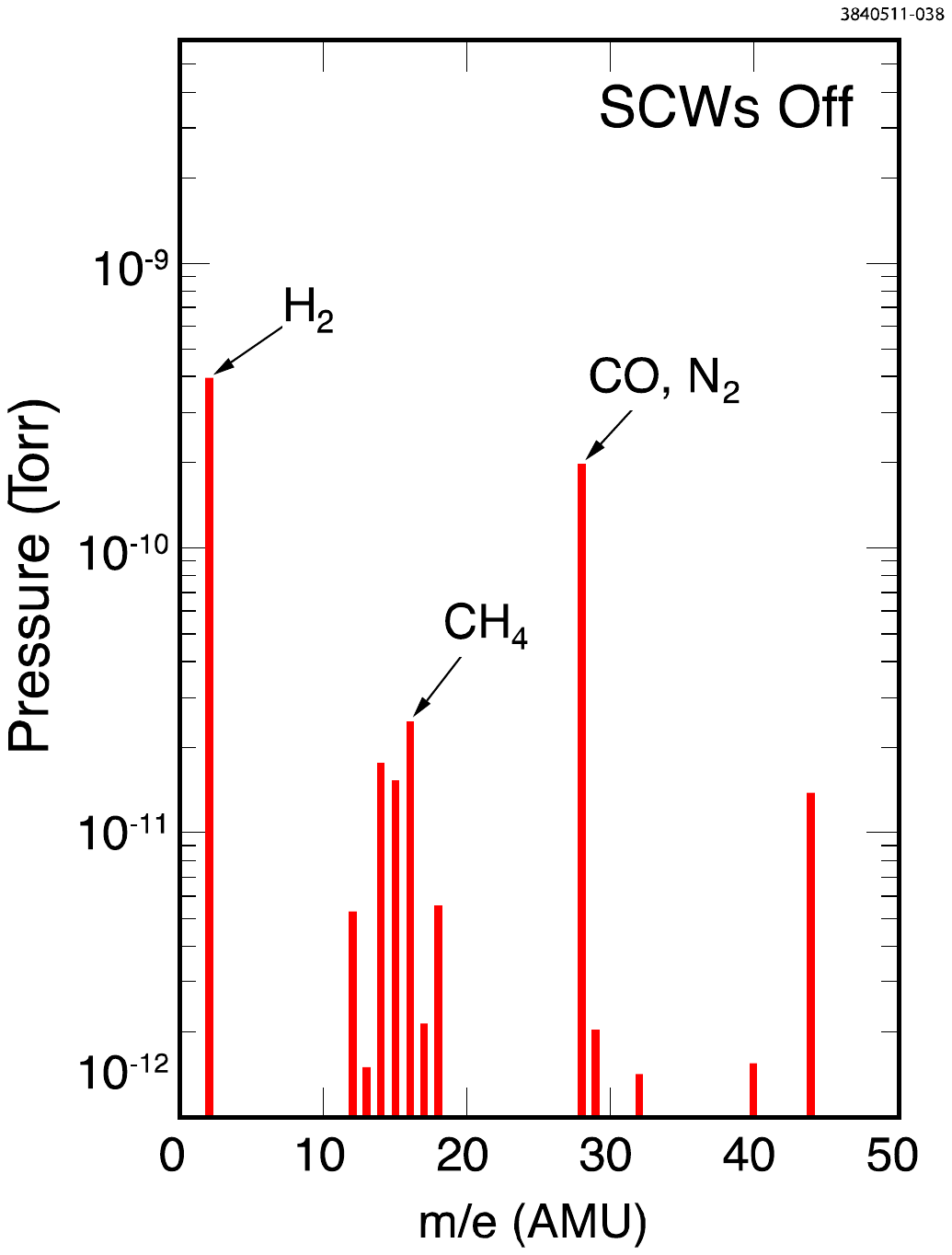}\\
\end{tabular}
    \caption[L0 Mass Spectra]{RGA spectra at L0 Center, with 120~mA positron beam at 4~GeV with SCWs ON (left) and OFF (right).  The relative peak heights of signals at $m/e = $28 and 14 clearly indicated desorption of N$_2$ with SCWs on, due to higher energy SR photons.
\label{fig:cesr_conversion:l0_rga1}}
\end{figure}

\begin{itemize}
    \item Very high pressure rises were observed initially at each startup, due to high SR-induced desorption yield on exposed surfaces.  The beam-induced pressure rise drops quickly with accumulated beam dosage ($D$), typically following ${dP/dI} \propto D^{-\alpha}$, with $\alpha=0.6$ to $1.0$.
    \item The vacuum level in the region usually becomes acceptable for beam operation after accumulating 30~to ~50~A~$\cdot$~hr beam dose at each startup.
    \item As seen in Figure~\ref{fig:cesr_conversion:vac_l0_pumping} the vacuum in the west side of L0 (L0 West), measured by C00W, C01W and C02W, is always significantly worse than the symmetrical locations in the east side of L0 (L0 East) during beam-operations. Note that the base pressures are comparable throughout the entire L0 region.  L0 West hosts three RFA-equipped SCWs, while L0 East hosts three CESR-c SCWs.  Higher SR-induced outgassing rates from TiN coated beam pipes at L0 Center and SCWs is a likely reason for the poorer vacuum pressures in the L0 West.
    \item RGA data at L0 Center indicated desorption of nitrogen, most likely from TiN coated surfaces.  As shown in Figure~\ref{fig:cesr_conversion:l0_rga1} the desorption of N$_2$ is even more evident with higher energy SR photons generated by the SCWs with a 4~GeV positron beam.
\end{itemize}

\begin{figure}
    \centering
    \includegraphics[width=0.9\textwidth, angle=0]{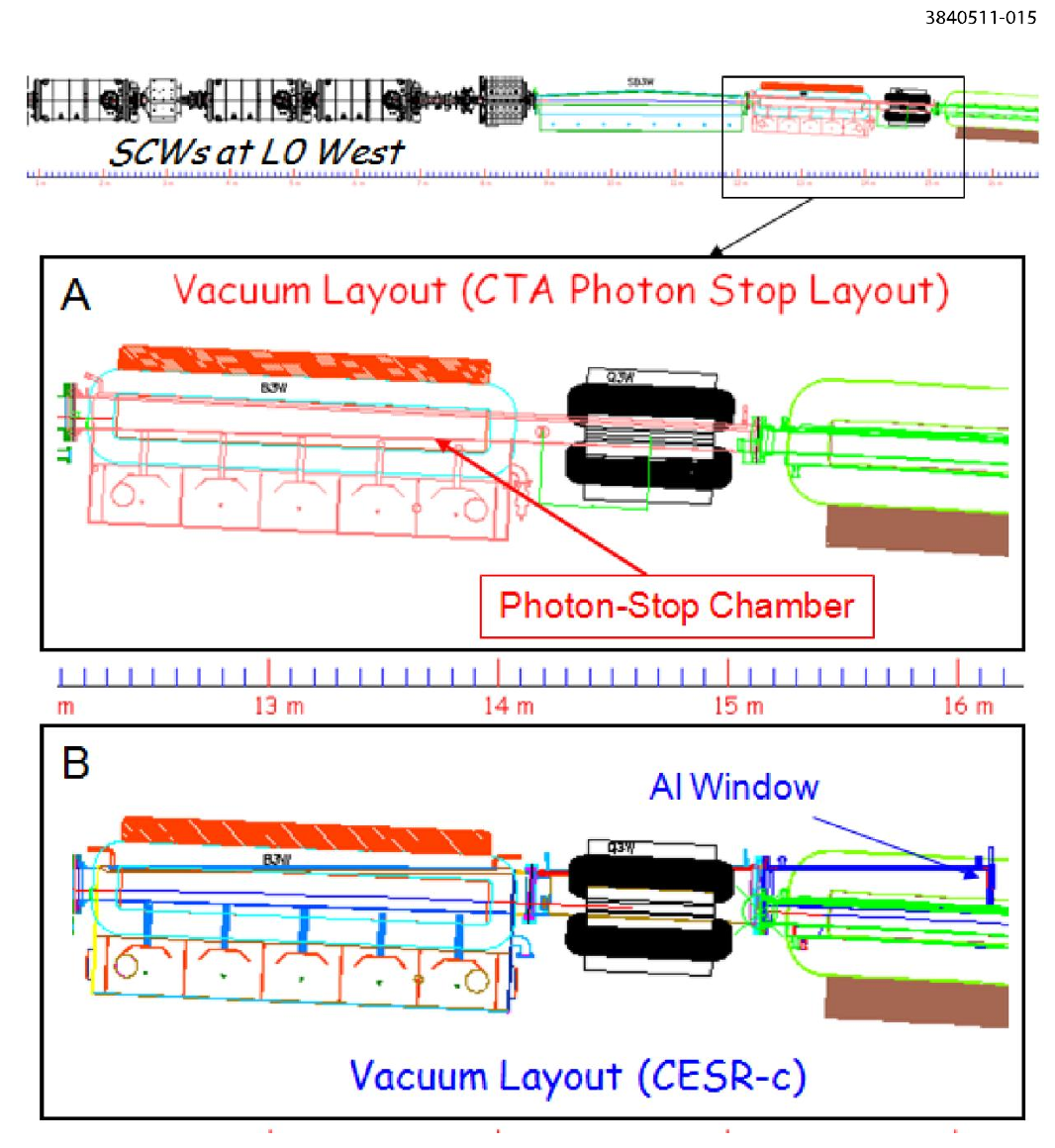}
    \caption[B3W Photon Stopper Layout]{A specially designed photon-stop chamber (A) replaced CESR-c era FLM chambers (B), enabling operations of L0 SCWs with a positron beam at 5 GeV.
    \label{fig:cesr_conversion:photon_stop_vc}}
\end{figure}

\begin{figure}
    \centering
    \includegraphics[width=1.0\textwidth, angle=0]{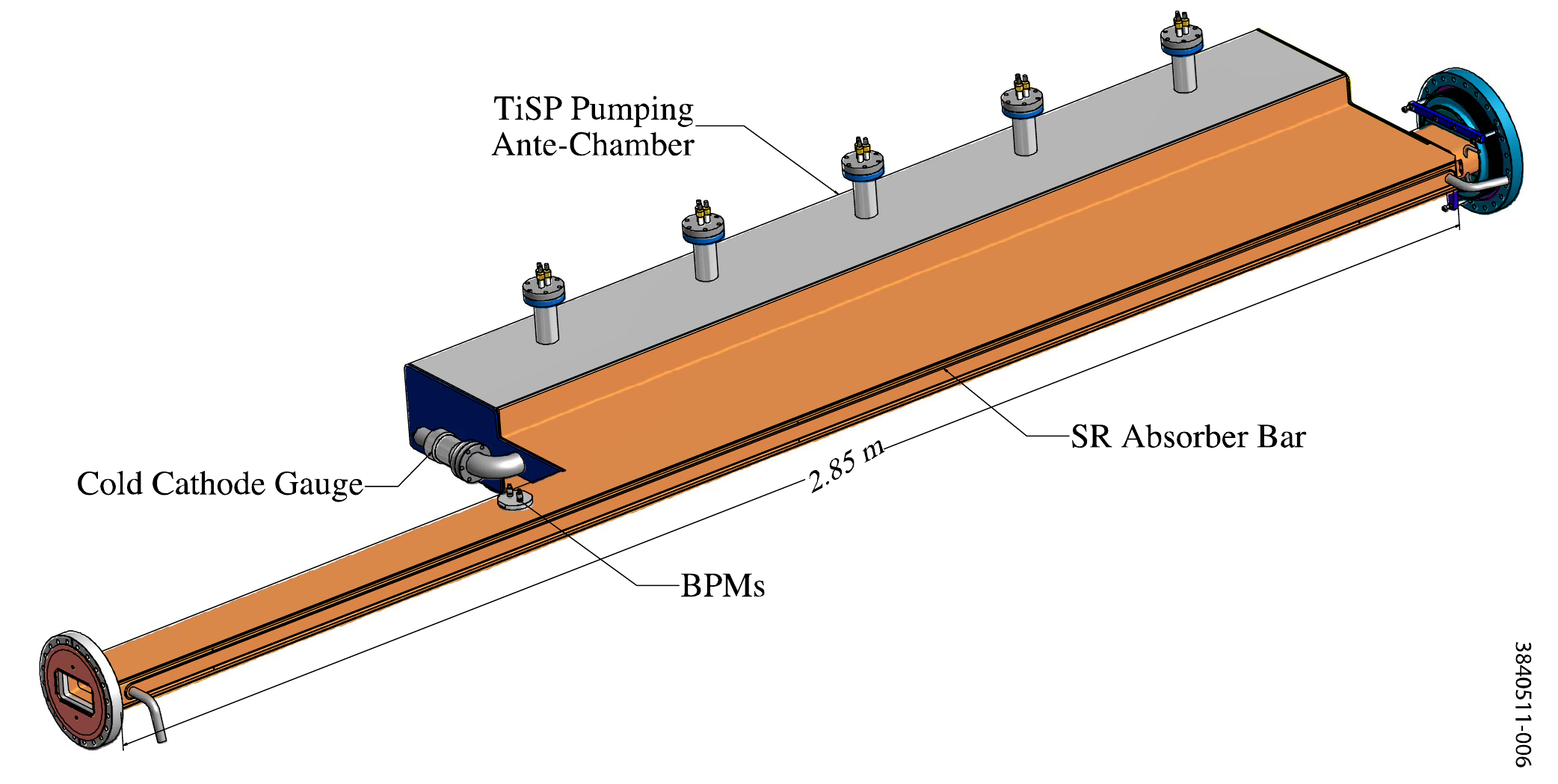}
\begin{tabular}{cc}
\includegraphics[width=0.45\textwidth]{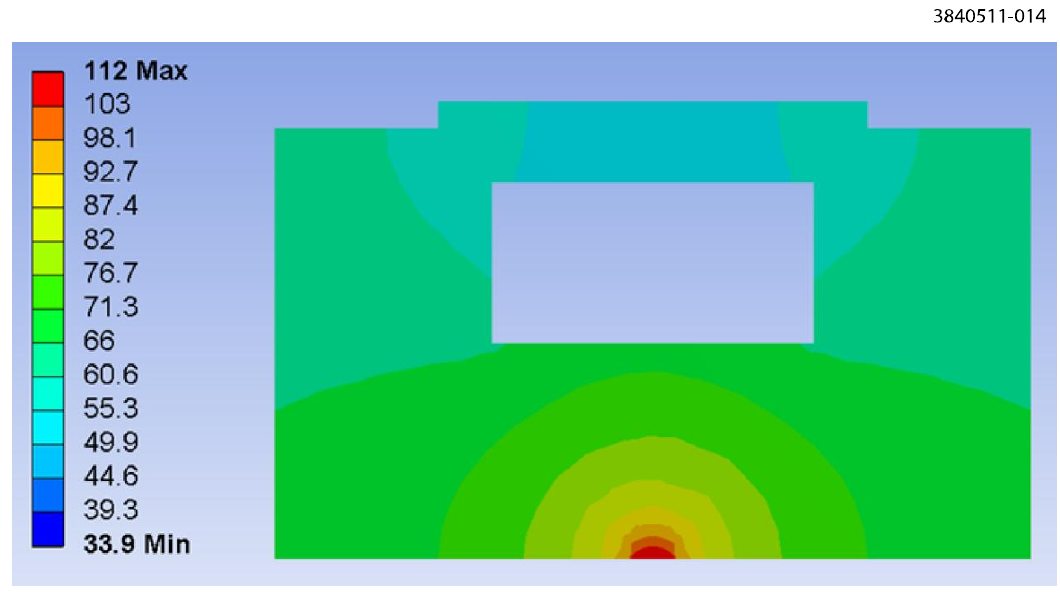} &
\includegraphics[width=0.45\textwidth]{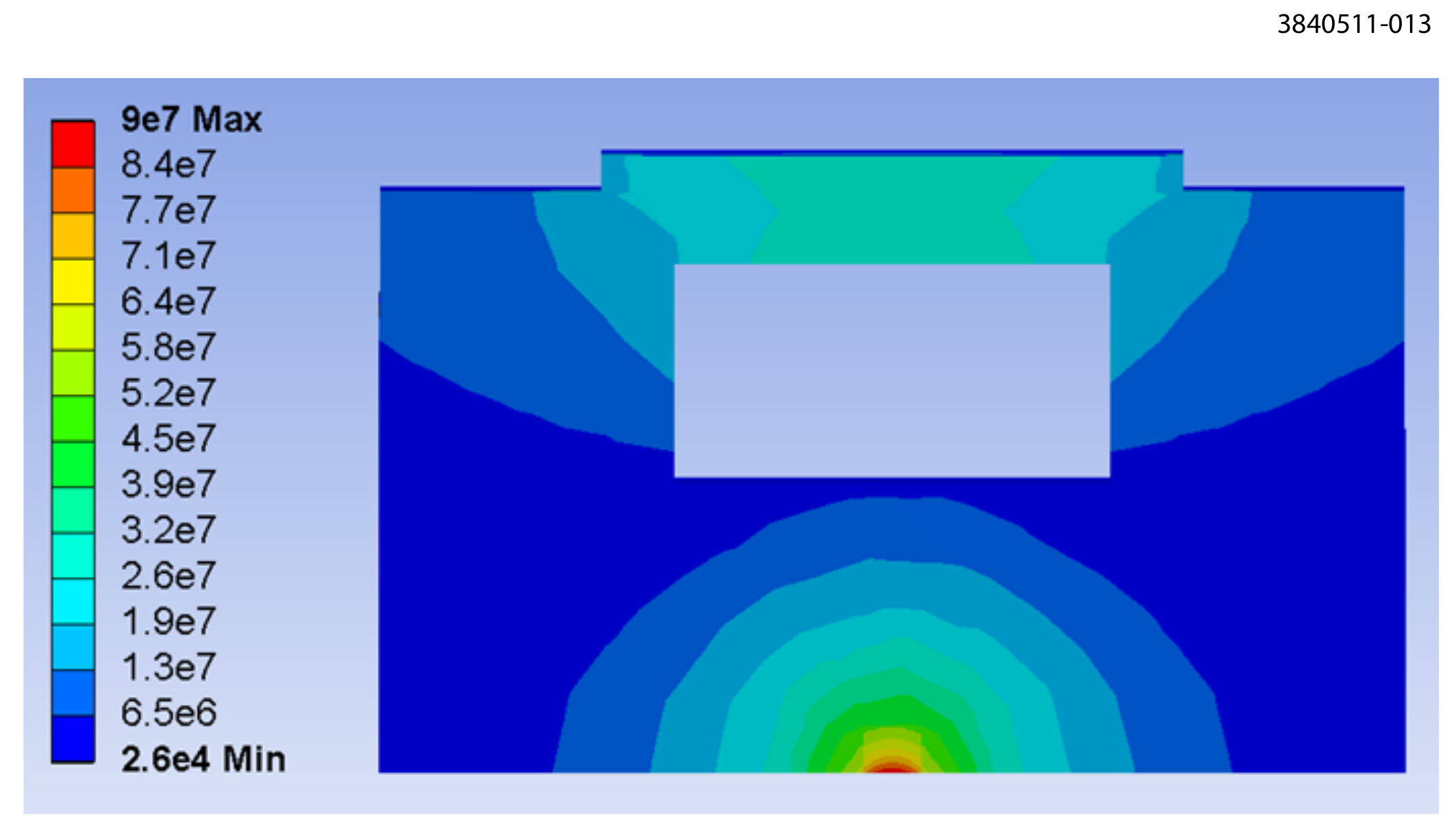}\\
\end{tabular}
    \caption[L0 Photon Stopper Chamber]{Drawing of the photon stop chamber (top); ANSYS calculation of temperature rise on the cooling bar in degrees Celsius (bottom left) and thermal stress in Pascals (bottom right).
    \label{fig:cesr_conversion:photon_stopper}}
\end{figure}

It is critical to study EC buildup and suppression in the wiggler magnetic field for ILC Positron Damping Ring conditions, i.e. with significant positron beam currents at 4~to~5~GeV.  However, the string of six SCWs at the L0 region will generate up to 40~kW synchrotron radiation (SR) power with 100~mA positron beam at the beam energy of 5~GeV.  To deal with this intense SR power, vacuum chambers on the west side of the L0 region (i.e., down-stream for the positron beam relative to the SCW-string) had to be modified.  Figure~\ref{fig:cesr_conversion:photon_stop_vc} shows the layout of the area.  Specifically, a CESR-c era aluminum vacuum chamber, that was a part of CLEO-c fast luminosity monitor (FLM)~\cite{PAC05:RPPE044}, could not safely stop this high SR power from the SCW-string.  A new photon-stop (PS) chamber was constructed to replace FLM chambers.  The PS chamber, as shown in Figure~\ref{fig:cesr_conversion:photon_stopper}, is made of OFHC copper, with a 2.85~m long water-cooled bar to intercept the main SR power from the SCW-string and with a large Ti-sublimation pump ante-chamber to deal with expected gas-load from SR-induced desorption.  To determine the effect of the 40~kW of SR power from the L0 SCWs, the longitudinal distribution of this power was calculated on the absorber for this worst-case condition using the simulation code, SYNRAD\cite{synrad:manual}, and this yielded peak incident surface power from SR of 6~W/mm$^2$ at 5~GeV.  The ANSYS code was utilized to perform thermal analysis for this case (also shown in Figure~\ref{fig:cesr_conversion:photon_stopper}) and it verified the design has a thermal stress less than one third of the yield stress of the absorber's material.  This factor of three for safety allows the absorber to handle the expected SR power.  To ensure proper cooling, eight thermocouples (TCs) are attached to the PS chamber.  The measured temperature rises by these TCs agreed reasonably well with the values as predicted by the ANSYS calculation.

%==================================================================%

\subsubsection{Arc Test Regions}
\label{sssec:cesr_conversion.vac_system.ec_exp_regions.Arc}
\begin{figure}
    \centering
    \includegraphics[width=1.0\textwidth, angle=0]{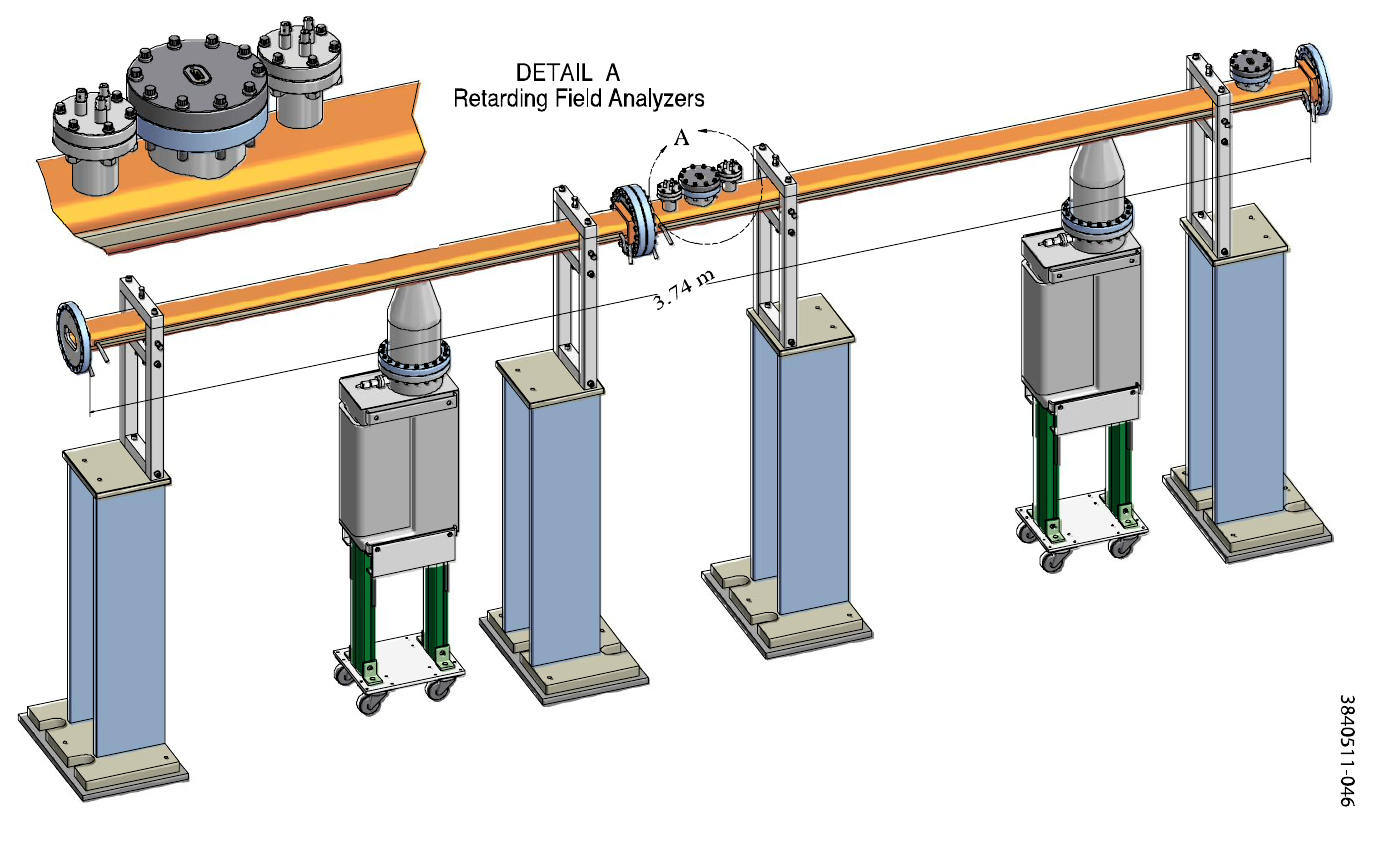}
    \caption[Q14 Test Chambers]{EC experiment section created by the removal of SCW-doublets in CESR.\label{fig:cesr_conversion:q14_scw_replacement}}
\end{figure}

Upon the removal of the CESR-c SCWs from CESR arcs, two EC experimental sections were created on both east and west sides of CESR (see Figure~\ref{fig:cesr_conversion:vac_fig1}).  At the former locations of CESR-c SCW-doublets, a pair of copper beam pipes were installed for each SCW-doublet, as shown in Figure~\ref{fig:cesr_conversion:q14_scw_replacement}.  The longer (2.075~m in length) replacement copper beam pipe of these two was coated with TiN thin film for half of its length (while the other half remained bare copper). Two segmented RFAs were installed at each end of this EC test chamber to compare EC-intensity on TiN coated copper to the bare copper.  Since these EC-test chambers reside in long vacuum sectors in CESR, they are not intended to be frequently replaced.

\begin{figure}
    \centering
    \includegraphics[width=0.9\textwidth]{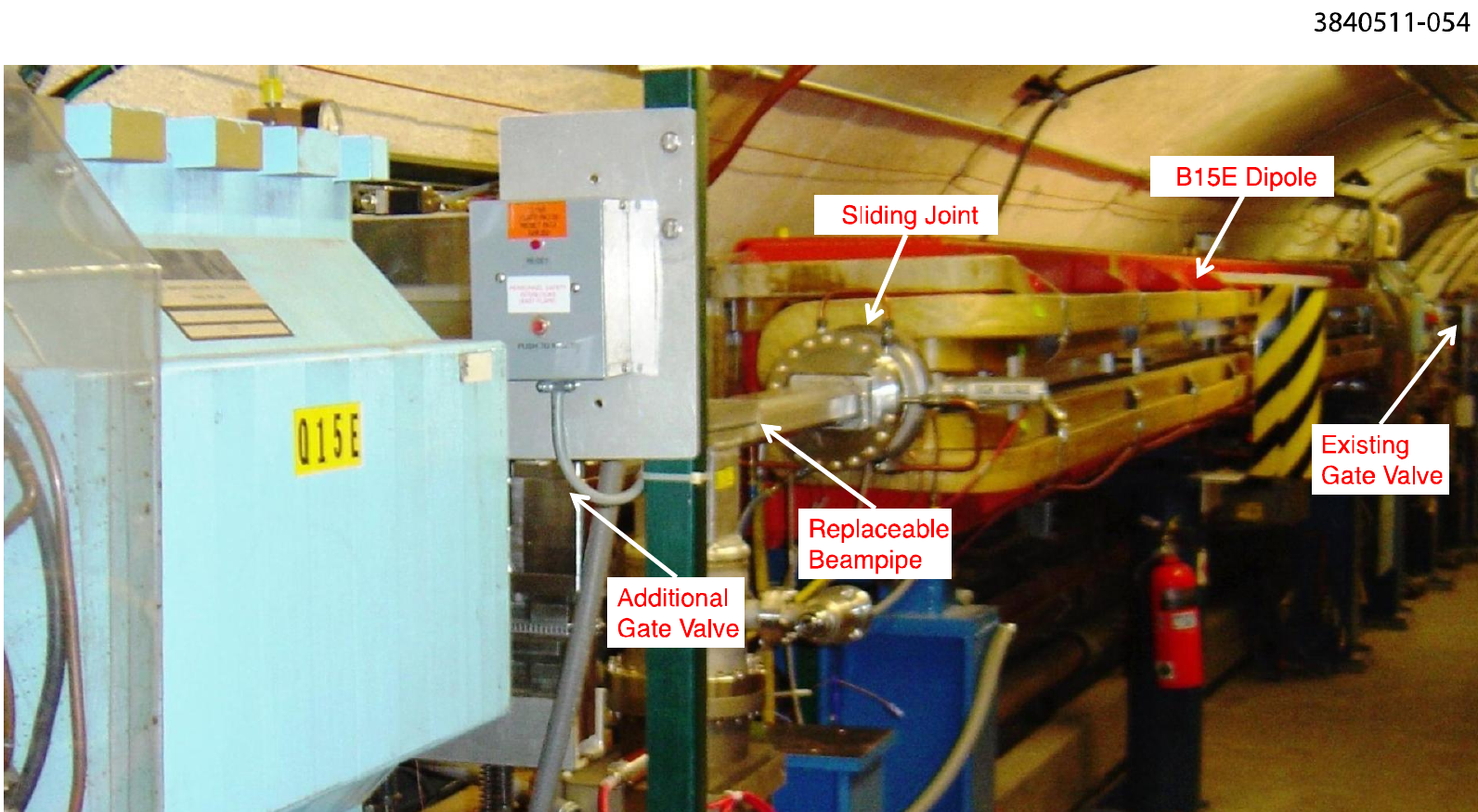}
\begin{tabular}{cc}
\includegraphics[width=0.45\textwidth]{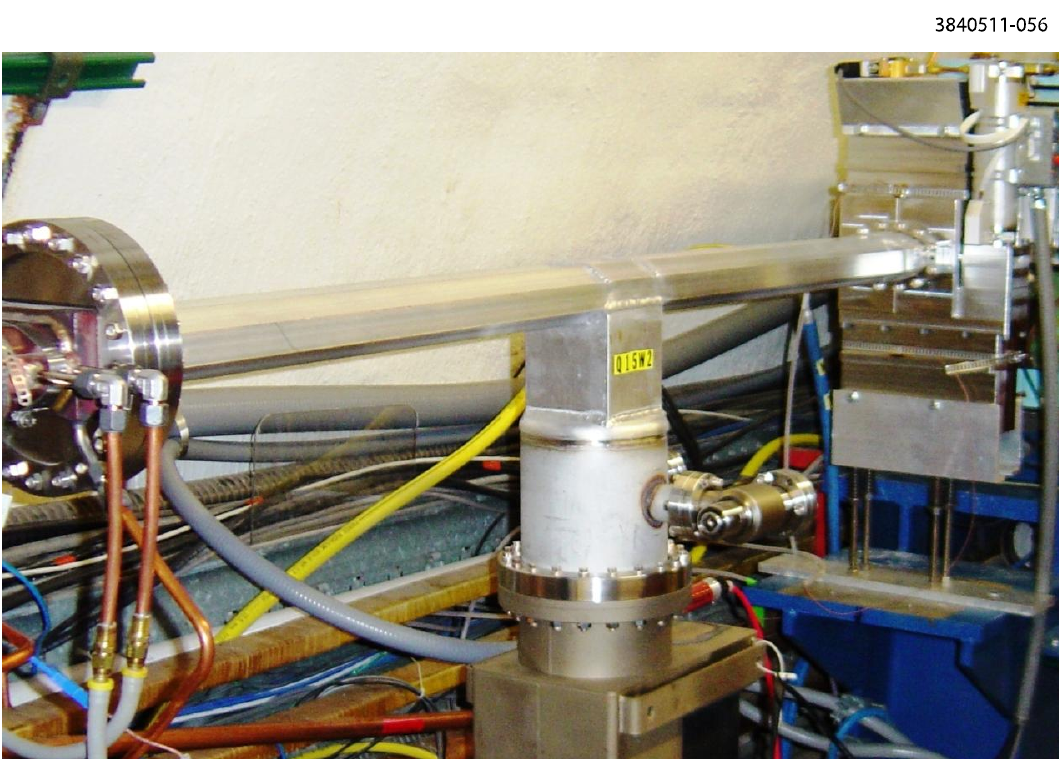} &
\includegraphics[width=0.45\textwidth]{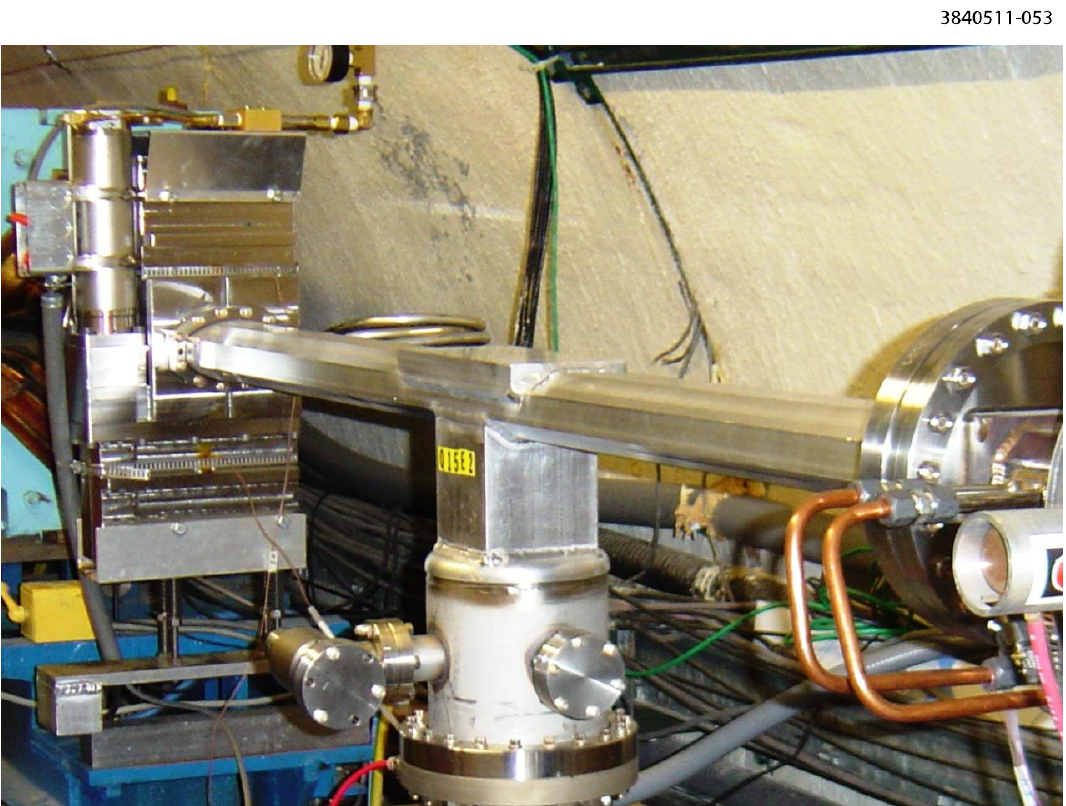}\\
\end{tabular}
    \caption[Q15 EC Test Region]{Two very short EC experimental sections were created in the CESR arcs. Top: photograph showing the chamber adjacent to Q15E, including the downstream bending magnet.  Bottom: close-up photographs of the interchangeable test chambers in Q15W (left) and Q15E (right).
    \label{fig:cesr_conversion:q15_test_sector}}
\end{figure}

To allow frequent exchanges of EC test chambers and to minimize the impact to accelerator operations, two very short experimental regions were created in place of two CESR-c SCWs near the Q15W and Q15E locations in the arcs.  Additional RF-shielded, UHV gate valves were installed in these two regions, so that only very small portions (approximately 8.2~m in length) of the CESR vacuum may be vented to N$_2$ in order to replace a test chamber in these short straight sections.  Each vacuum sector includes only one dipole bending chamber and a short straight section.  Figure~\ref{fig:cesr_conversion:q15_test_sector} shows photographs of the Q15 experimental regions, as they were created during the summer 2008 CESR shutdown.  Many test chambers were rotated through the Q15 test regions during the {\cesrta} program, as described in Section~\ref{ssec:cesr_conversion.vac_system.exp_chambers}.

%=================================================================%
\subsubsection{L3 Test Region}
\label{sssec:cesr_conversion.vac_system.ec_exp_regions.L3}

\begin{figure}
    \centering
    \includegraphics[width=\textwidth]{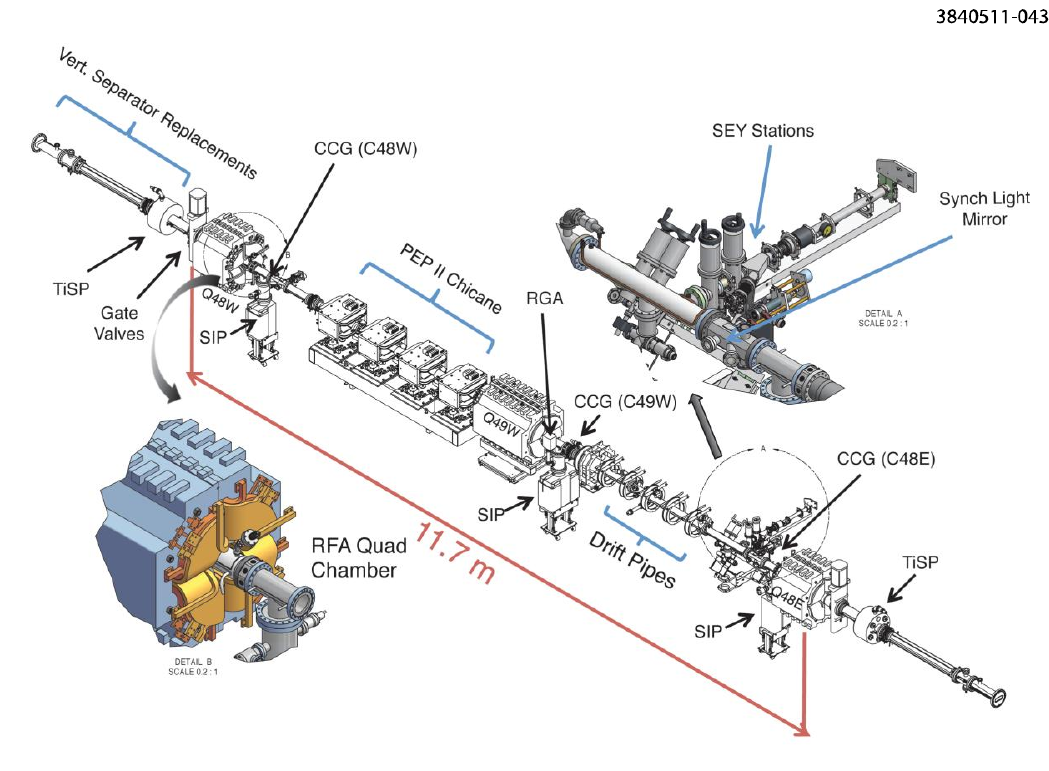}
    \caption{L3 {\cesrta} EC experimental region, showing sputter ion pumps (SIP), Titanium sublimation pumps (TiSP), cold cathode gauges (CCG), a residual gas analyzer (RGA), secondary emission yield (SEY) stations  and a quadrupole retarding field analyzer (RFA). \label{fig:cesr_conversion:vac_l3}}
\end{figure}

The CESR L3 long straight section (see Figure~\ref{fig:cesr_conversion:vac_fig1}) was the site of another area of major modification of CESR vacuum system.  To accommodate the design and fabrication of the vacuum components and to be compatible with the availability of the technical resources of the laboratory, this {\cesrta} EC experimental region (illustrated in Figure~\ref{fig:cesr_conversion:vac_l3}) was constructed during two major CESR shutdowns, as summarized below.

\begin{itemize}
    \item During a 3-month shutdown in the summer 2008, a pair of electrostatic vertical separators were replaced with beam pipes originally used in the CLEO IR region.  These CLEO IR region beam pipes had adequate cooling and vacuum pumping to handle SR from the hard bend (high field) dipole magnets to either side of the L3 region.  These chambers also facilitated smooth transitions of beam pipe cross-section from the rectangular shape at the end of the soft bends to the required 90~mm diameter round pipe.  A pair of RF-shielded all-metal gate valves were also installed in preparation for the planned EC experimental region configuration.
    \item The L3 center EC experimental region (17.94~m in length) was installed during a 2-week long shutdown in the spring 2009.  With gate valve isolation from the rest of CESR, it is sufficiently flexible to allow the venting of this experimental section for {\cesrta} EC studies.  As its location is furthest away from CHESS area, the impact to the CHESS operations is minimal.  Normally, approximately 10~A$\cdot$hr of beam processing is sufficient after each venting of this region.
\end{itemize}

\begin{figure}
    \centering
    \includegraphics[width=0.75\textwidth]{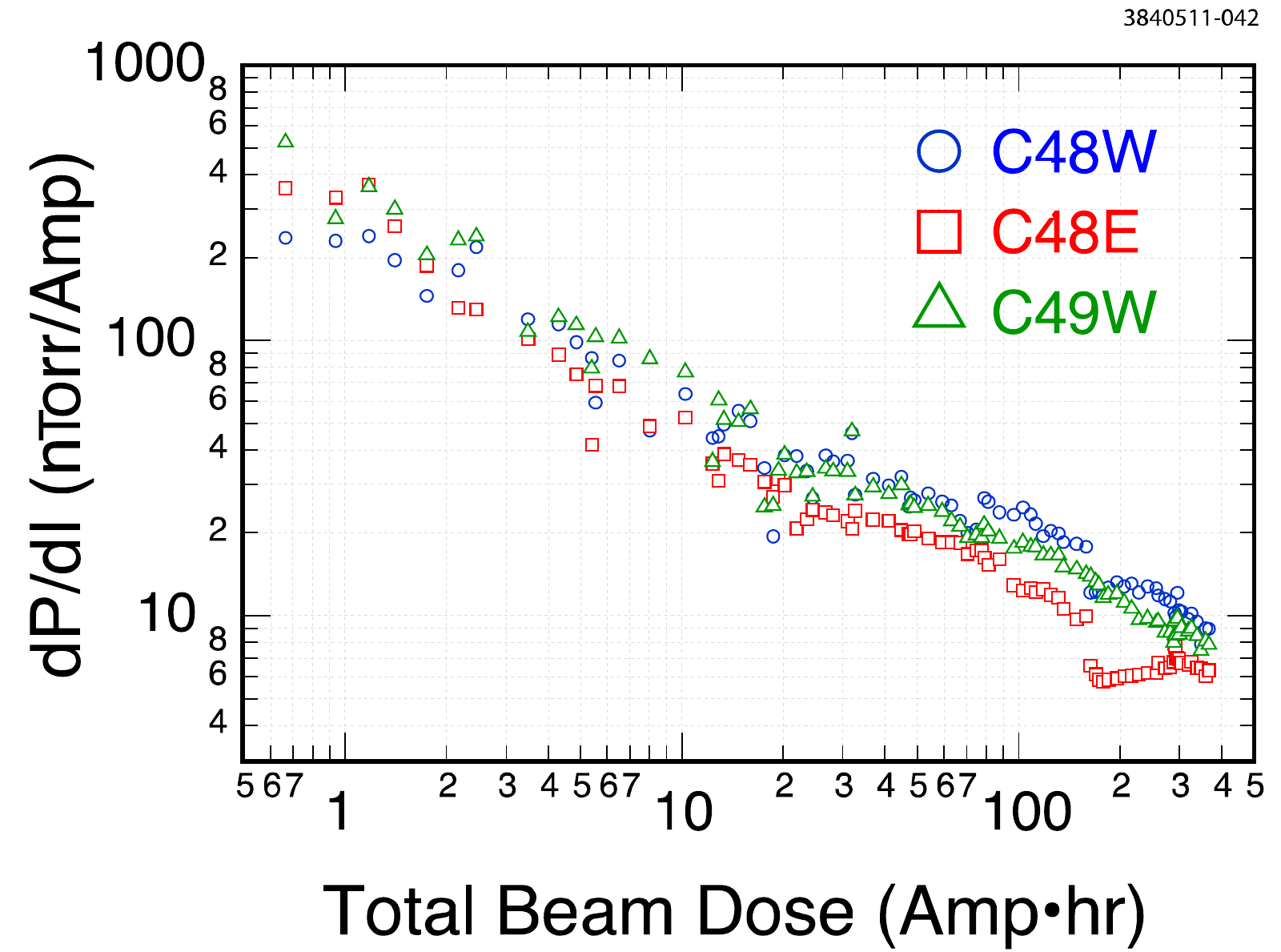}
    \caption[Vacuum beam conditioning in L3 EC experimental region]{Typical vacuum beam conditioning performance for the L3 experimental region \label{fig:cesr_conversion:l3_dpdi}}
\end{figure}

As shown in Figure~\ref{fig:cesr_conversion:vac_l3} vacuum pumping in the L3 experimental region is accomplished by three sputter ion pumps (SIPs) and two Ti-sublimation pumps (TiSPs). Vacuum system performance in the region is monitored by three CCGs, evenly spaced within the straight section (see Figure~\ref{fig:cesr_conversion:vac_l3}), and a RGA at the center.  A typical vacuum beam conditioning performance for the region after venting the L3 experimental region is displayed in Figure~\ref{fig:cesr_conversion:l3_dpdi}.

Many EC studies have been conducted in the L3 EC experimental region with several types of beam- and EC-diagnostic instruments on test chambers in the presence of different configurations of magnetic fields, as listed below.  The details of these test chambers and EC diagnostics are given in Section~\ref{ssec:cesr_conversion.vac_system.exp_chambers}.

\begin{figure}
    \centering
    \includegraphics[width=0.75\textwidth]{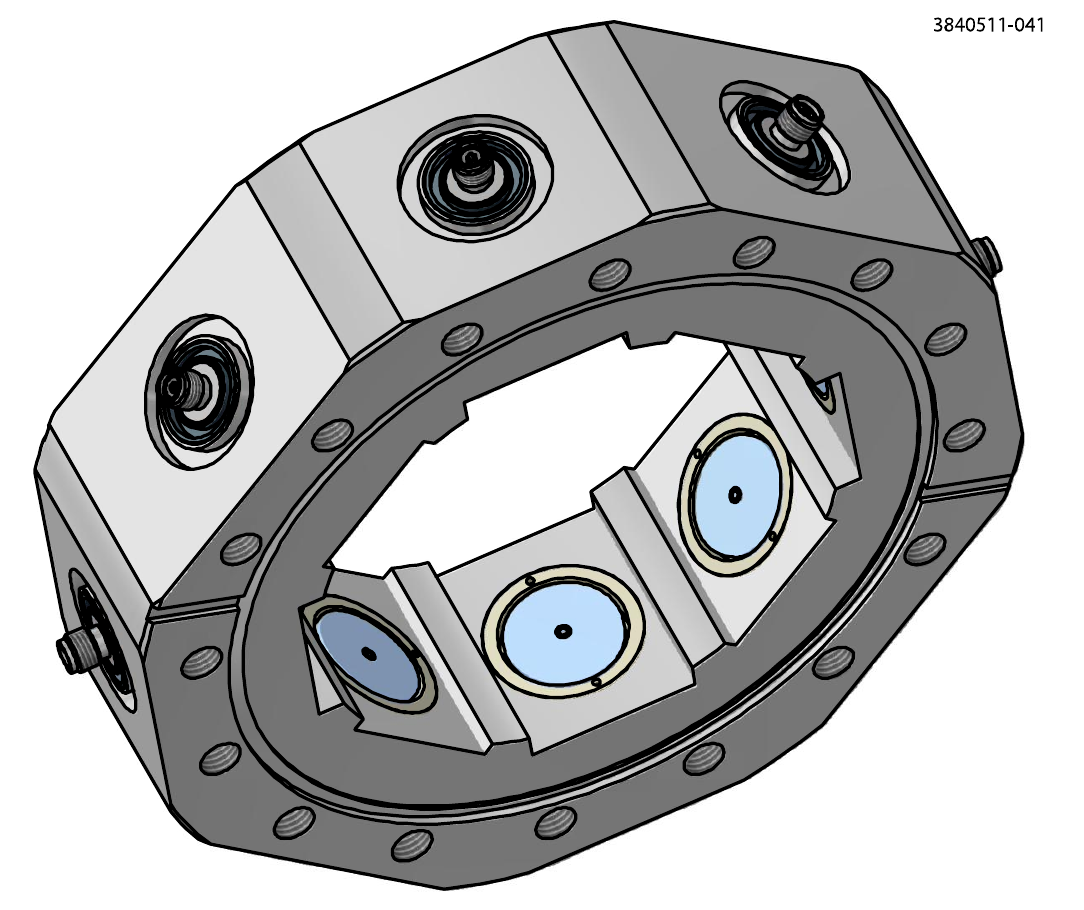}
    \caption[8-Button BPM and TE-wave assembly]{Special 8-button assemblies are used in L3 Experimental region.  The vertical and horizontal buttons were used as TE-wave transmitters/receivers, while the other four are used as BPM pickups. \label{fig:cesr_conversion:bpm_tw_button}}
\end{figure}

\begin{itemize}
    \item Diagnostics chambers in dipole magnets (formerly the PEP-II Chicane).  Chambers, which were tested in this location, included extruded aluminum chamber with rectangular grooves and another with a TiN-coating.
    \item Drift test chambers, including an aluminum chamber with rectangular grooves (PEP-II chambers) and stainless chambers with NEG coating.
    \item A quadrupole chamber fitted with thin-style RFA at Q48W, first with bare aluminum surface and then with TiN-coated aluminum.
    \item Synchrotron light mirrors for independent electron and positron bunch length and transverse beam size measurements.  The highly polished beryllium mirrors are retractable to avoid damage from high SR power during the CHESS operations.  Fully enclosed delivery systems guide the SR light to an optics table containing a Streak-camera and CCD camera.
    \item An in-situ SEY measurement system allows test samples to be exposed to SR during beam operations, primarily in CHESS operations, and then the SEY of the exposed samples are measured in scheduled accelerator accesses.  The in-situ SEY measurement system consists of a PEP-II beam pipe assembly housing  the SEY apparatus with two Cornell SEY measuring stations.
    \item Special beam-buttons assemblies (with 8 pickup buttons) served as a TE-wave measurement system for EC studies, as shown in Figure~\ref{fig:cesr_conversion:bpm_tw_button}.
    \item An RF-Shielded pickup provided by the LBNL team.
\end{itemize}

% --- Subsection: Experimental Vacuum Chambers
\subsection{Experimental Vacuum Chambers}
\label{ssec:cesr_conversion.vac_system.exp_chambers}

During {\cesrta} vacuum system reconfiguration and throughout the
initial phases of the {\cesrta} program, many new EC experimental
vacuum chambers were cycled through CESR. EC diagnostics and EC
suppression techniques were integrated into these experimental
chambers.  The EC diagnostics included RFAs, SPUs, beam-buttons (for
both BPMs and TE-wave).  The EC suppression techniques included
coatings, grooved surfaces and a clearing electrode.  In this
section we will describe the design, construction and vacuum
characteristics for these experimental chambers.  The functionality
and the associated EC measurement performance of the EC diagnostics
and efficacy of the EC suppression techniques are presented
elsewhere \cite{NIMA770:141to154}.

%% MGB corrected reference here.

In the following sub-sections the EC experimental chambers are cataloged by the type of magnetic fields where the chambers resided: the field-free drifts, the dipole fields and the quadrupole field.

\subsubsection{Drift Chambers}
\label{sssec:cesr_conversion.vac_system.exp_chambers.drift}

A variety of {\cesrta} vacuum chambers were installed in field-free drifts to complete the beam transport beamlines during the vacuum system conversion.  To maximize the usage of available space in the drifts, many functional components were integrated into these new chambers, including vacuum pumps and gauges, beam instrumentation (BPMs), EC diagnostics (RFAs, SPUs, etc.) and EC-suppression coatings (TiN).  In this sub-section only the EC experimental chambers, which were deployed in the dedicated test regions, are described in detail.  For the other chambers, in most cases they were installed only one time (and then not interchanged with other chambers) in the longer vacuum sectors in CESR, where there would have been a major impact to CESR and CHESS operations with additional vacuum interventions.  A list of experimental chambers within these longer vacuum sectors is presented here for completeness.

\begin{itemize}
    \item Copper chambers at Q14W and Q14E locations in CESR, as shown in Figure~\ref{fig:cesr_conversion:q14_scw_replacement}, as replacement beam pipes for the CESR-c SCW-doublets.  Insertable segmented RFAs (a Cornell thin-style design, using flexible printed circuits as electron detectors) were installed in these chambers in the early stage of the {\cesrta} program to test and verify the thin-style RFA design.  These are described in one the accompanying papers.
    \item Almost all the copper chambers in the L0 experimental region (see Figure~\ref{fig:cesr_conversion:vac_l0_center}) are fitted with beam buttons (as BPMs and as TE-wave transmitters/receivers) and with the insertable RFAs.
\end{itemize}

% \paragraph{CESR Chambers} %

\paragraph{Q15E and Q15W Regions' Test Chambers}

These regions have been used extensively for the study of various passive coatings, including TiN, amorphous carbon (a-C) and diamond-like carbon (DL-C), in order to evaluate their EC-suppression effectiveness, as well as their vacuum performance in an intense SR environment.  These studies are in collaboration with the CLIC/CERN and the KEK groups.

\begin{figure}
    \centering
    \includegraphics[width=1.0\textwidth, angle=0]{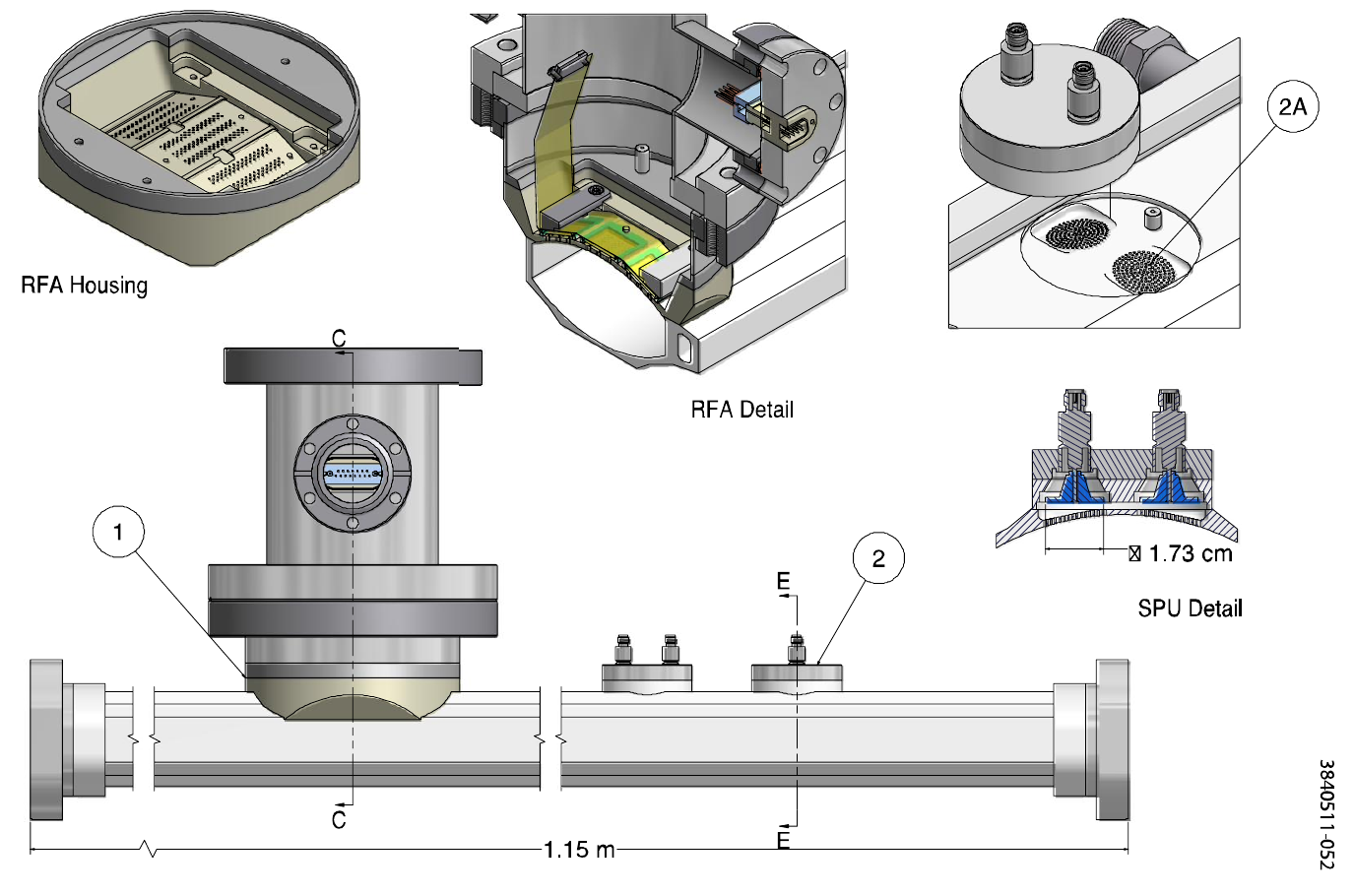}
    \caption{Q15 EC Test Chamber, equipped with a RFA (1) and 4 SPUs (2) \label{fig:cesr_conversion:Q15_vc}}
\end{figure}

To fulfill the above research goals, an experimental chamber design was developed to allow the characterization of the EC growth and decay and its transverse distribution within the vacuum chamber for different wall surfaces and during beam-processing over the course of time.  The design of this EC experimental chamber is illustrated in Figure~\ref{fig:cesr_conversion:Q15_vc}.  The beam pipe is machined from a standard CESR aluminum (Type 6063-T6 alloys) extrusion. For EC measurements an RFA port and a set of 4 SPUs are added to the chamber. The two sets of SPUs are directly welded to the top of the beam pipe, using explosion-bonded aluminum-to-stainless steel blocks. Arrays of small holes (0.75~mm in diameter) that connect the SPUs to the beam space are directly drilled vertically through the beam pipe top wall (approximately 2.5~mm in thickness), as shown. Backing plates made of similar materials were used in drilling these tiny holes to reduce the incidence of burrs.  The RFA housing is machined from a separate block of explosion-bonded aluminum-to-stainless steel material, and is welded to the cutout on top of the beam pipe.  The lower face of the RFA housing matches the curvature of the beam pipe aperture. Small holes are drilled through the three milled-flat surfaces, connecting the RFA port to the beam space.  These RFAs, having holes similar in geometrical dimensions as the SPU holes (0.75~mm in diameter and approximately 2.5~mm in thickness,) are grouped into three `segments' on each flat with each segment containing 44 holes (giving a total of 9$\times$44~=~396 holes).  The segmented hole pattern allows the sampling of the transverse distribution of the EC in the beam pipe.  The dimensions of the RFA and SPU holes are chosen to ensure no significant leakage of the beam's RF fields into the SPU and RFA, while allowing as much transmission of cloud electrons from the beam pipe into the SPU and RFA as possible.

\begin{figure}
    \centering
    \includegraphics[width=0.75\textwidth]{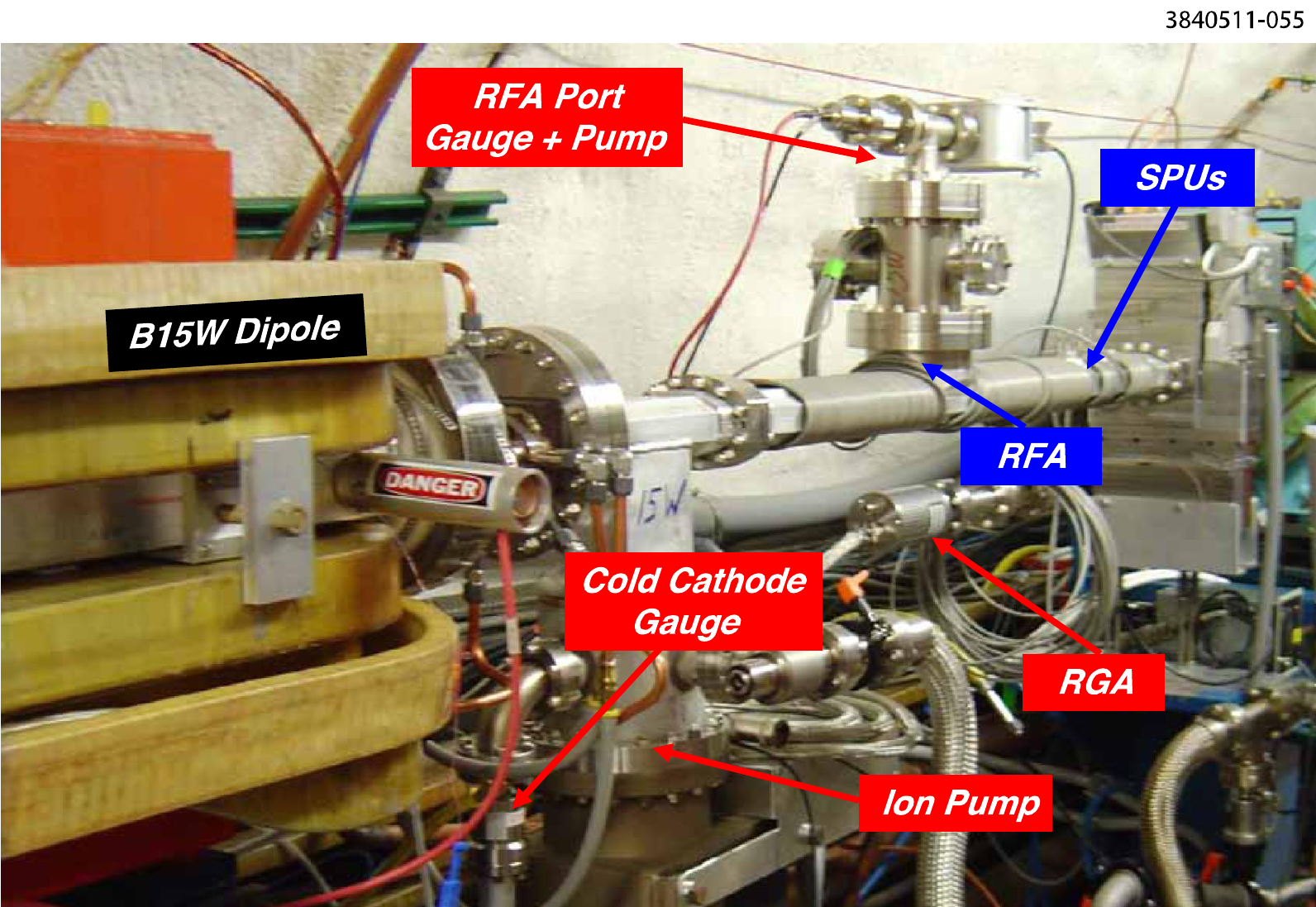}
    \caption{A Q15 EC experimental chamber installed at Q15W in CESR\cite{NIMA760:86to97}. \label{fig:cesr_conversion:Q15W_installed}}
\end{figure}

\begin{figure}
    \centering
    \includegraphics[width=0.75\textwidth]{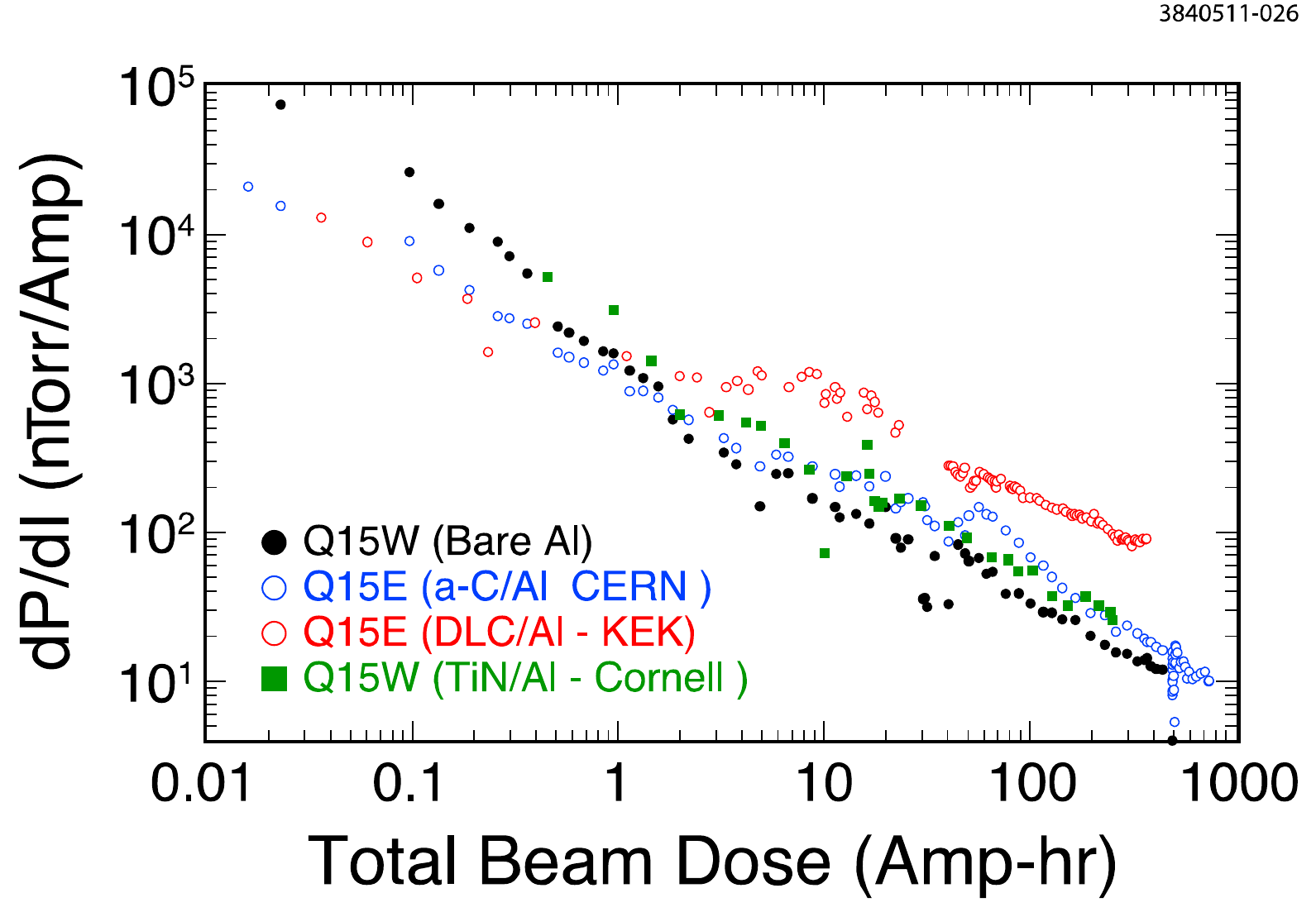}
    \caption{History of beam-induced vacuum conditioning of four Q15 EC test chambers with different surfaces. The synchrotron radiation for this region has a critical energy of 4.73~keV for 5~GeV beam energy.\label{fig:cesr_conversion:Q15_dPdI}}
\end{figure}

\begin{figure}
    \centering
\begin{tabular}{cccc}
    \includegraphics[width=0.229\textwidth]{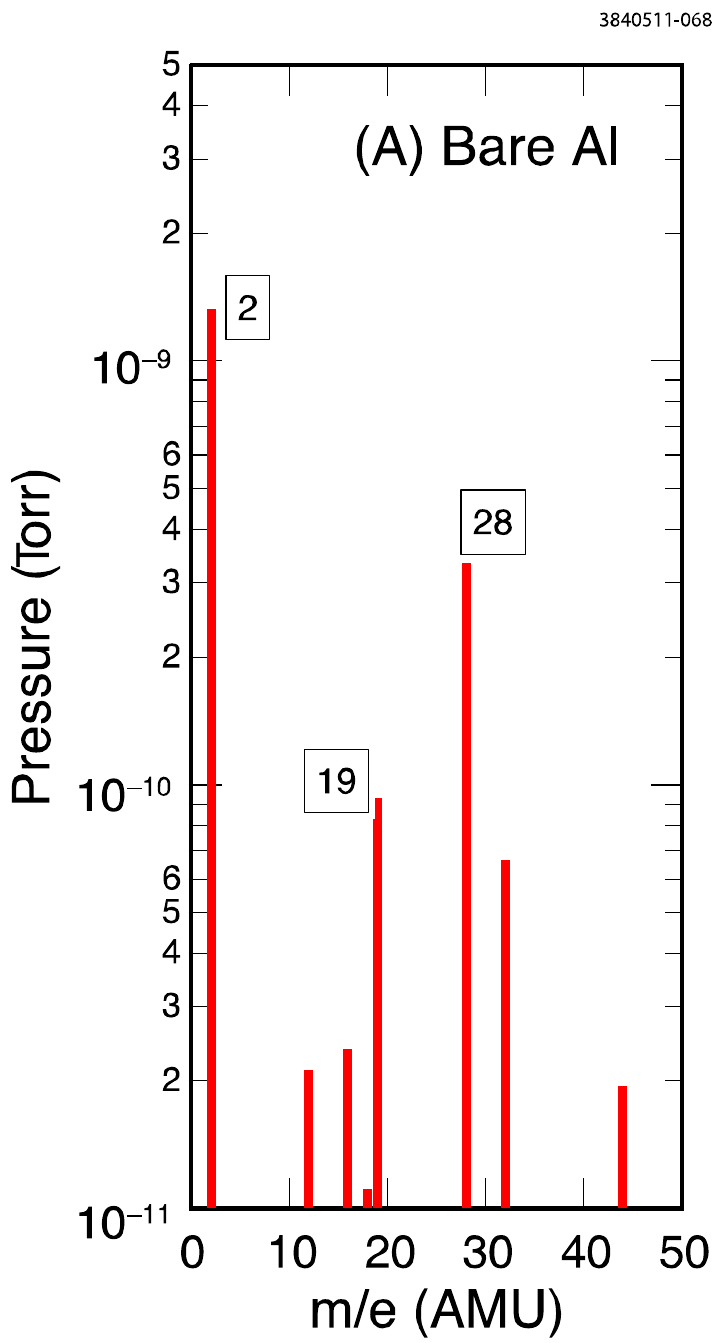} &
    \includegraphics[width=0.241\textwidth]{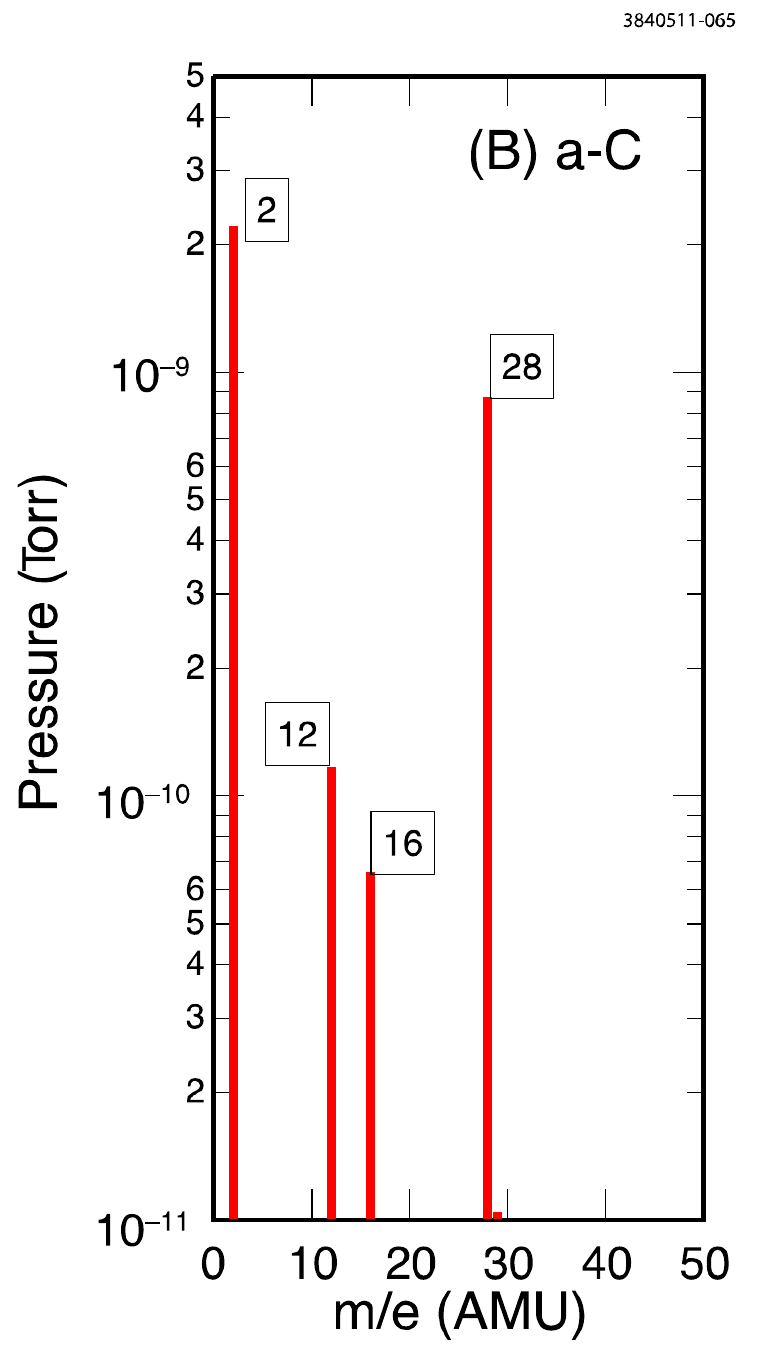} &
    \includegraphics[width=0.238\textwidth]{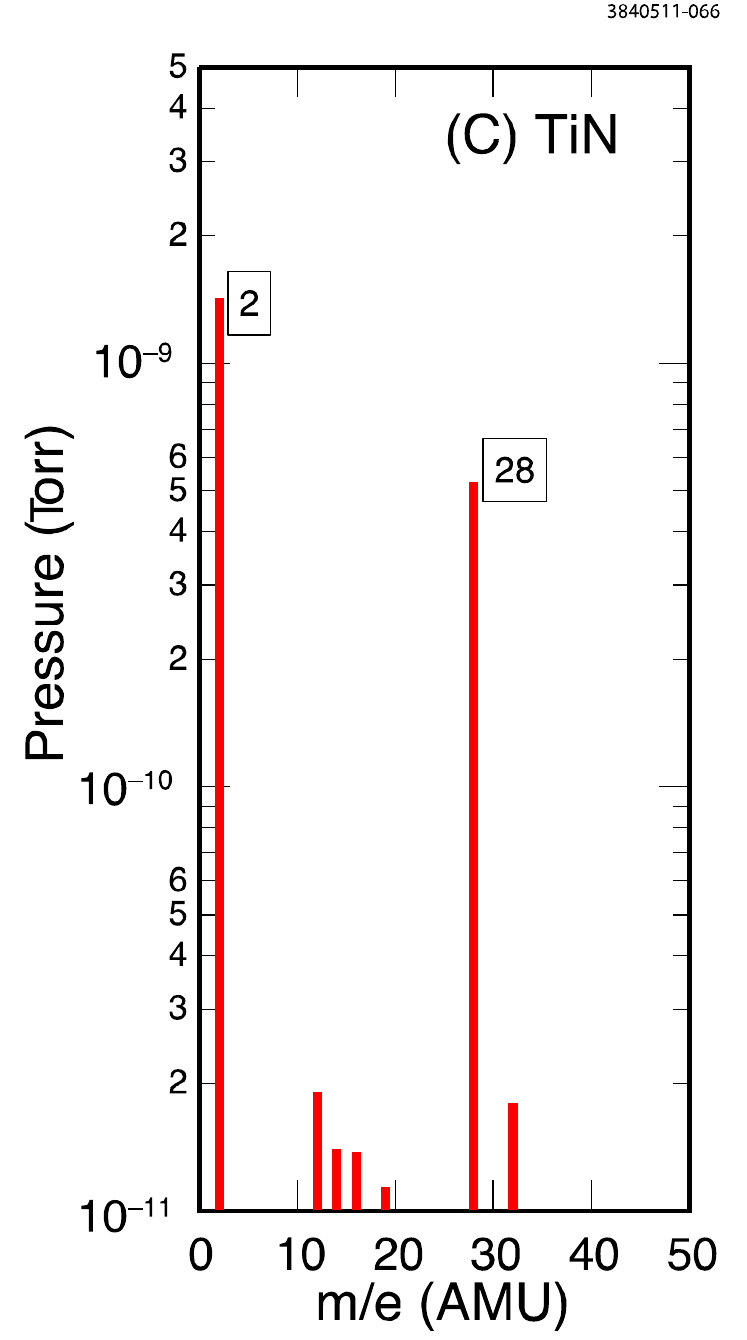} &
    \includegraphics[width=0.233\textwidth]{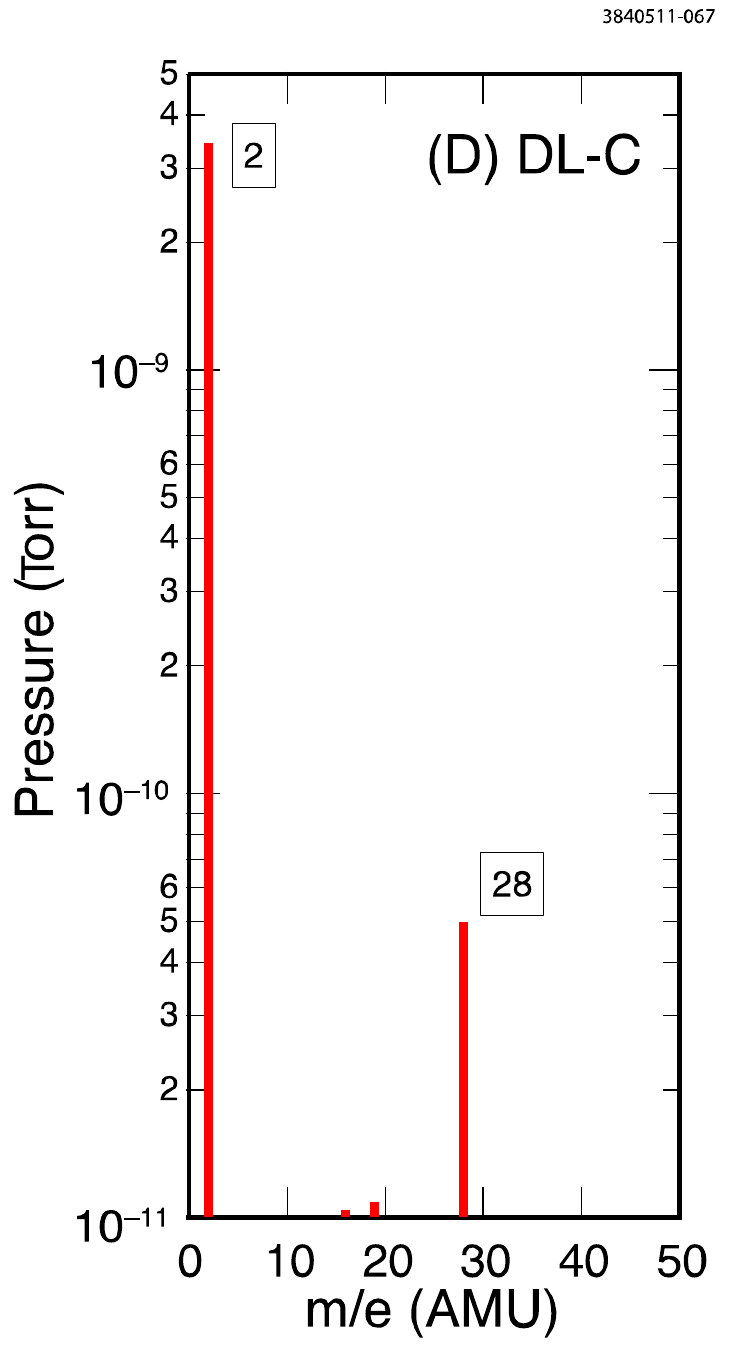} \\
\end{tabular}
    \caption[RGA Spectrum of Q15 Test Chambers]{Typical RGA spectra for four Q15 EC test chambers with different surfaces, as labeled.  All RGA spectra were recorded during CHESS operations with approximately 400~mA stored beam current in CESR and after significant beam processing (with the beam dose over 400~A$\cdot$Hr).  The same RGA sensors (Model MicroVision 2, MKS Instruments) were used at Q156W and Q15E locations with factory calibrations.  (Refer to Table~\ref{tab:Q15_chamber_table} for EC-chamber test periods and locations.)
\label{fig:cesr_conversion:Q15_RGA}}
\end{figure}

Over the first 3-year {\cesrta} program, five Q15 experimental chambers were fabricated and tested in the Q15W and Q15E regions.  Among these experimental chambers four types of interior surfaces and two types of RFA designs were tested.  The four types of tested surfaces are: bare aluminum (as it was originally extruded), amorphous carbon coatings (coated by CERN/CLIC), TiN coating (by Cornell) and diamond-like carbon coating (by KEK).  Table~\ref{tab:Q15_chamber_table} summarizes these test chambers.  Figure~\ref{fig:cesr_conversion:Q15W_installed} shows a typical installation of these experimental chambers.  The vacuum pumping of the test chamber is by a 110-l/s noble-diode ion pump, and the two adjacent distributed ion pumps at B13W/E and B15W/E.  Since the gas conductance between the beam space and the RFA port is very limited, a small ion pump (8-l/s) was installed for the RFA port.  The vacuum performance of each test chamber is monitored by a cold-cathode ion gauge (CCG) and an RGA during the beam runs.  SR-induced gas desorption from the chamber surfaces dominates the gas load.  As for all newly installed vacuum chambers, very high SR-induced pressure rises were initially measured from these experimental chambers, but the SR-induced desorption decreases rapidly with the accumulated beam dose.  In Figure~\ref{fig:cesr_conversion:Q15_dPdI} the beam conditioning characteristics of the four surfaces are compared.  To make the SR-induced desorption measurements from the four types of surfaces, all the data points shown were taken during CHESS operations, when there are roughly equal stored electron and positron beam currents, making similar total SR fluxes at both the Q15W and Q15E locations.  The data in Figure~\ref{fig:cesr_conversion:Q15_dPdI} indicate:
\begin{itemize}
    \item All coatings, except the DL-C, have similar beam conditioning characteristics when compared to the bare aluminum surfaces.  By contrast the DL-C coating indicates significantly higher outgassing rates.
    \item However, the RGA data in Figure~\ref{fig:cesr_conversion:Q15_RGA} show that DL-C coating has a much `cleaner' desorbed gas composition, i.e. it is dominated by hydrogen.
\end{itemize}

\begin{table}
    \caption{Summary of Q15W and Q15E Experimental Vacuum Chambers (VCs) \label{tab:Q15_chamber_table}}
    \begin{minipage}{\textwidth}
    \begin{center}
        \begin{tabular}{|c|c|c|c|c|p{0.220\textwidth}|}\hline
VC &    Surface &   RFA Style & Test Period &   Location &  Note \\
    \hline
1 & Aluminum &  Thin &  2010.04 to &    Q15W &  Reference \\
 & & & 2011.01 & & surface \\
    \hline
2 & TiN &   Thin &  2009.12 to &    Q15E &  TiN coating via\\
 & Run\#1 & & 2010.01 & & DC sputtering \\
 & & & & & at Cornell \\
    \cline{2-6}
  & TiN &   Thin &  2010.08 to &    Q15W &  Same chamber \\
 & Run\#2 & & 2011.01 & & as above \\
    \cline{2-6}
  & TiN &   Insertable &    2011.02 to &    Q15W & Cross-compar- \\
 & Run\#3 & & 2011.07 & & ison of two \\
 & & & 2010.04 & & RFA designs \\
    \hline
3 & a-C\#1 &    Thin &  2009.12 to &    Q15W & a-C coating via \\
 & & & & & DC sputtering \\
 & & & & & at CERN\footnote{A 150$^\circ$C pre-installation bakeout was performed as a standard CESR practice after the installation of the thin RFA.  An a-C coated coupon went through the bakeout and was sent back to CERN for analysis.  The peak SEY value of approximately 1.5 was measured on the coupon, much higher than the other coupons ($\delta\leq$~1.1) at CERN\@.  It was suspected a sub-monolayer trace of silicon on the a-C coating was responsible for the higher SEY\@.  One possible source of the silicon was the silicone adhesive on the UHV-Kapton tape used on the thin RFA assembly.} \\
    \hline
4 & a-C\#2 &    Thin &  2010.04 to &    Q15E &  a-C coating via \\
 & Run\#1 & & 2011.01 & & DC sputtering \\
  & & & & & at CERN\footnote{During this round of RFA installation, efforts were made to reduce adhesive on the Kapton tape by at least 90$\%$, and the preinstallation bakeout temperature was reduced to 120$^\circ$C\@.  However, higher $\delta$ and higher trace levels of Si were still observed in the witness coupon!} \\
    \cline{2-6}
  & a-C\#2 &    Insertable &    2011.07 to &    Q15W &  Cross-compar- \\
 & Run\#2 & & present & & ison of two \\
 & & & & & RFA designs \\
    \hline
5 & DL-C &  Insertable &    2011.02 to &    Q15E &  DL-C coating \\
 & & & present  & & via pulsed DC \\
 & & & & & plasma-CVD, \\
 & & & & & KEK \\

    \hline
        \end{tabular}
    \end{center}
    \end{minipage}
\end{table}

Two generations of RFA designs were used on the Q15 experimental chambers.  The first generation was adapted from the thin RFA design used for a CESR dipole chamber (see Section~\ref{sssec:cesr_conversion.vac_system.exp_chambers.dipole} and Figure~\ref{fig:cesr_conversion:B12W_RFA}). As listed in Table~\ref{tab:Q15_chamber_table} this thin-style design was used in the first four test chambers, including a bare aluminum chamber, two amorphorous carbon coated chambers and a TiN coated chamber (in Runs \#1 and \#2).

\begin{figure}
    \centering
\begin{tabular}{ccc}
\includegraphics[width=0.4\textwidth]{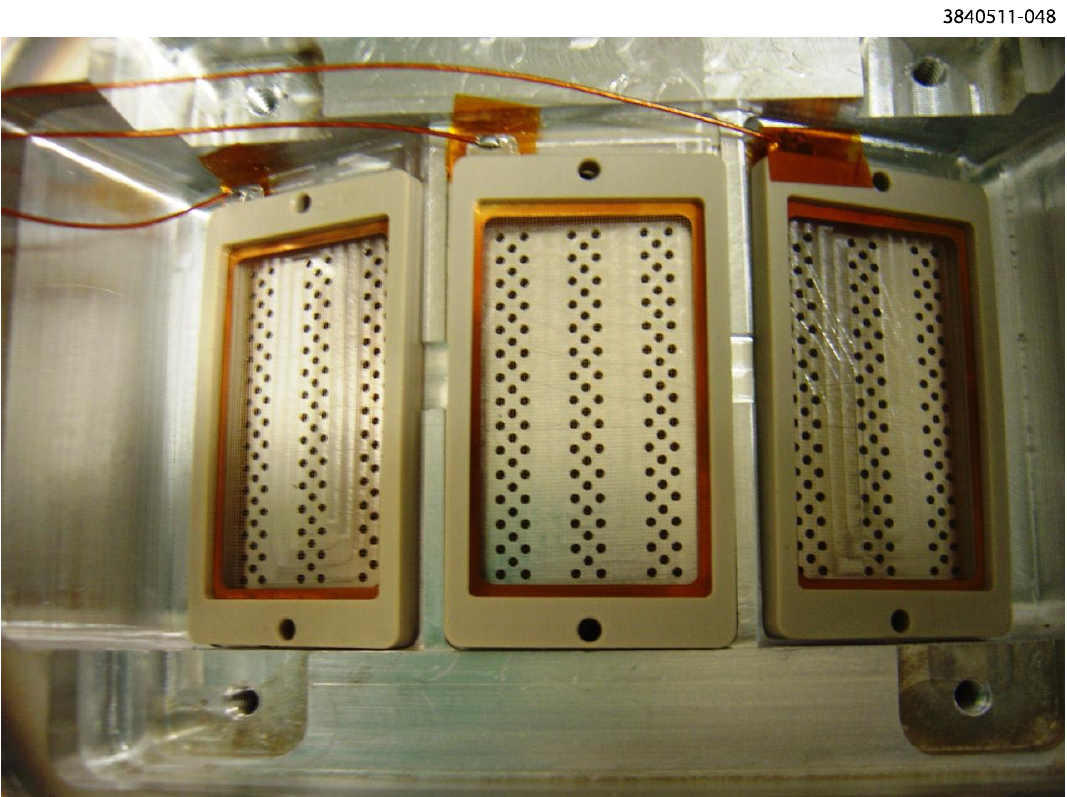} &
\includegraphics[width=0.4\textwidth]{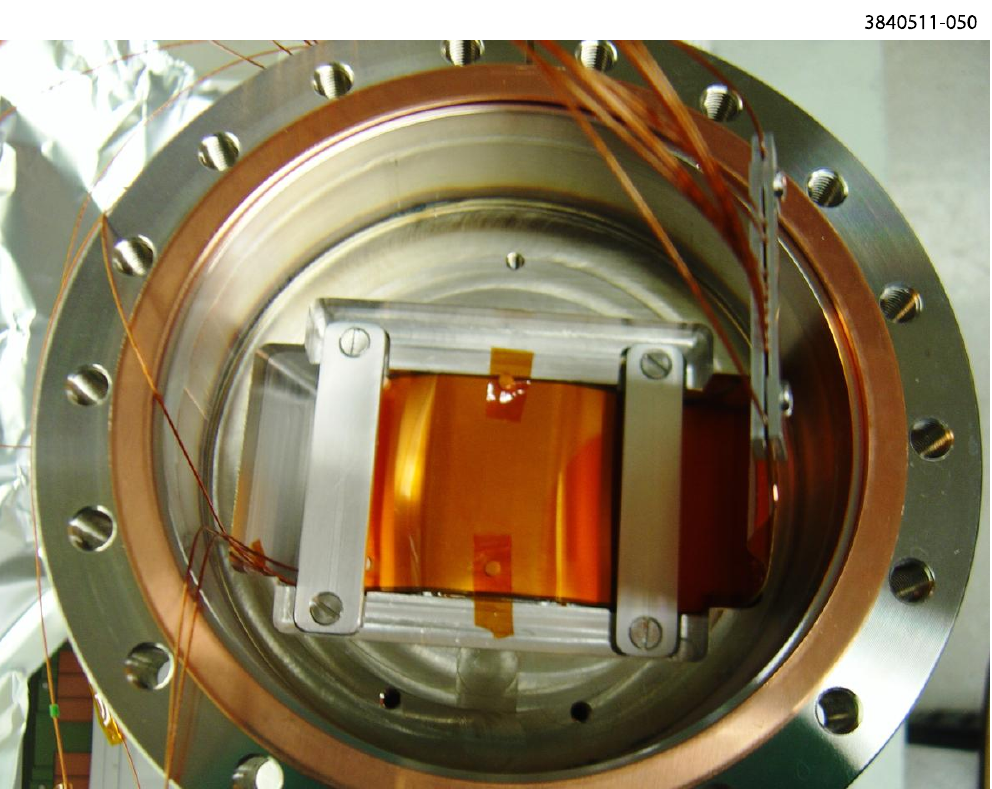} \\
\end{tabular}
    \caption[Cornell Dipole thin-style RFA installed in Q15 Experimental chamber]{Photos of the Cornell Dipole thin-style RFA taken while it was being installed into the Q15 experimental chamber.  Left:  the three high-transparency retarding grids after installation onto the beam pipe.  The beam pipe holes are clearly visible through the fine meshes of the grids.  Right: installation of the collector circuit, which is clamped down with aluminum bars. \label{fig:cesr_conversion:q15_thin_rfa}}
\end{figure}

Photos of a thin-style RFA are shown in Figure~\ref{fig:cesr_conversion:q15_thin_rfa}. In the thin-style design UHV-compatible Kapton tape with Silicone adhesive (Model \#~KAP-TP-36-2S from Accu-Glass Products, Inc.) was used to electrically isolate the flexible RFA collector circuit.  We performed independent vacuum evaluation of the Kapton tape by measuring vacuum total pressure as well as the RGA spectrum of the tape at 230$^\circ$C.  The vacuum tests indicated no unusual outgassing from the tape, thus qualifying their applications in CESR vacuum system.  However, traces (approximately 6$\%$ mono-layer) of silicon were measured on the a-C coated samples that were present during the 150$^\circ$C bakeout of the first a-C coated RFA chamber.  This trace of silicon contamination may have contributed to a much higher measured SEY than was observed on the witness sample.  In the second a-C coated chamber, although the amount of Kapton tape was reduced by more than 90$\%$, an even higher level of silicon contamination and higher SEY was still measured on the witness a-C coated coupons!

\begin{figure}
    \centering
    \includegraphics[width=0.75\textwidth]{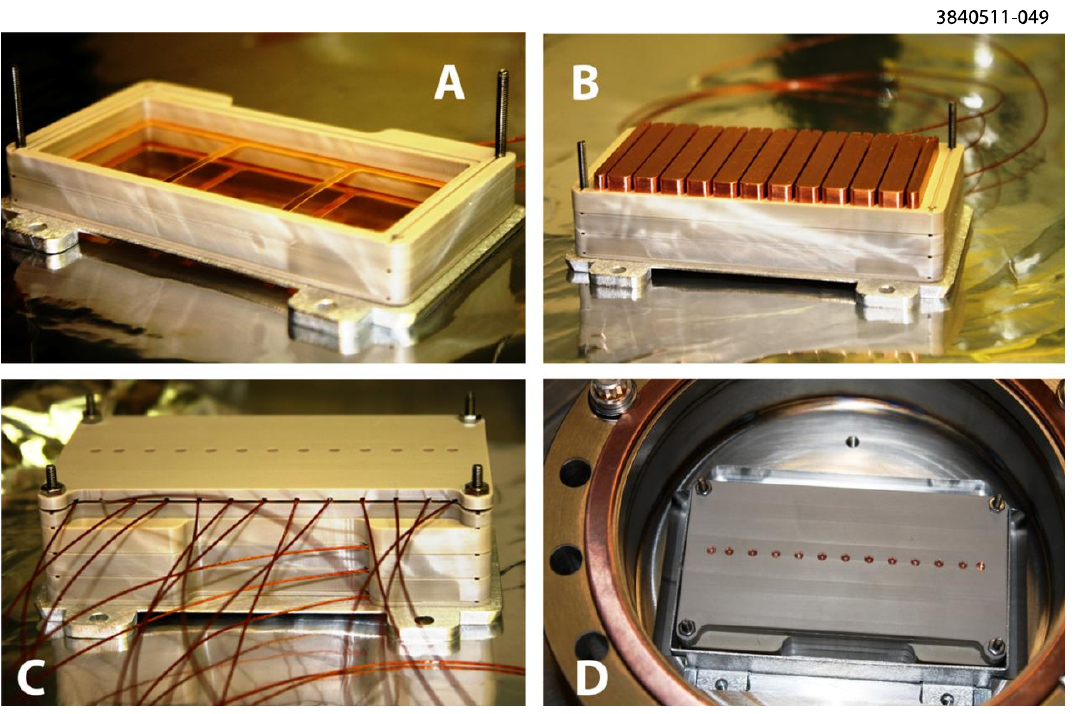}
    \caption[Photographs  of insertable RFA used in Q15 experimental chambers]{Photographs of insertable RFA used in Q15 experimental chambers.  (A)~High-transparency gold-coated copper meshes after mounting in PEEK frames.  (B)~Copper bar collectors mounted above the meshes.  (C)~RFA assembly with PEEK top cap, after soldering all connections (including 2 grids and 13 collectors).  (D)~Insertable RFA in the vacuum port of a test chamber (for clarity, wires are not shown.) \label{fig:cesr_conversion:Q15_insertable_RFA}}
\end{figure}

Thus a second generation of the Q15 RFA design was developed to be completely adhesive-free.  This fully insertable RFA assembly is illustrated in Figure~\ref{fig:cesr_conversion:Q15_insertable_RFA}. The insertable RFA consists of three high-transparency copper meshes, with the bottom mesh grounded and second and third meshes permitted to be individually biased.  These meshes are nested in frames made of PEEK and connected through Kapton-coated wires.  The flexible circuit RFA collector was replaced with copper bars.  The insertable RFA was installed in the fifth Q15 test chamber, having a diamond-like carbon coating.  To provide cross calibration between the two RFA designs, the thin-style RFA in the TiN and a-C coated test chambers were replaced with the insertable one.  (Refer to Table~\ref{tab:Q15_chamber_table}.)

\paragraph{L3 Chambers}

In the east side of the L3 experimental region (see Figure~\ref{fig:cesr_conversion:vac_l3}), a field-free section (of approximately 2~m in length) is allocated for EC studies.  Three types of beam pipes were tested in this section, as listed below.

\begin{itemize}
    \item  Aluminum beam pipe with smooth interior wall, provided by PEP-II
    \item  Aluminum beam pipe with rectangular fins (grooves) provided by PEP-II
    \item  NEG-coated beam pipes
\end{itemize}

\subparagraph{PEP-II Chambers}

\begin{figure}
    \centering
    \includegraphics[width=0.75\textwidth]{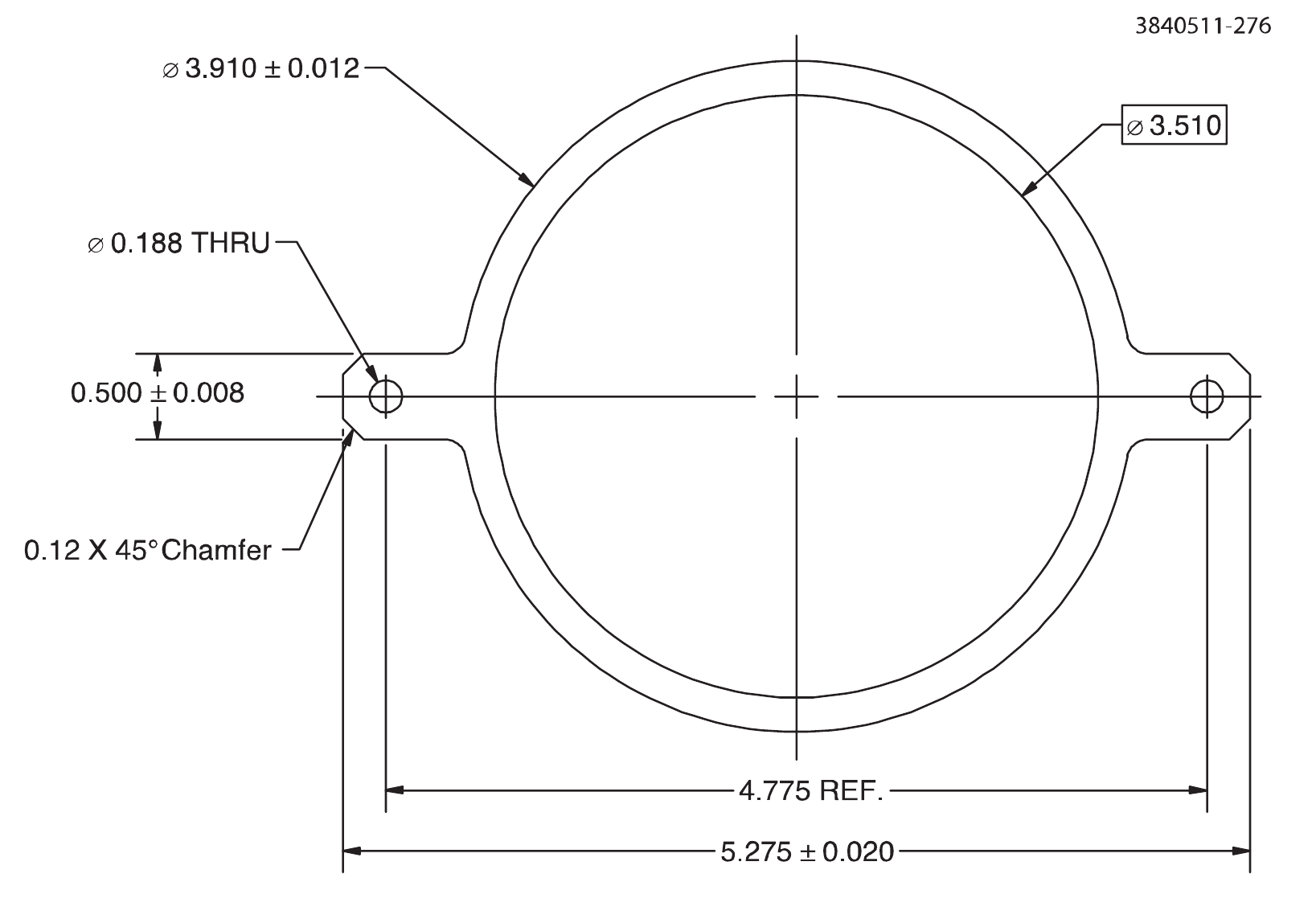}
    \caption[PEP-II Aluminum Chamber with smooth interior wall]{Cross section of PEP-II aluminum beam pipe with a smooth interior wall.  The beam pipe was extruded with type 6063-T6 aluminum alloy. SLAC/PEP-II photo courtesy of M. Pivi.  (All dimensions are in inches.) \label{fig:cesr_conversion:SLAC_L3_pipe_smooth}}
\end{figure}

\begin{figure}
    \centering
\begin{tabular}{cc}
    \includegraphics[width=0.75\textwidth]{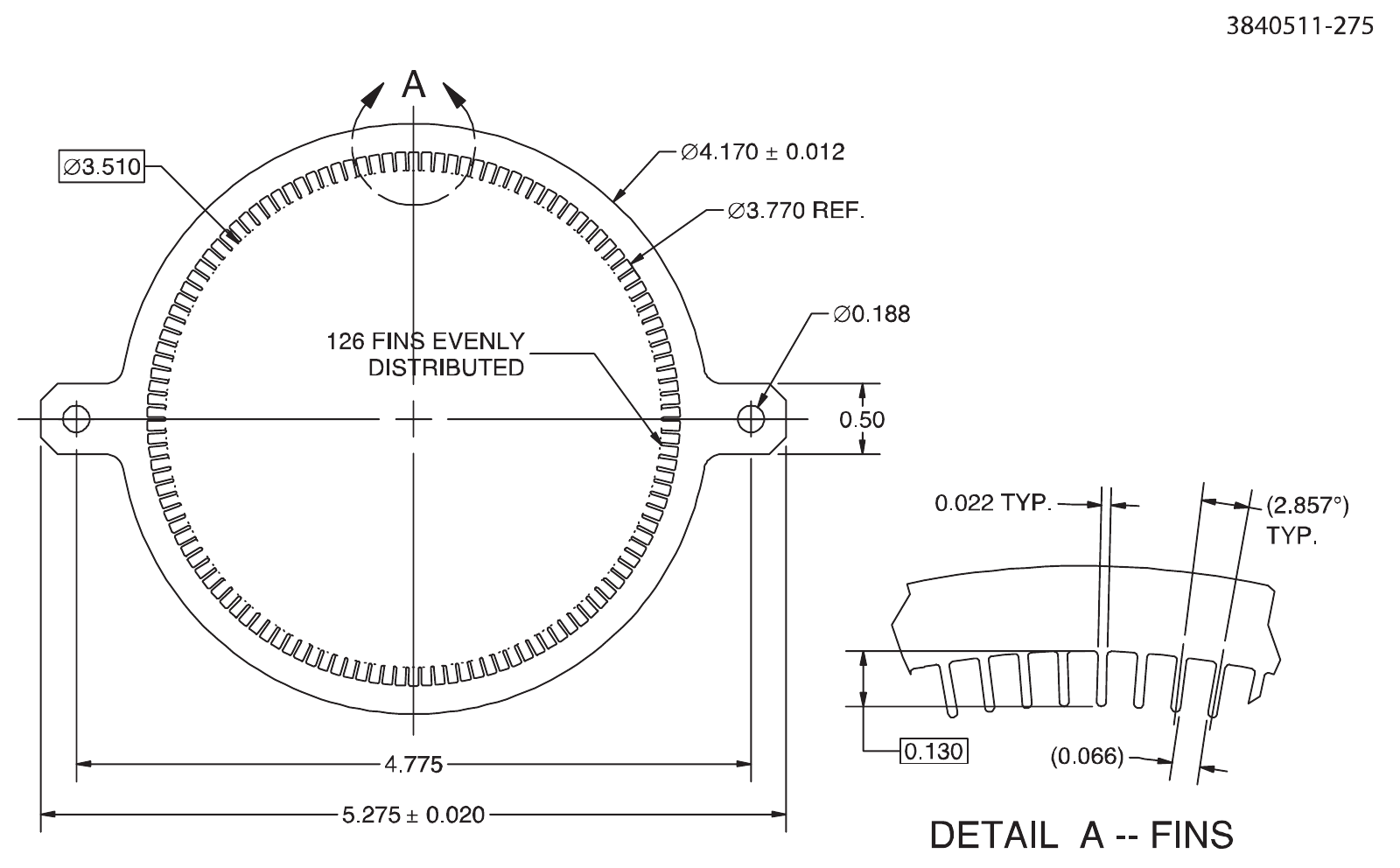}\\
     \includegraphics[width=0.75\textwidth]{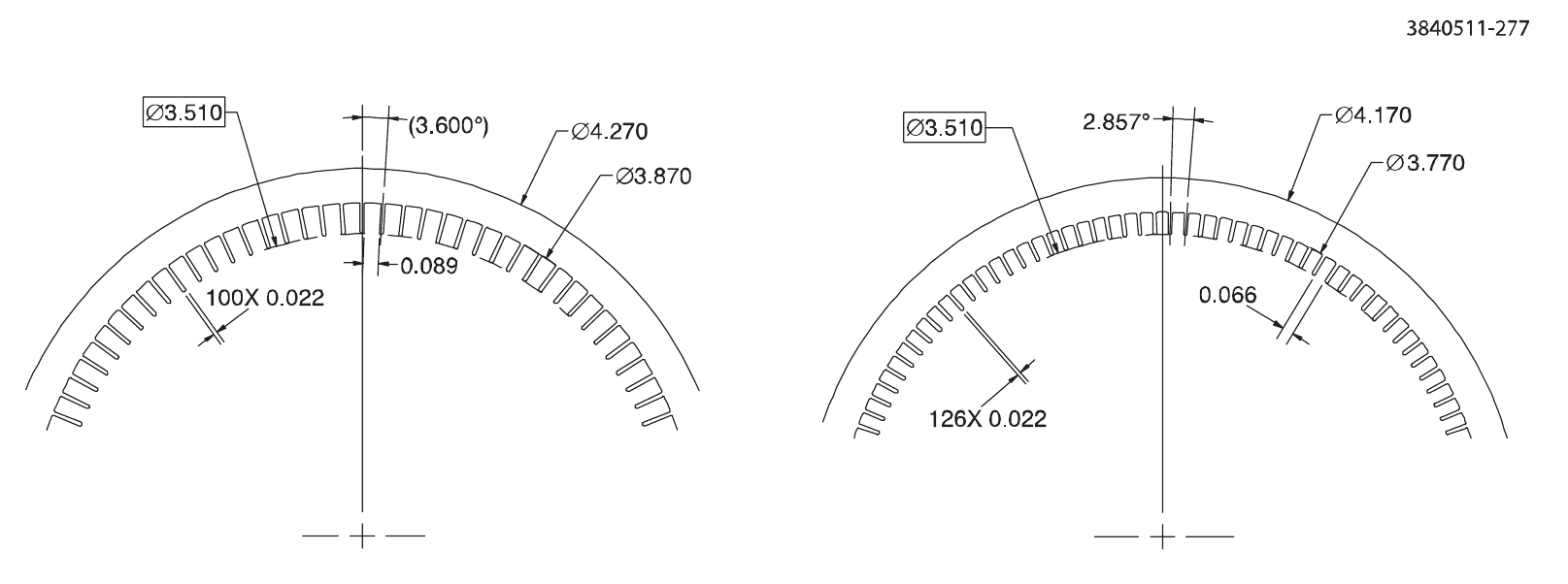}\\
\end{tabular}
    \caption[PEP-II Aluminum Chamber with 3.3~mm rectangular grooves]{(Top) Cross section of PEP-II aluminum beam pipe with a 3.3~mm deep rectangular grooves.  The beam pipe was extruded with type 6063-T6 aluminum alloy. (Bottom) Details of rectangular grooves of two different groove depths on PEP-II aluminum beam pipes.  The left shows grooves with depth of 4.5~mm and right with depth of 3.3~mm.  The beam pipes were extruded with type 6063-T6 aluminum alloy. (All dimensions are in inches.)\label{fig:cesr_conversion:SLAC_L3_pipe_3_3mm_fin}}
\end{figure}

Direct measurements and previous simulation work on rectangular
groove samples have shown an SEY $\delta_{max}$ well below unity, as low as
approximately 0.6 \cite{ECLOUD04:139to141,
JAP104:104904, VACUUM73:195to199, NIMA571:588to598}.  Following the
successful tests on several groove samples with different materials
and coatings, four aluminum (type 6063-T6 alloy) test chambers were
manufactured.  Two of the chambers had smooth interior surfaces, as
shown in Figure~\ref{fig:cesr_conversion:SLAC_L3_pipe_smooth}.  The
other two chambers were constructed with rectangular grooves, having
two different groove depths, as detailed in
Figures~\ref{fig:cesr_conversion:SLAC_L3_pipe_3_3mm_fin}.  All these aluminum
extrusions have effective inner diameters of 89 mm, identical to
both the stainless steel chambers located in the PEP-II straight
section and the nominal CESR L3 beam pipes.

\begin{figure}
    \centering
\begin{tabular}{cc}
\includegraphics[width=0.60\textwidth]{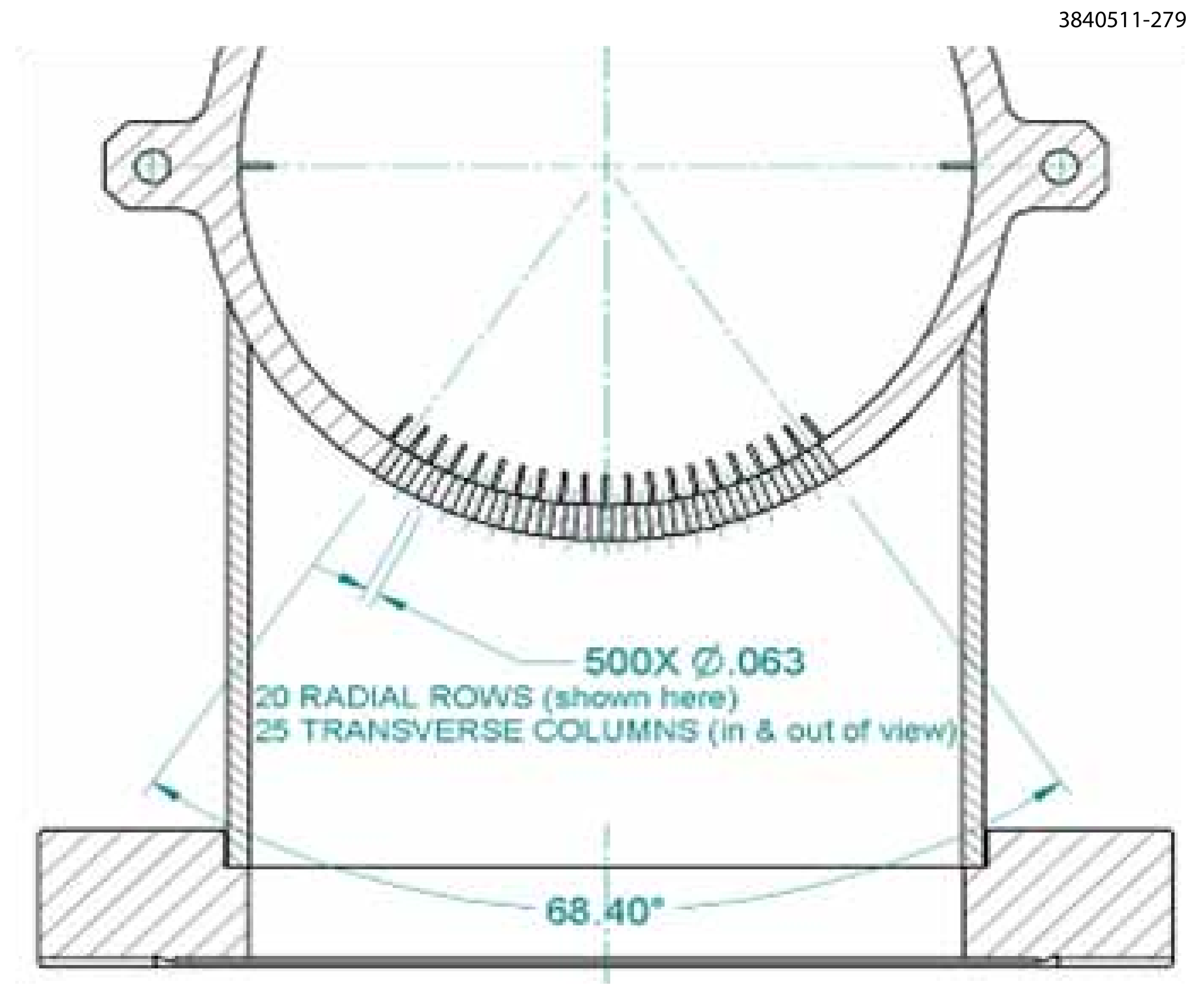} & \\
\includegraphics[width=0.45\textwidth]{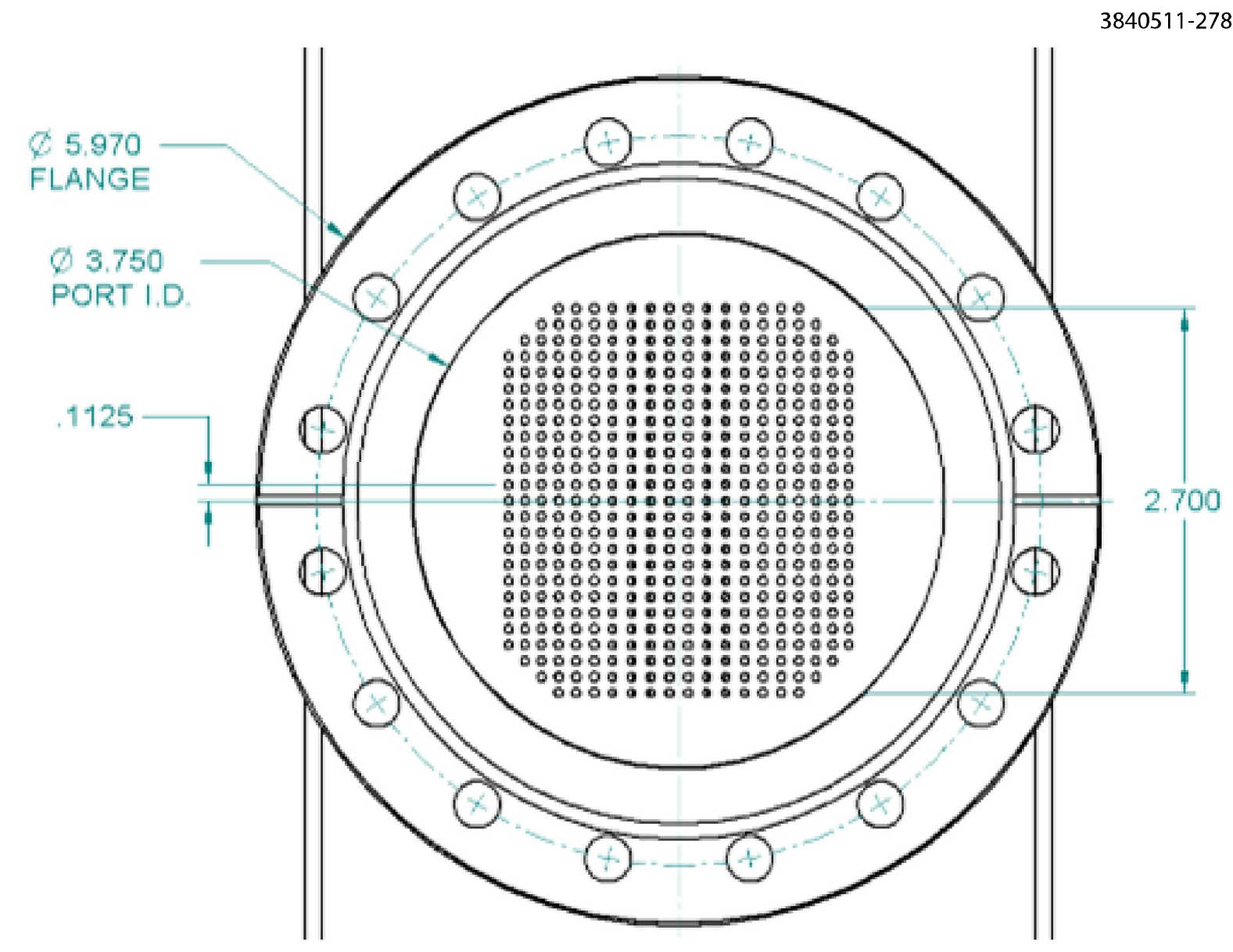} &
 \includegraphics[width=0.45\textwidth, angle=90.]{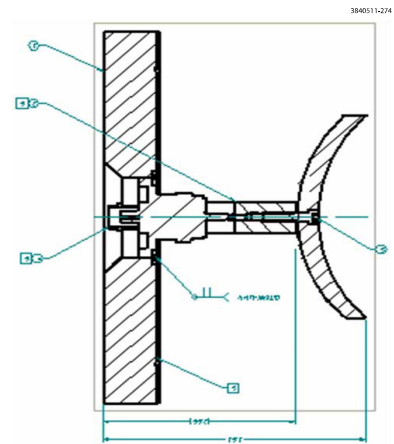} \\

\end{tabular}
    \caption[PEP-II Drift Pipe Detector Port Views]{(Top) Beam pipe holes for the large electron detector ports built into each of the four test chambers.  The front sectional view.  (Bottom) Right: The bottom sectional
view of the beam pipe holes.  Left:  Electron collector plate in the larger port of PEP-II EC chambers.  (All dimensions are in inches.)
    \label{fig:cesr_conversion:SLAC_drift_pipe_detector_port}}
\end{figure}

\begin{figure}
    \centering
    \includegraphics[width=0.75\textwidth]{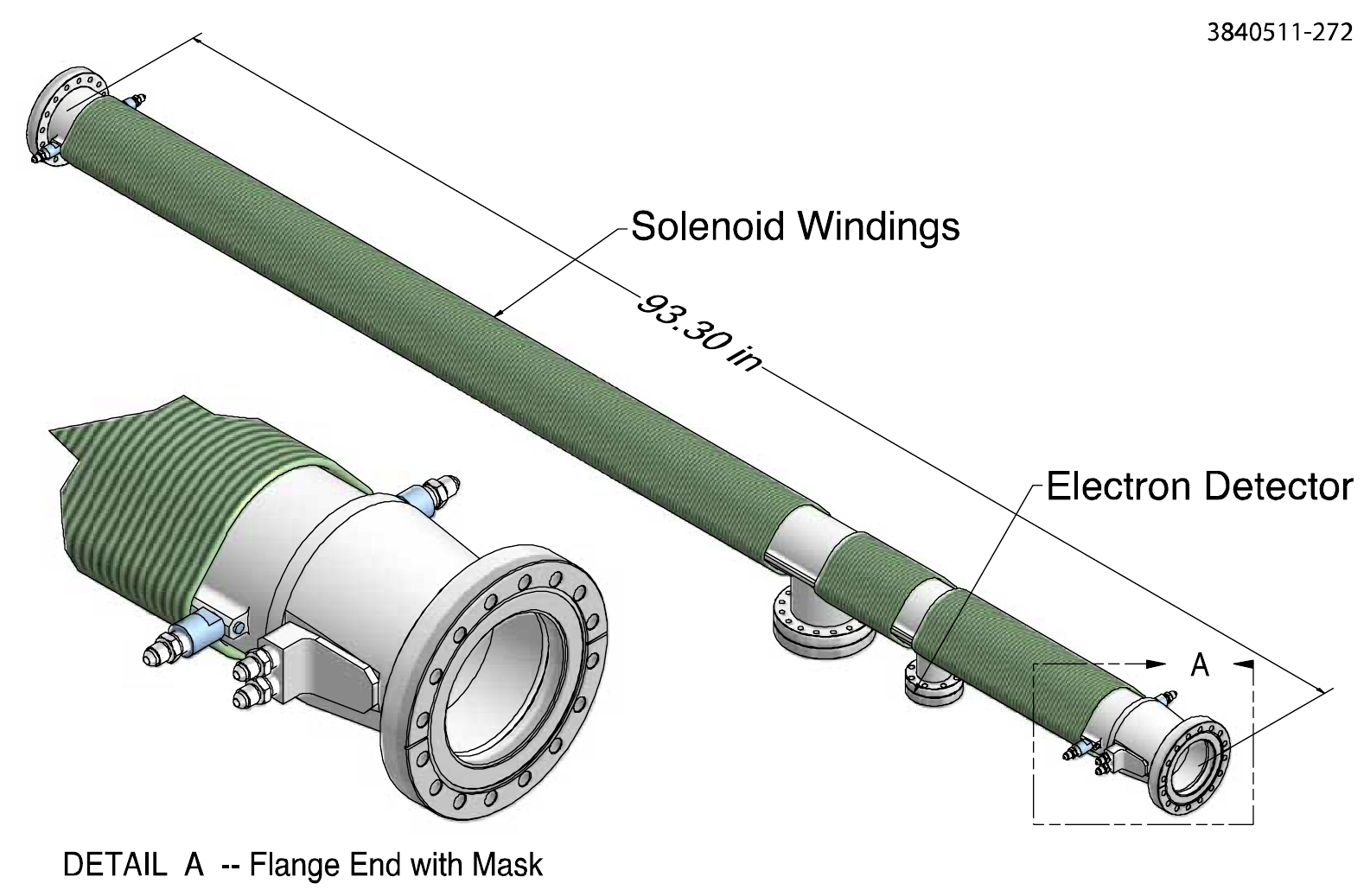}
    \caption[L3 PEP-II EC chambers in the Drift]{PEP-II EC vacuum chambers tested in the L3 experimental region. \label{fig:cesr_conversion:SLAC_L3_drift_pipe}}
\end{figure}

All four test chambers include two vacuum ports (one with 100~mm diameter and one with 38~mm diameter) on the bottom for insertion of electron collectors.  In the larger port, 500 1.6~mm diameter holes (in a 20~x~25 pattern) connect the beam space to the collector, as shown in Figure~\ref{fig:cesr_conversion:SLAC_drift_pipe_detector_port}.  Similarly, 50 holes with 1.6~mm diameters (in a 5~x~10 pattern) connect the beam space to the collector in the smaller port.  The diameter of these holes is chosen to meet requirement of diameter-to-wall-thickness ratio of 1:3 in order to limit the penetration of EM fields, generated by the passing bunch, into the electron collector ports.  The electron collector, used in the large port, is shown in Figure~\ref{fig:cesr_conversion:SLAC_drift_pipe_detector_port}.  The beam pipe assembly is illustrated in Figure~\ref{fig:cesr_conversion:SLAC_L3_drift_pipe}, which includes a re-entrant SR mask on one end of the chamber.  All four EC chambers were then coated with a TiN thin film.

\begin{figure}
    \centering
    \includegraphics[width=0.5\textwidth]{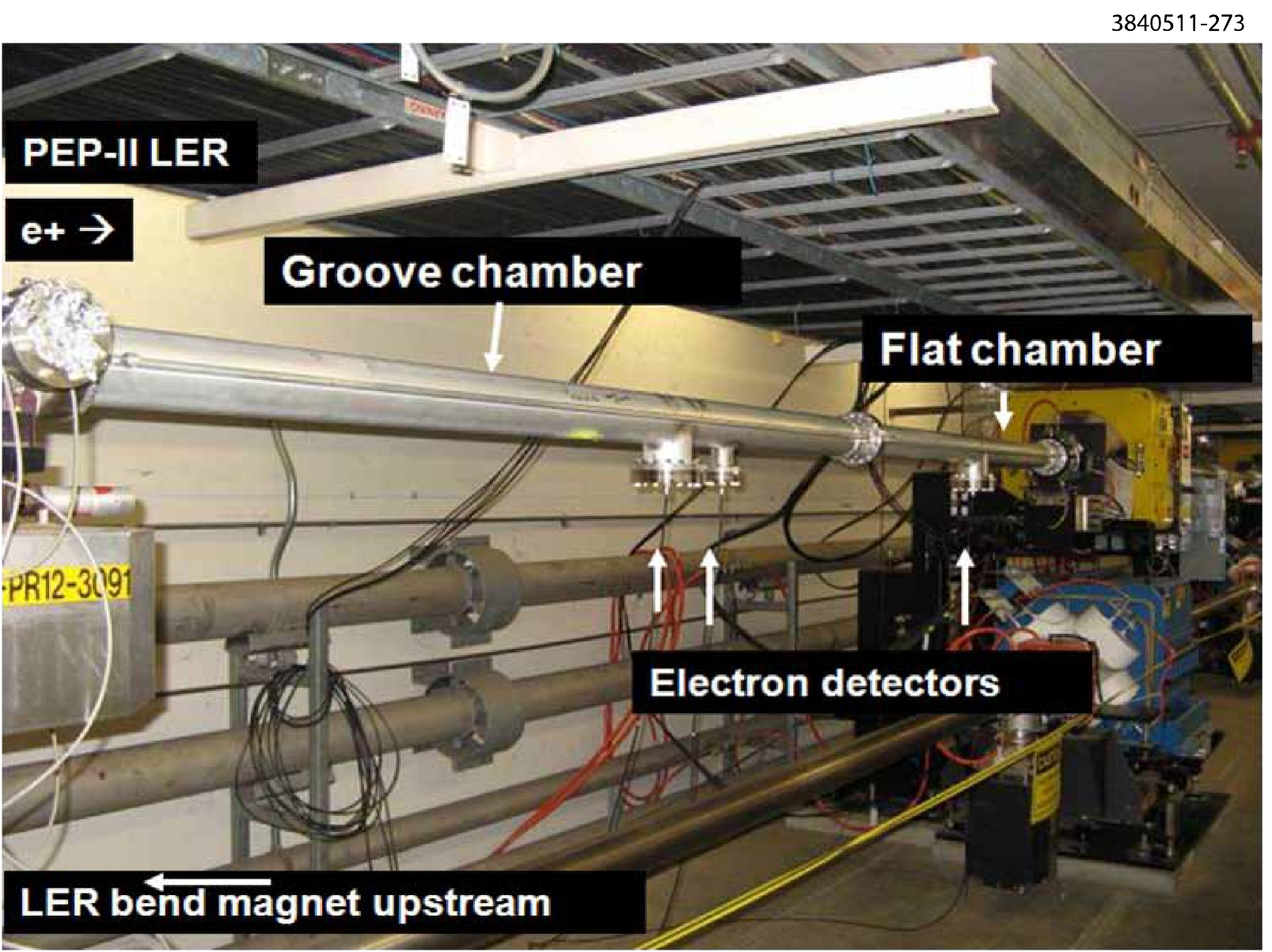}
    \caption[Photo of PEP-II EC Test chambers in PEP-II LER]{The four PEP-II drift EC chambers were tested in PEP-II LER, prior to deployment in {\cesrta} L3 EC Region. \label{fig:cesr_conversion:SLAC_chambers_in_PEP_II}}
\end{figure}

These four EC chambers were originally tested in the PEP-II LER (see Figure~\ref{fig:cesr_conversion:SLAC_chambers_in_PEP_II}).  Following the tests in the PEP-II, these chambers were re-deployed in the CESR's L3 EC drift section for continuation of the EC mitigation studies.

%\begin{figure}
   % \centering
    %\includegraphics[width=0.75\textwidth]{pub_figures/SLAC_L3_Pipes_Two_Fins.pdf}
    %\caption[PEP-II Aluminum Chambers, detail of grooves with two different depths]{Details of rectangular grooves of two different groove depths on PEP-II aluminum beam pipes.  The left shows grooves with depth of 4.5~mm and right with depth of 3.3~mm.  The beam pipes were extruded with type 6063-T6 aluminum alloy. (All dimensions are in inches.)\label{fig:cesr_conversion:SLAC_L3_pipe_two_fins}}
%\end{figure}

%\begin{figure}
   % \centering
   % \includegraphics[width=.45\textwidth, angle=90.]{pub_figures/SLAC_Electron_Collector1.pdf}
   % \caption[Electron collector plate in the larger port of PEP-II EC chambers]{Electron collector plate in the larger port of PEP-II EC chambers. \label{fig:cesr_conversion:SLAC_Electron_Collector1}}
%\end{figure}

\subparagraph{NEG Test Section}

\begin{figure}
    \centering
\begin{tabular}{c}
    \includegraphics[width=0.75\textwidth]{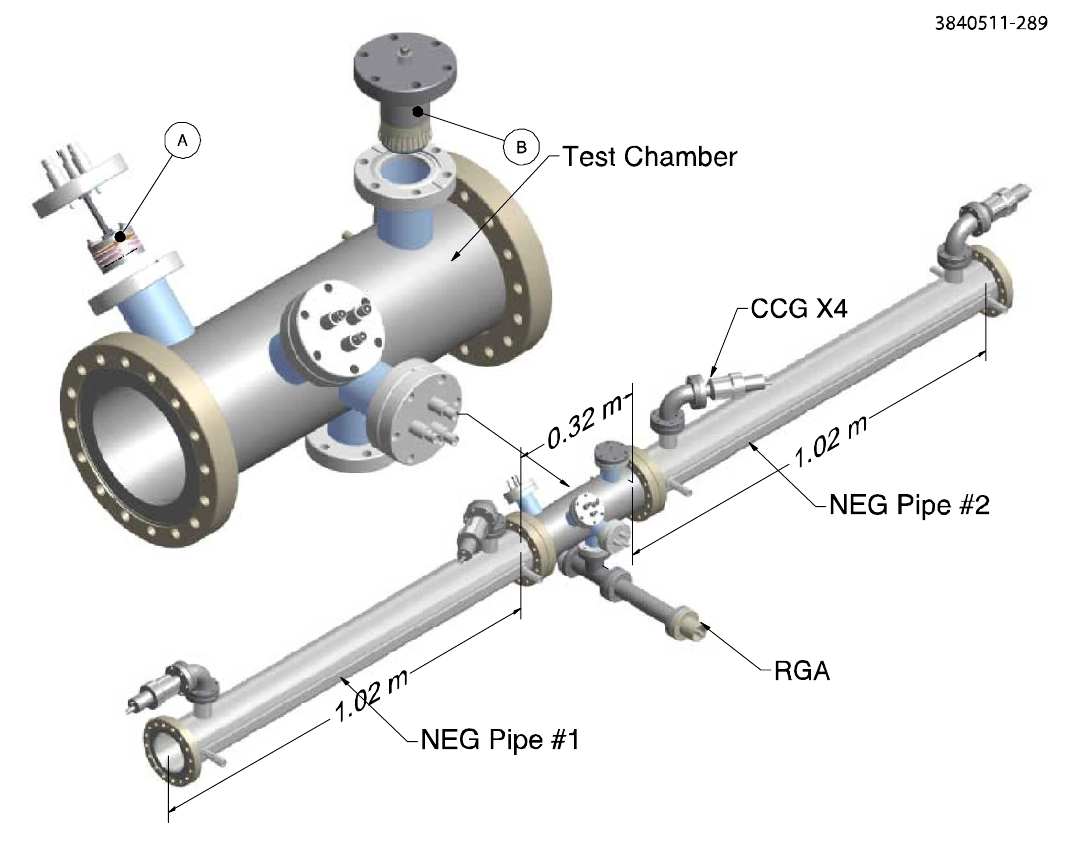}\\
    \includegraphics[width=0.75\textwidth]{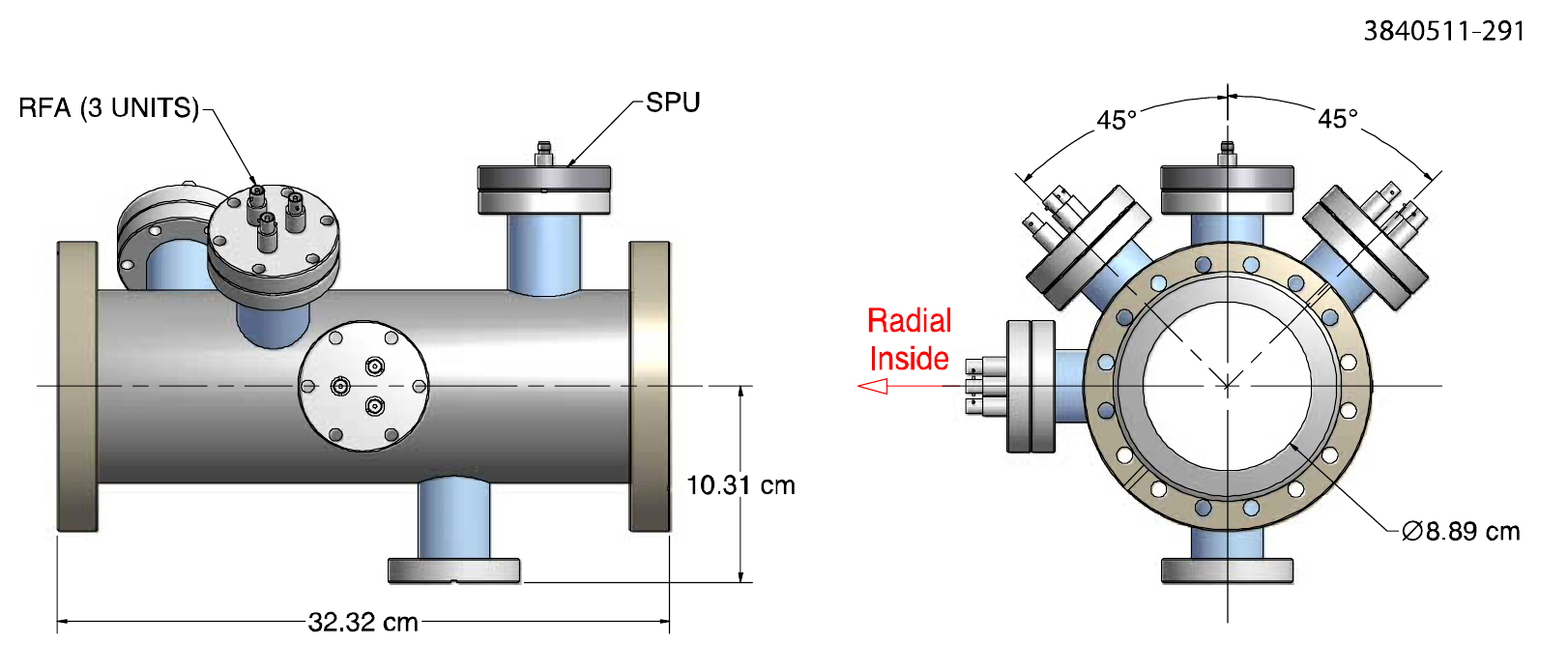}\\
   \end{tabular}

    \caption[NEG coated drift pipes in L3]{(Top) Vacuum chambers with NEG thin film coating in the drift section of the L3 experimental region. The test chamber included: (A) 3 units of APS-style RFAs; (B) RF-shielded pickup assembly supplied by LBNL. (Bottom) EC Diagnostic chamber with NEG thin film coating\cite{NIMA760:86to97}. \label{fig:cesr_conversion:NEG_pipes}}
\end{figure}

A Ti-Zr-V non-evaporable getter (NEG) thin film
\cite{NIMA554:92to113, PAC11:THOBS6} has been shown to have a low
SEY after its activation at elevated temperatures under vacuum.
The activated NEG coating also has the benefit of providing vacuum
pumping.  A NEG-coated test chamber, equipped with EC diagnostics,
was built and tested in the drift section of the L3 experimental
region.  To prevent rapid saturation of the activated NEG thin film
from residual gases in the surrounding beam pipes, the test chamber
was sandwiched between two 1~m long NEG coated beam pipes, as shown
in Figure~\ref{fig:cesr_conversion:NEG_pipes}.  The EC test chamber
was equipped with three APS-style RFAs at three angles and an
RF-shielded pickup on the top (again see
Figure~\ref{fig:cesr_conversion:NEG_pipes}).  All three
chambers were made of stainless steel (Type 304L).

The NEG thin film deposition was done by SAES Getter Inc. via a DC magnetron sputtering method using twisted wires of Ti, Zr and V as the sputtering cathode.  The thickness of the NEG thin film is approximately 2~$\mu$m.  During the coating process all diagnostic instruments [RFAs and shielded pickups (SPUs)] were removed.  A 24~hr, 150$^\circ$C  bakeout was carried out to the NEG-coated beam pipe string with RFAs and SPUs inserted prior to the installation in the L3 experimental region.  These NEG-coated beam pipes replaced the PEP-II EC text chambers in the drift section of L3 region.  This follows an established procedure for NEG installation and activation\cite{PAC03:WOAA0xx}.

\begin{figure}
    \centering
    \includegraphics[width=0.75\textwidth]{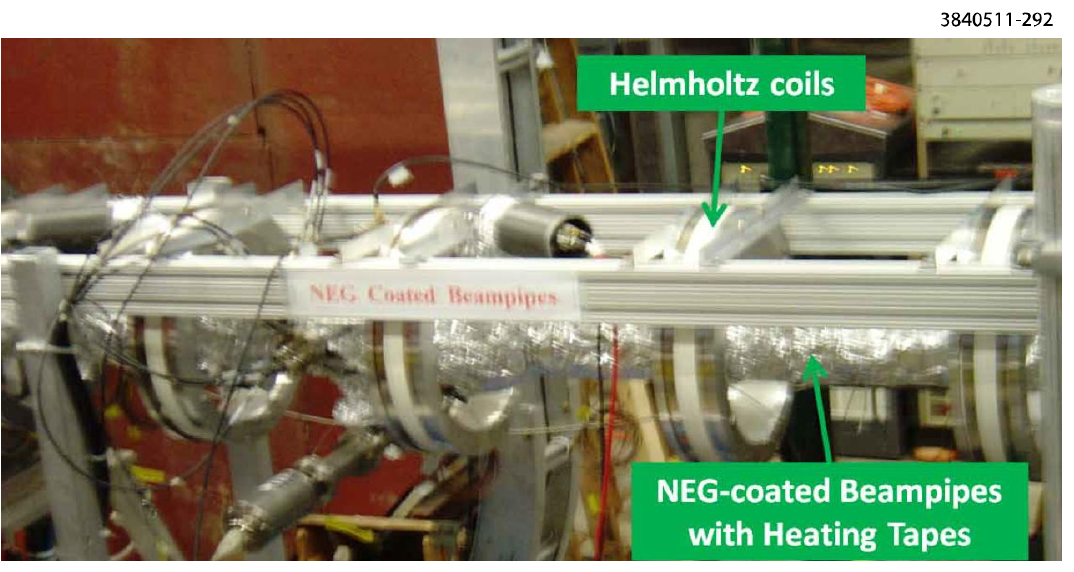}
    \caption[NEG coated beam pipes Installed in L3 with Helmholtz Coils]{NEG-coated beam pipes in L3 after installation of Helmholtz Coils. \label{fig:cesr_conversion:NEG_pipe_and_solenoid}}
\end{figure}

Fiberglass insulated heating tapes were wrapped around and along the NEG beam pipe string for the activation.  Six large bore (11-inch diameter) Helmholtz coils, evenly spaced along the beam pipe string (shown in Figure~\ref{fig:cesr_conversion:NEG_pipe_and_solenoid}) replace the normal solenoid winding, found on most of the CESR beam pipes, since the solenoid windings are incompatible with the high temperature heating required during the  activation of the NEG coating.  The NEG coating was activated at 250$^\circ$C for a duration of 24 to 48~hours.  After each venting of the L3 region and to preserve pumping capacity of the NEG thin film, the activation was normally carried out following the initial period of beam conditioning of the beam pipes.

\begin{figure}
    \centering
    \includegraphics[width=0.75\textwidth]{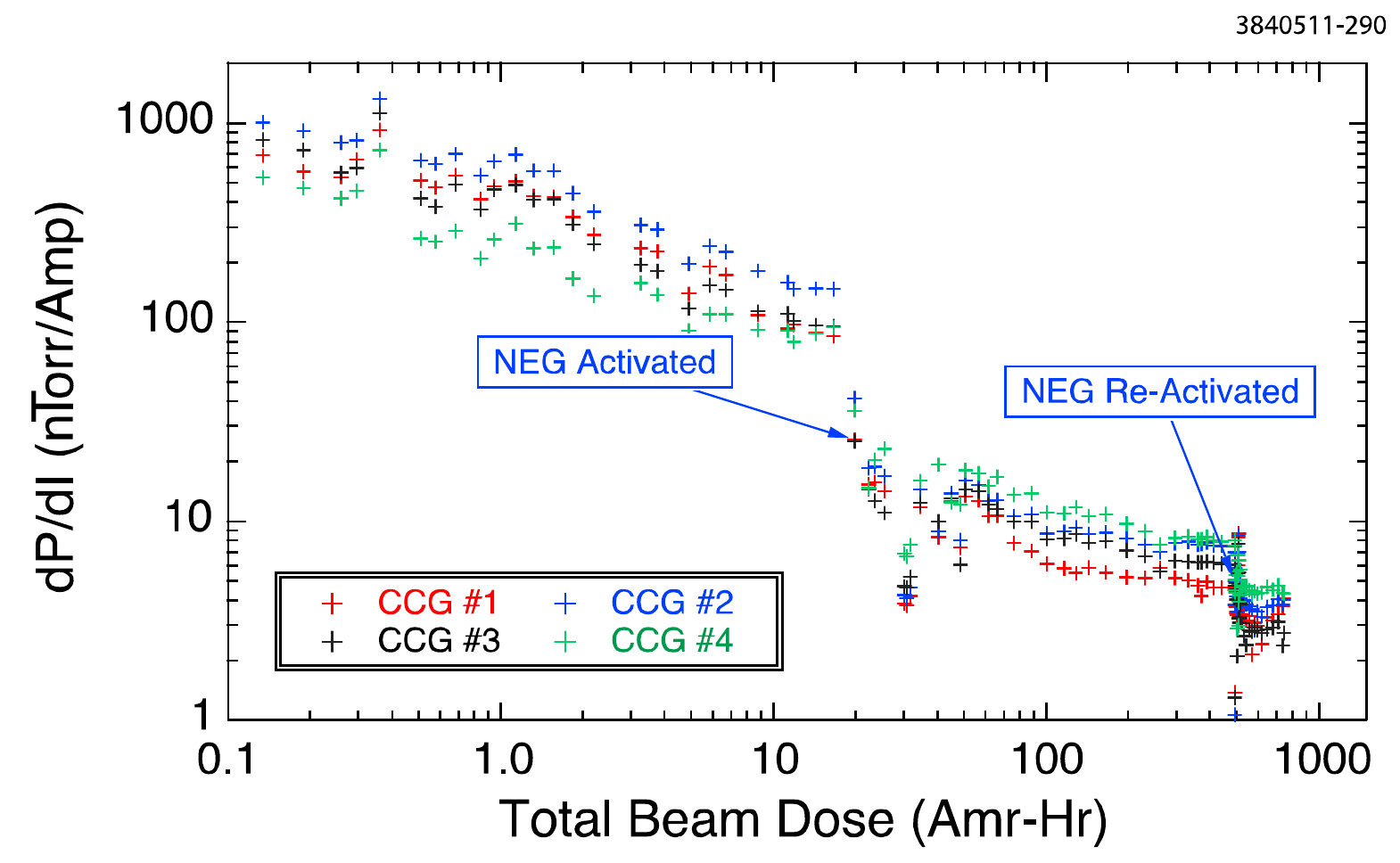}
    \caption[Vacuum performance of NEG coated beam pipes in L3]{Beam conditioning of the NEG coated beam pipes in L3. \label{fig:cesr_conversion:NEG_pipe_dPdI}}
\end{figure}

\begin{figure}
    \centering
    \includegraphics[width=0.75\textwidth]{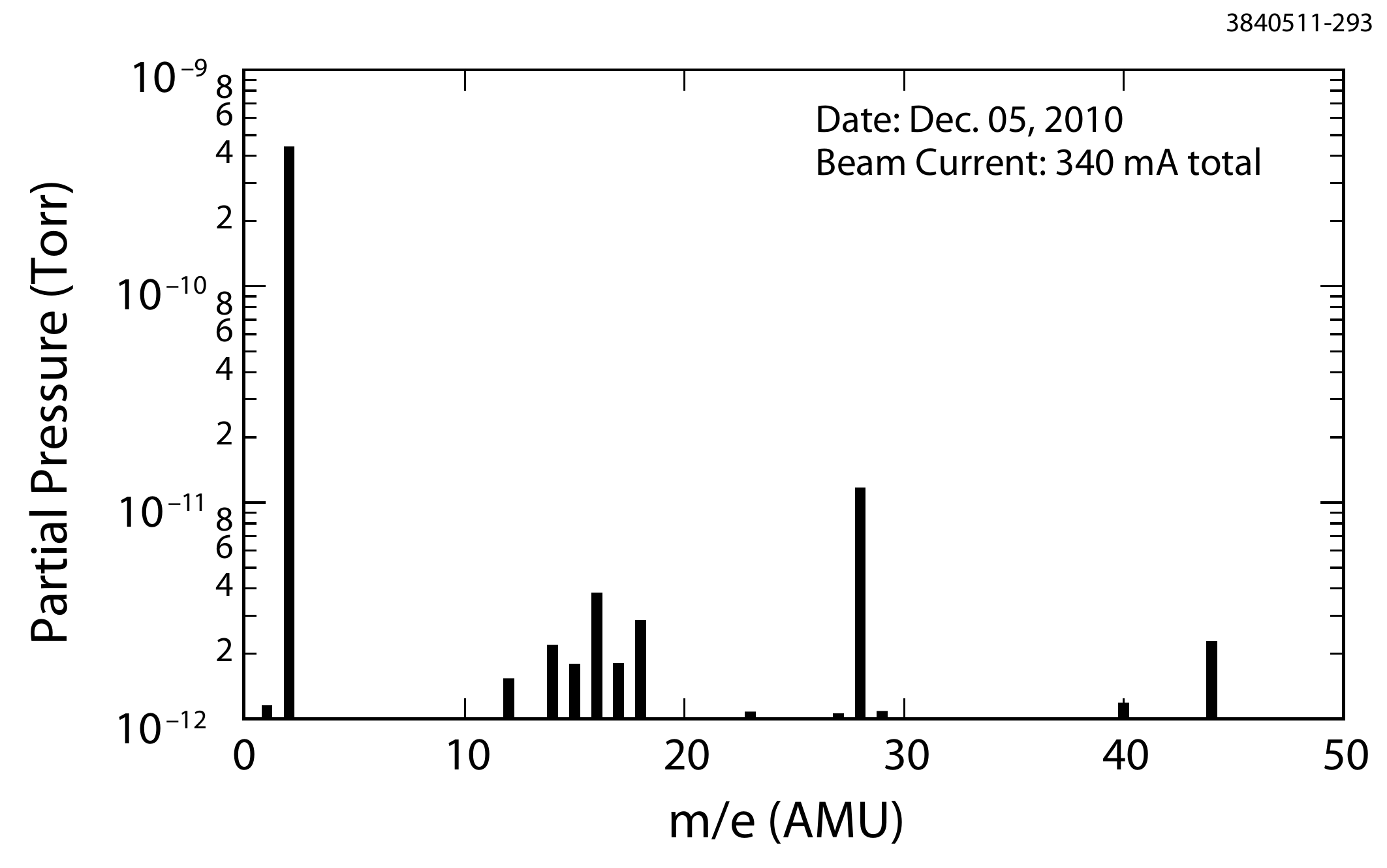}
    \caption[Typical RGA Spectrum of NEG coated beam pipes in L3]{A typical RGA spectrum taken with full stored beams in CESR with accumulated beam dose of approximately 900 A$\cdot$hour.\label{fig:cesr_conversion:NEG_pipe_RGA}}
\end{figure}

Vacuum performance of the NEG coated beam pipes was monitored by four CCGs and an RGA (see Figure~\ref{fig:cesr_conversion:NEG_pipes}).  Figure~\ref{fig:cesr_conversion:NEG_pipe_dPdI} plots the measured SR-induced pressure rises ($\frac{dP}{dI}$) as a function of accumulated beam dose.  As expected, the RGA showed a residual mass spectrum (Figure~\ref{fig:cesr_conversion:NEG_pipe_RGA}) that was dominated by hydrogen.

%\begin{figure}
%    \centering
%    \includegraphics[width=0.75\textwidth]{pub_figures/NEG_EC_Chamber.pdf}
%    \caption[NEG coated EC Chamber in L3]{EC Diagnostic chamber with NEG thin film coating\cite{NIMA760:86to97}. \label{fig:cesr_conversion:NEG_pipe_diagnostics}}
%\end{figure}

\subsubsection{Dipole Chambers}
\label{sssec:cesr_conversion.vac_system.exp_chambers.dipole}

\paragraph{CESR Arc Chamber}

A thin-style RFA design was developed and implemented in a standard CESR dipole chamber at the B12W location.  The major design challenge of the RFA design is the very limited vertical space available in order to fit the RFA-equipped chamber within the existing CESR dipole magnet's iron poles.

\begin{figure}
    \centering
    \includegraphics[width=0.75\textwidth, angle=0, width=6.0in]{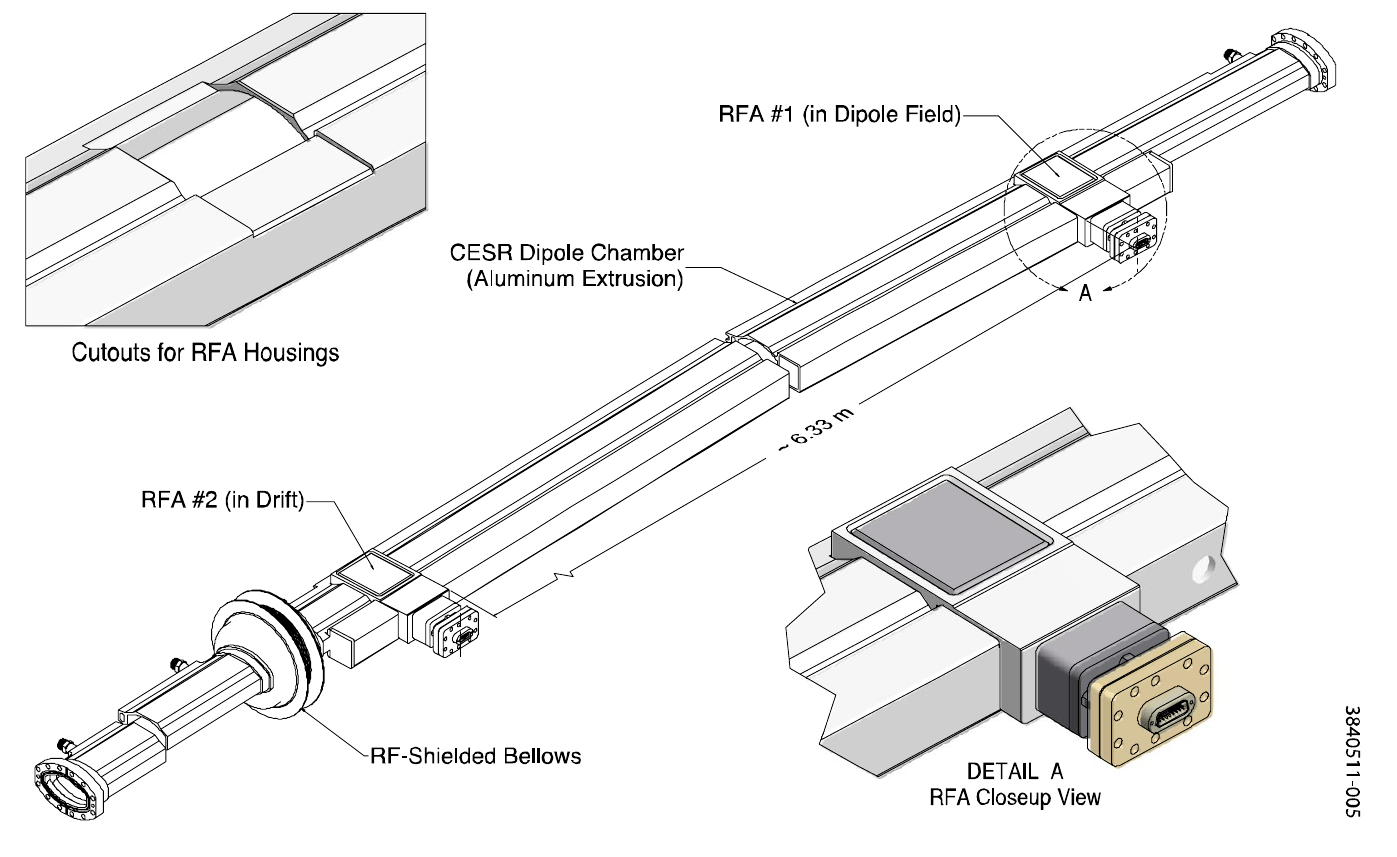}
    \caption{A CESR dipole chamber with 2 RFAs. \label{fig:cesr_conversion:B12W_with_2RFAs}}
\end{figure}

\begin{figure}
    \centering
    \includegraphics[width=0.75\textwidth, angle=0, width=6.0in]{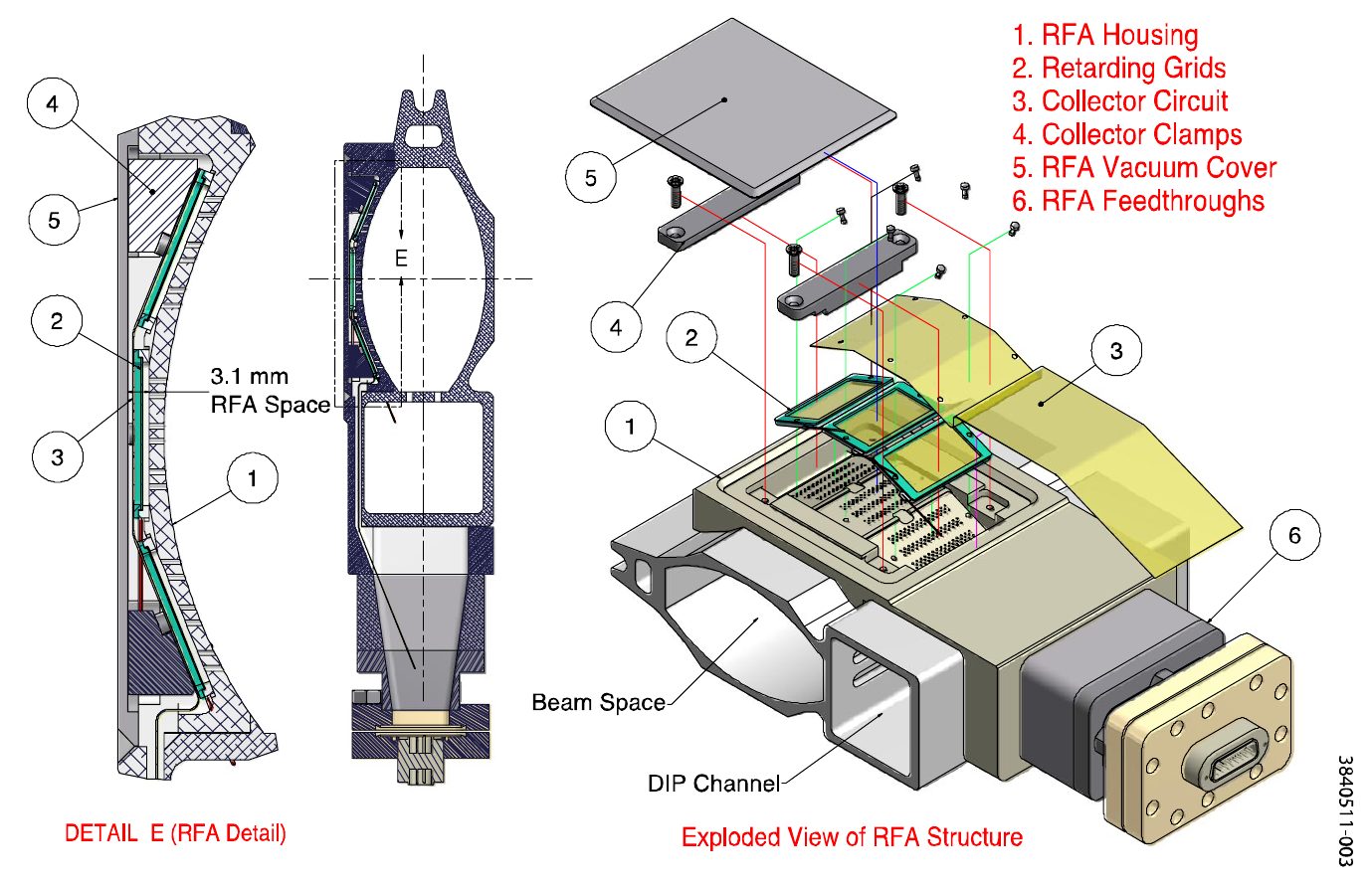}
    \caption{RFA design detail for a CESR dipole chamber\cite{NIMA770:141to154}. \label{fig:cesr_conversion:B12W_RFA}}
\end{figure}

\begin{figure}
    \centering
    \includegraphics[width=0.8\textwidth]{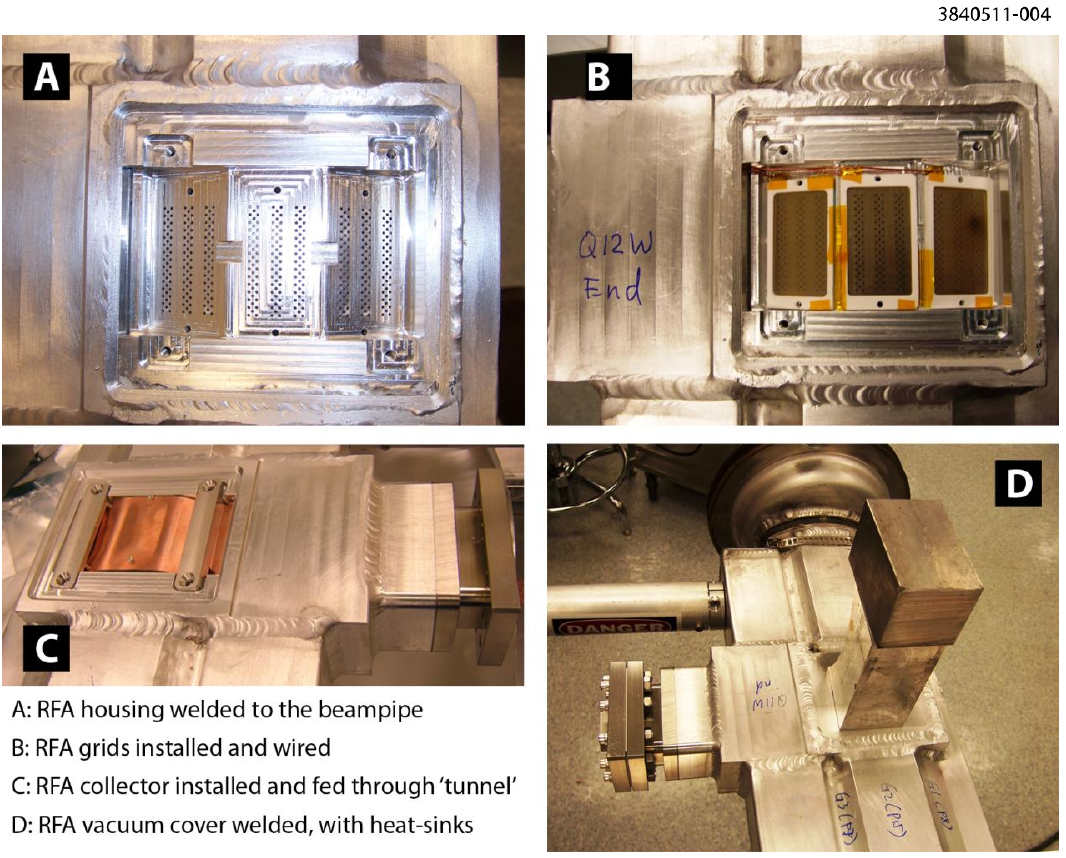}
    \caption{CESR dipole RFA assembly and welding photos. \label{fig:cesr_conversion:B12W_RFA_welding}}
\end{figure}

\begin{figure}
    \centering
    \includegraphics[width=0.75\textwidth, angle=0, width=6.0in]{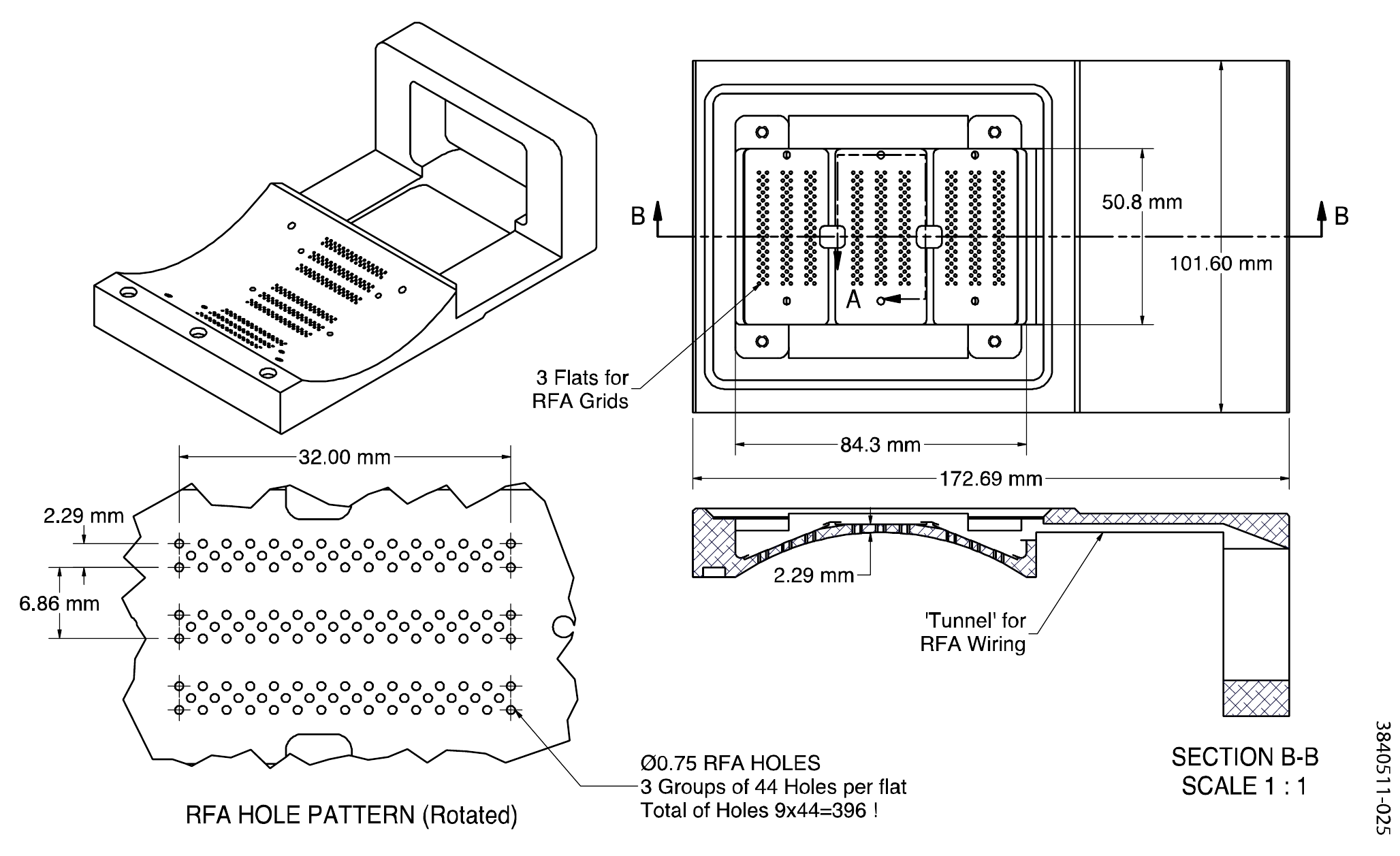}
    \caption{RFA Housing block for a CESR dipole chamber. \label{fig:cesr_conversion:B12W_RFA_Housing}}
\end{figure}

\begin{figure}
    \centering
    \includegraphics[width=0.8\textwidth, angle=0, width=6.0in]{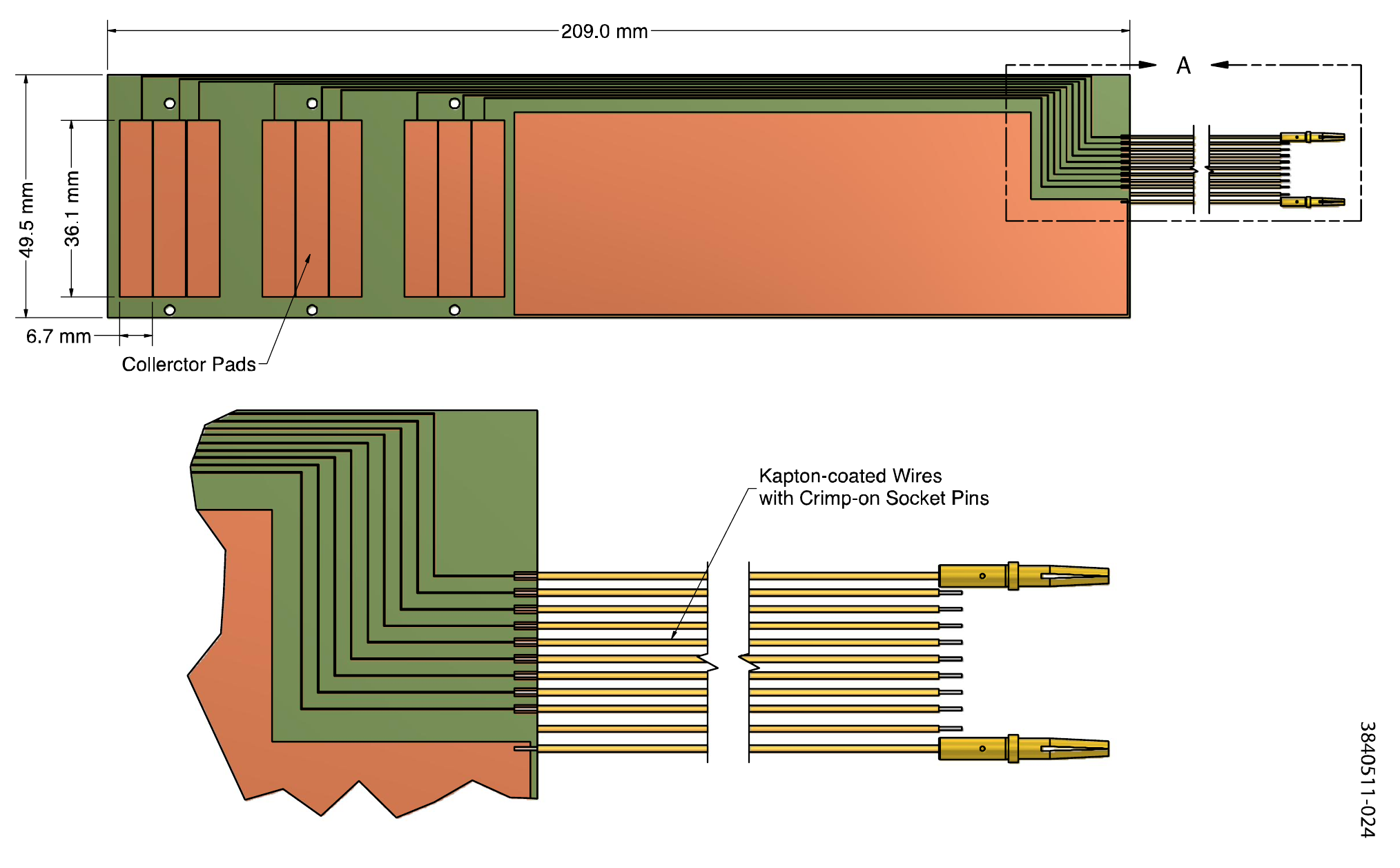}
    \caption{RFA flexible circuit collector for a CESR dipole chamber. \label{fig:cesr_conversion:B12W_RFA_Circuit}}
\end{figure}

Figure~\ref{fig:cesr_conversion:B12W_with_2RFAs} gives an overview of the CESR dipole chamber with the two RFA assemblies.  During installation one of the RFA is placed in the dipole field, while the other in the field-free drift. Figure~\ref{fig:cesr_conversion:B12W_RFA} illustrates the structure of the dipole RFA.  As shown in the cross-section view, the entire RFA structure is fitted within approximately 3~mm vertical space.  Some of the key steps of the dipole RFA construction are summarized as follows:

\begin{enumerate}
    \item Two RFA openings (see top-left insert in Figure~\ref{fig:cesr_conversion:B12W_with_2RFAs}) were machined into the top of a CESR dipole chamber.  The dipole chamber, used for this project, was a fully functional spare beam pipe, containing the inserted distributed ion pump elements in the ante-chamber and a RF-shielded bellows assembly.  The machining of the RFA openings were done completely dry (i.e.without any cutting fluids) with a cleaned numerically controlled-milling machine and cleaned cutting tools.  In addition special fixtures were designed and used to contain metal chips and particulates within the opened section and to prevent them from entering the distributed ion pump (DIP) channels.
    \item RFA housing blocks made of 6061-T6 aluminum alloy were placed at the openings and welded to the beam pipe (see photo~A in Figure~\ref{fig:cesr_conversion:B12W_RFA_welding}).  The CAD model of the RFA housing block is shown in Figure~\ref{fig:cesr_conversion:B12W_RFA_Housing}.  Grouped into 9~segments on three flats, 396 small through-holes (0.75~mm diameter) allow electrons from the EC within the beam space to drift into the RFA space, while reducing the beam RF EMI.  A vacuum leak check was performed after welding the RFA housing to ensure the UHV quality.
    \item Three retarding grids, made of stainless steel mesh and sandwiched into ceramic frames, were mounted on the three flats (see Photo B in Figure~\ref{fig:cesr_conversion:B12W_RFA_welding}).  The three grids were individually wired, using Kapton-coated copper wires, which were fed through a `tunnel' (formed by the RFA housing block) to the connection port.  Small amounts of Kapton tape with silicone adhesive were used to keep the grids in place.
    \item Thin (approximately 0.15~mm thick) copper-coated-Kapton flexible circuits were used as the RFA collectors.  As shown in Figure~\ref{fig:cesr_conversion:B12W_RFA_Circuit} they have 9 segmented copper patches that match with the 9 groups of RFA holes on the housing block.  Kapton-coated wires are soldered onto the circuit.  After soldering each flexible circuit was cleaned and vacuum baked (at 150$^\circ$C) before assembling into the RFA.  The flexible circuit was precisely placed on top of the retarding grid frames with ceramic head-pins (see Photo C in Figure~\ref{fig:cesr_conversion:B12W_RFA_welding}).  The pinned circuit was further clamped down with two stainless steel bars, also utilized to avoid over-heating of the circuit in the final step of the assembly, i.e. the welding of the RFA vacuum cover.  The end of the circuit, containing the external electric connections for the circuit, was fed through the `tunnel' into the connection port with the 9-pin D-type vacuum feedthrough.
    \item After a thorough electrical checkout of the RFA assembly, an aluminum (6061-T6 alloy) cover was welded to seal the RFA.  Heat-sinks were used to prevent over-heating of the RFA collector circuit as shown (see Photo D in Figure~\ref{fig:cesr_conversion:B12W_RFA_welding}).
\end{enumerate}

The RFA dipole chamber was then leak checked and a 150$^\circ$C pre-installation bakeout was carried out.  The chamber was successfully installed in CESR at B12W location during a one-week shutdown in May 2008.  Both RFAs on the dipole chamber have been continuously functioning since their installation.

\paragraph{PEP-II Chicane Chambers}

\begin{figure}
    \centering
    \includegraphics[width=0.9\textwidth]{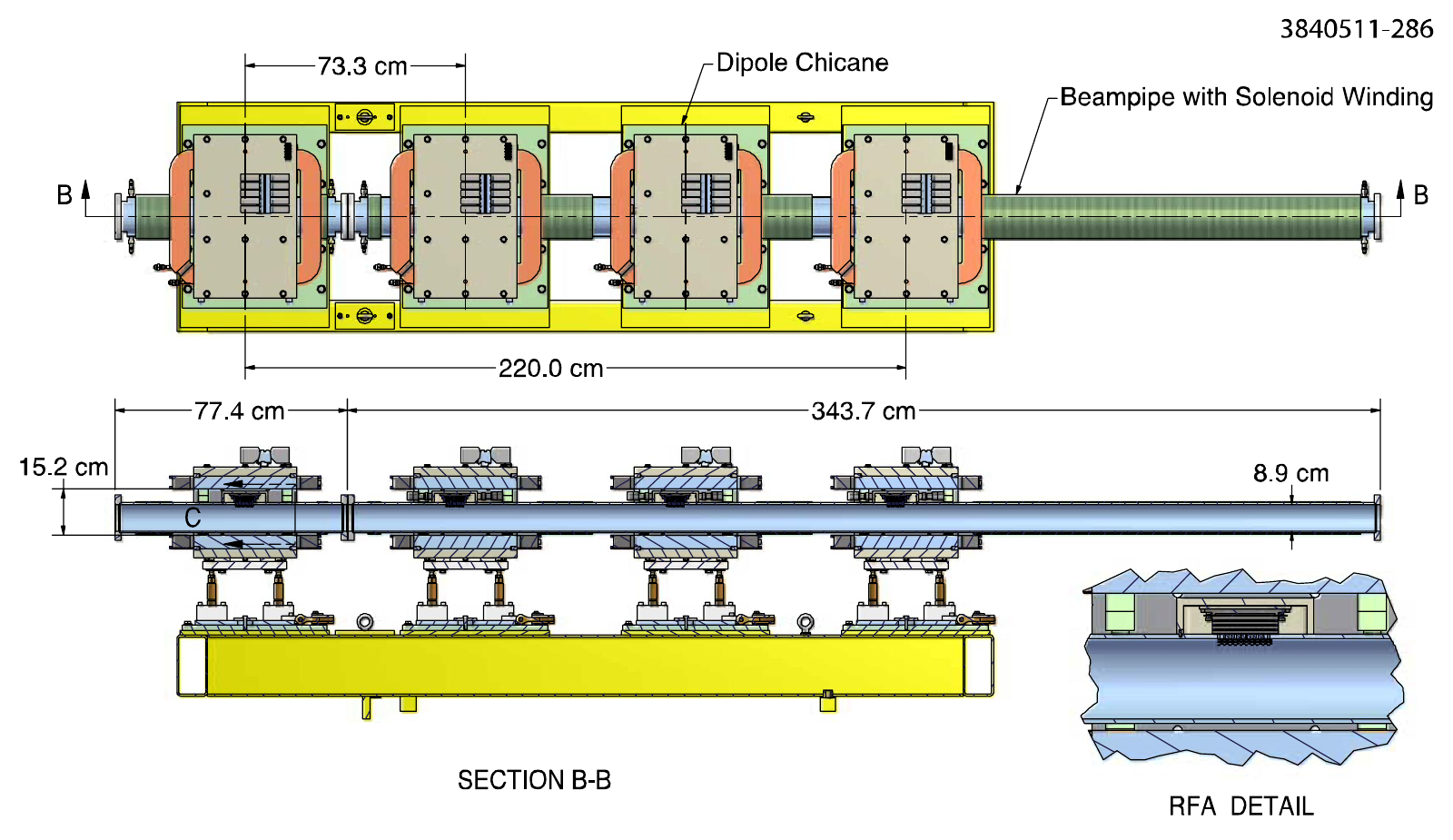}
    \caption{PEP-II 4-dipole magnet chicane and RFA-equipped EC chambers\cite{NIMA770:141to154}. \label{fig:cesr_conversion:SLAC_Chicane}}
\end{figure}

\begin{figure}
    \centering
\begin{tabular}{cc}
\includegraphics[width=0.5\textwidth]{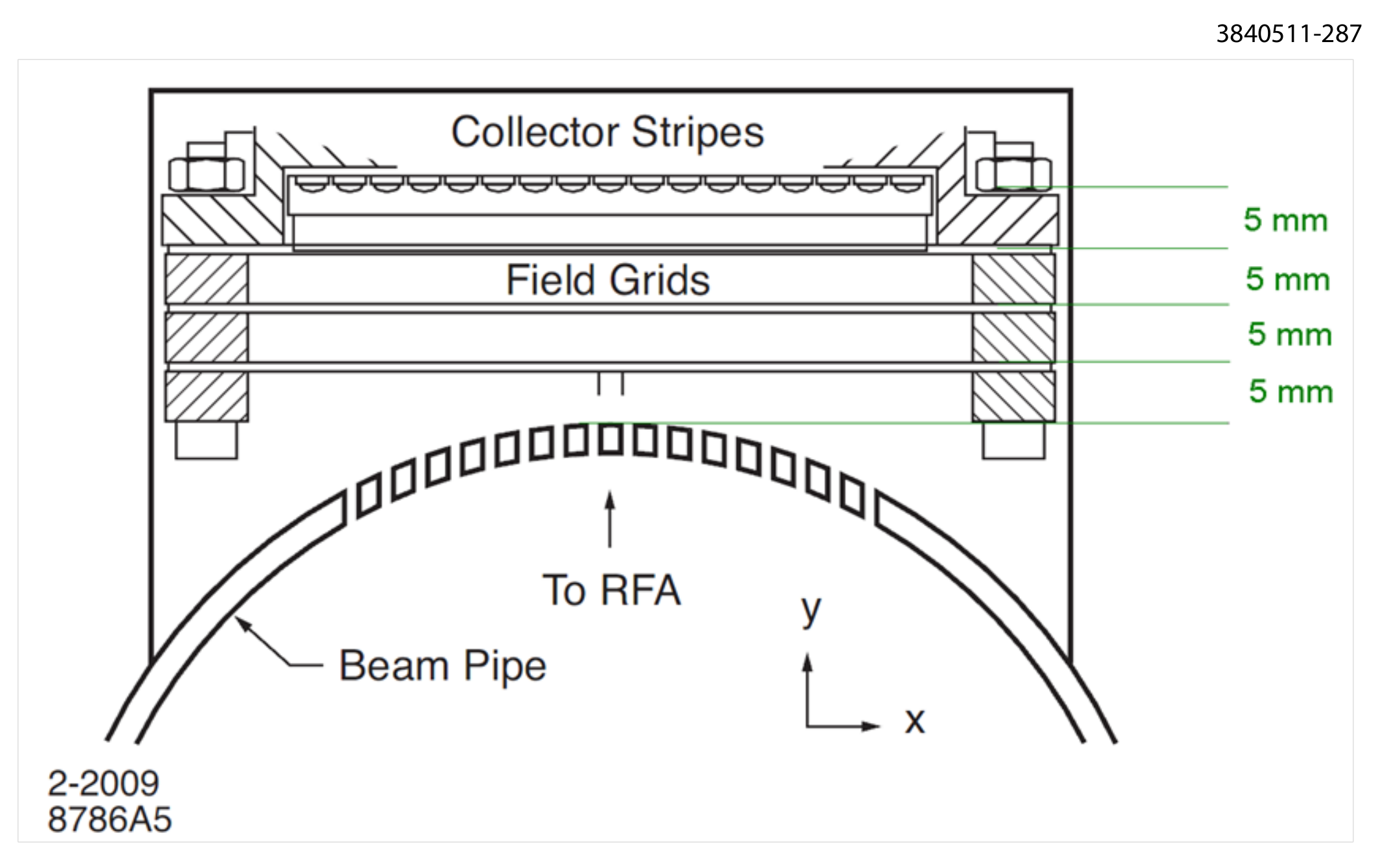} &
\includegraphics[width=0.40\textwidth]{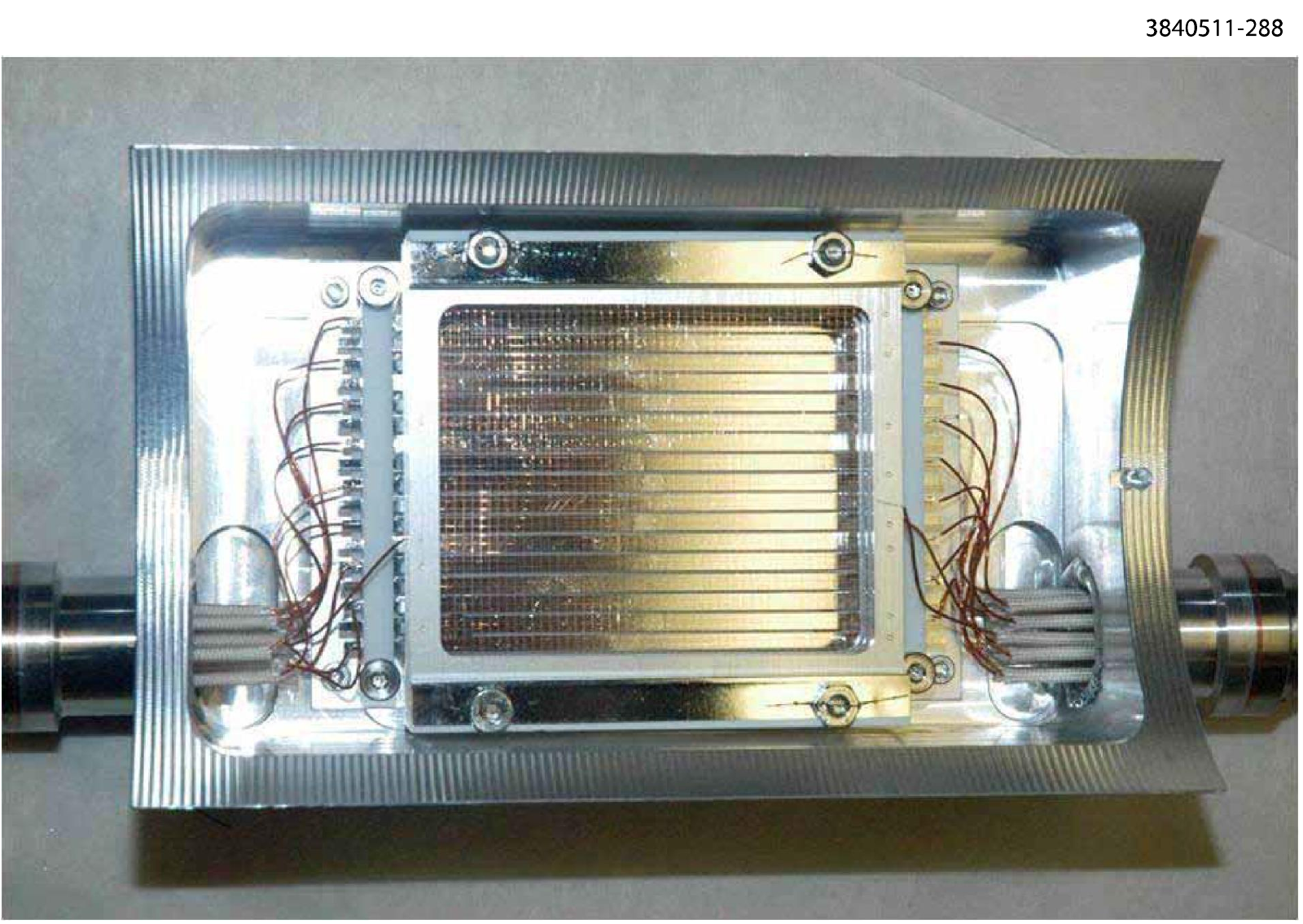}\\
\end{tabular}
    \caption[Segmented RFA in PEP-II Chicane chambers]{Four RFAs were welded onto the chicane beam pipes.  LEFT: Cross-section view showing the structure of these RFAs.  RIGHT: Photo showing the assembled RFA in its aluminum housing, welded on the top of the chicane beam pipes\cite{NIMA770:141to154}.
    \label{fig:cesr_conversion:Chicane_RFA}}
\end{figure}

A PEP-II 4-element dipole-magnet chicane was installed in the {\cesrta} L3 experimental region (see Figure~\ref{fig:cesr_conversion:vac_l3}) for the continuation of studies of EC in a dipole field. The field of the chicane dipoles can be varied over the range of 0 to 1.46~kG, the top limit corresponding to the nominal magnetic field strength of the ILC DR arc dipoles.  As shown in Figure~\ref{fig:cesr_conversion:SLAC_Chicane} the 4~dipoles are spaced 73~cm apart, using two aluminum beam pipes, for a total length of approximately 4.2~m.  The beam pipes have smooth inner surfaces with TiN coating.  Four RFAs were installed on these test chambers, with each of the RFAs located within the dipole magnets.  Figure~\ref{fig:cesr_conversion:Chicane_RFA} shows the structure of these RFAs.

%==========================================================%

%===============================================================%
\subsubsection{Quadrupole Chambers}
\label{sssec:cesr_conversion.vac_system.exp_chambers.quad}

\begin{figure}
    \centering
    \includegraphics[width=0.75\textwidth, angle=0, width=6.0in]{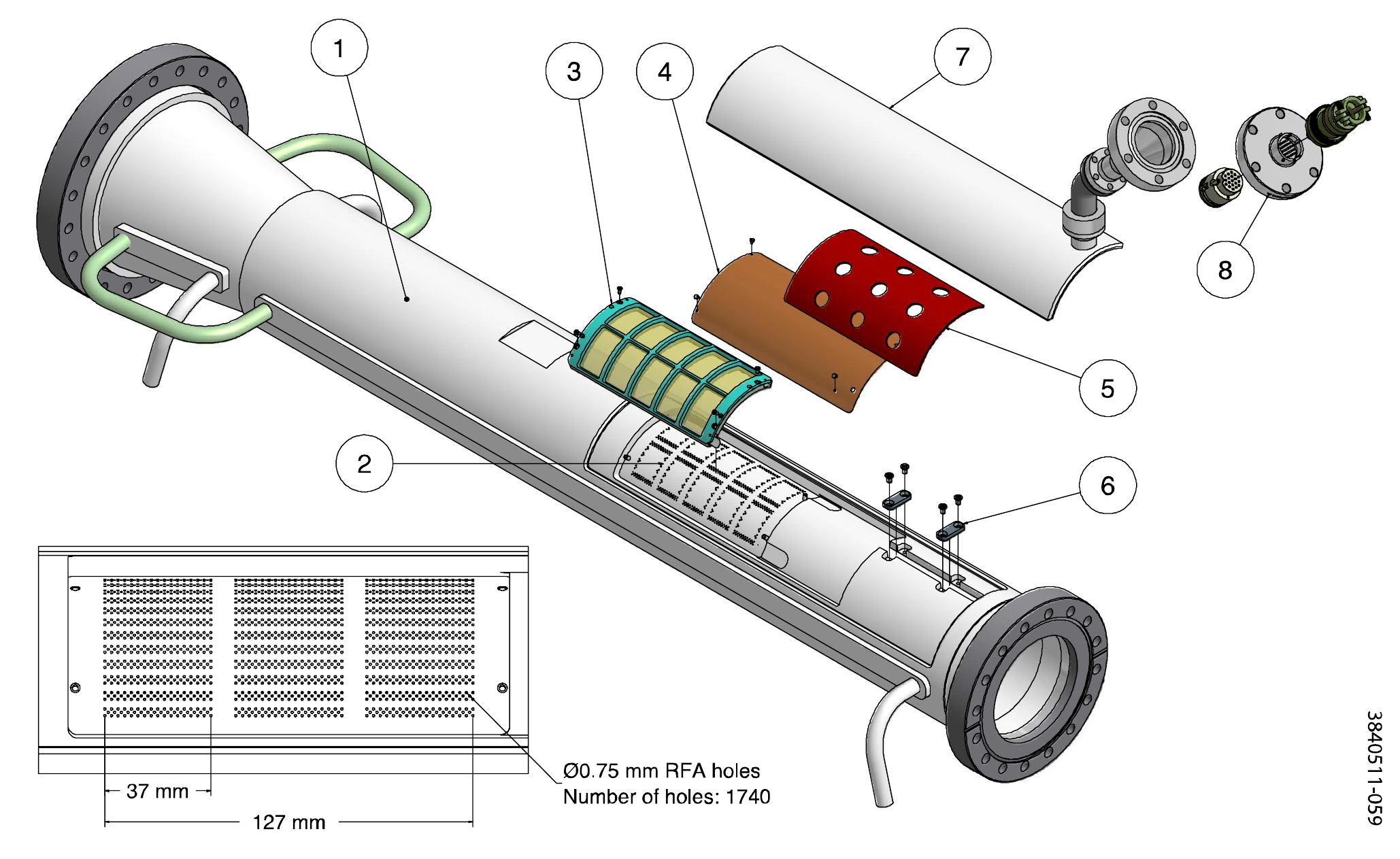}
    \caption[Exploded view of a RFA beam pipe in a CESR quadrupole magnet]{Exploded view of the structure of the RFA within a CESR quadrupole beam pipe. The major components of the RFA beam pipe include: (1) Aluminum beam pipe with cooling channels; (2) RFA housing and wiring channels; (3) Retarding grids, consisting of high-transparency gold-coated meshes nested in PEEK frames; (4) RFA collector flexible circuit; (5) Stainless steel backing plate; (6) Wire clamps; (7) RFA vacuum cover with connection port; (8) 19-pin electric feedthrough for RFA connector\cite{NIMA770:141to154}. \label{fig:cesr_conversion:Quad_RFA_Structure}}
\end{figure}

\begin{figure}
    \centering
    \includegraphics[width=0.75\textwidth, angle=0, width=6.0in]{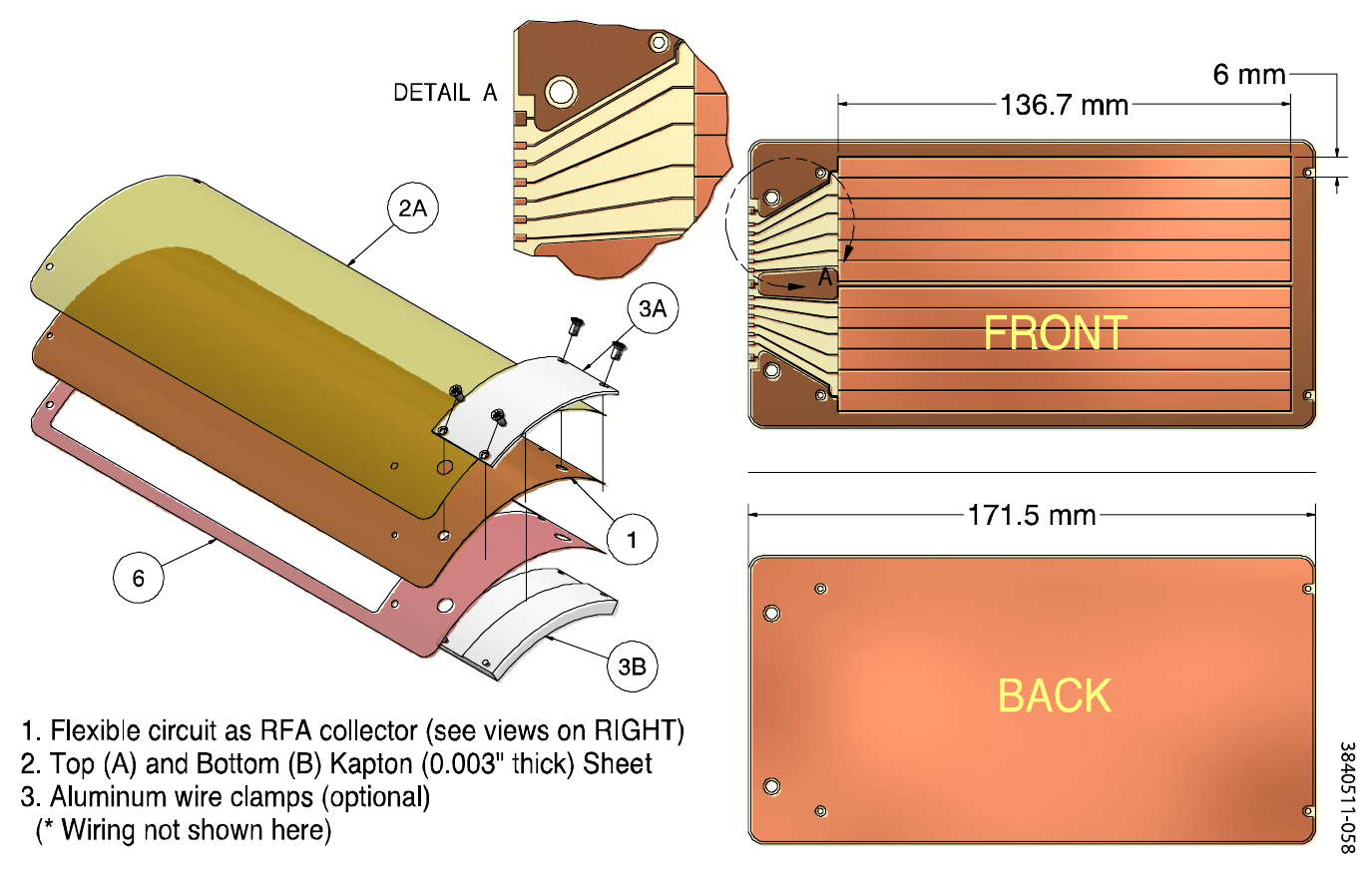}
    \caption{The flexible circuit used for the quadrupole RFA collector. \label{fig:cesr_conversion:Quad_RFA_Circuit}}
\end{figure}

\begin{figure}
    \centering
    \includegraphics[width=0.75\textwidth, angle=0, width=6.0in]{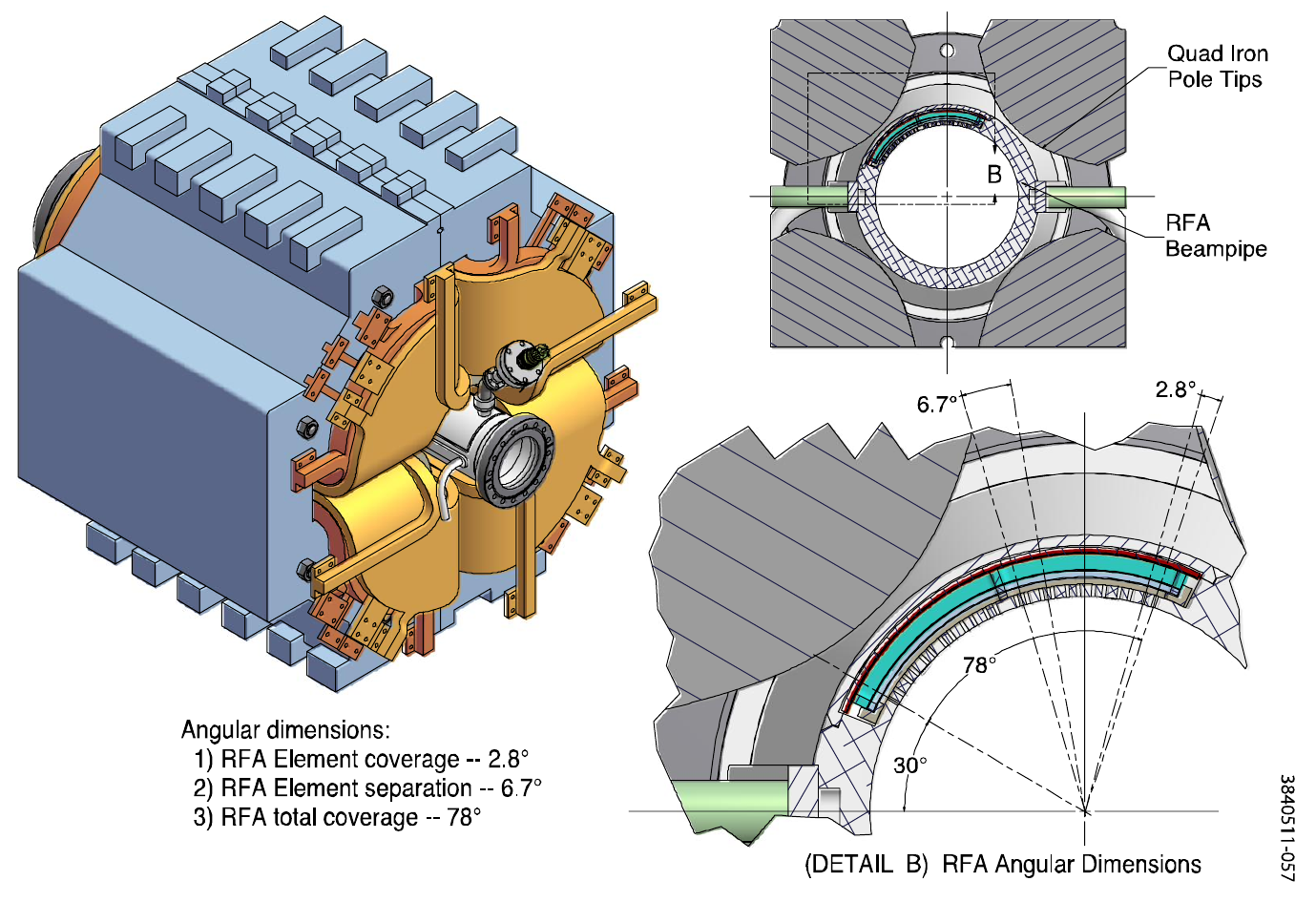}
    \caption{The RFA beam pipe in the Q48W quad (left). The RFA angular coverage (right)\cite{NIMA770:141to154}. \label{fig:cesr_conversion:RFA_in_quad}}
\end{figure}

Thin-style RFAs were implemented in beam pipes in a quadrupole magnet in CESR in the L3 experimental region (see Figure~\ref{fig:cesr_conversion:vac_l3}).  The design of the quadrupole RFA beam pipe is illustrated in Figure~\ref{fig:cesr_conversion:Quad_RFA_Structure}.  The beam pipe is constructed from thick wall aluminum tube (6061-T6 alloy, 4.5"~outside diameter, 3.5"~inside diameter), allowing room for a machined RFA housing pocket.  A channel was also machined as a part of the RFA housing pocket for wires connecting the grid and RFA collectors to the vacuum feedthrough port, which must be outside of the quadrupole magnet.  The structure of the RFA consists of high-transparency gold-plated copper meshes nested in PEEK frames and a segmented collector made of flexible circuit, similar to the RFA designs used in the CESR dipole and wigglers.  Many small holes (1740 of0.75~mm diameter holes) were drilled through the beam pipe to allow electrons to reach the RFA, while filtering out the beam-induced RF EMI.  These 1740 holes are grouped into 12~angular segments, matching the 12 RFA collector elements on the flexible circuit (Figure~\ref{fig:cesr_conversion:Quad_RFA_Circuit}).  The angular coverage and resolution of the RFA is shown in Figure~\ref{fig:cesr_conversion:RFA_in_quad}.

\begin{figure}
    \centering
    \includegraphics[width=0.5\textwidth]{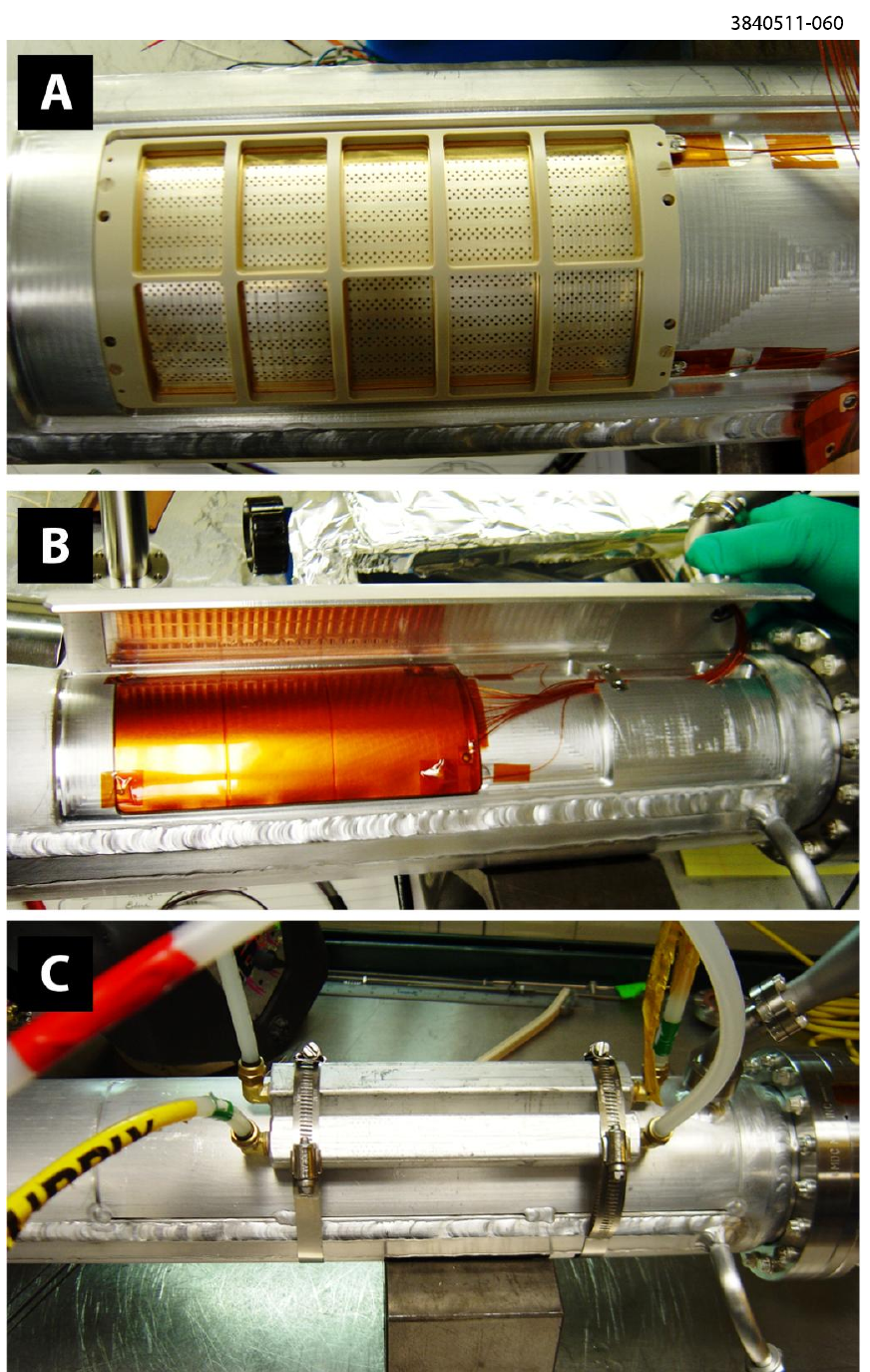}
    \caption[Photos of quadrupole RFA beam pipe construction]{Photos of quadrupole RFA beam pipe construction, showing key steps: (A) Gold-coated meshes in PEEK frames are mounted and wired; (B) Flexible collector circuit installed.  The circuit is electrically isolated with clean Kapton sheets; (C) Water-cooled bars were used during final welding of the RFA vacuum cover. \label{fig:cesr_conversion:Quad_RFA_photos}}
\end{figure}

Photos in Figure~\ref{fig:cesr_conversion:Quad_RFA_photos} show key steps in the RFA assembly process.  Before the installation of the RFA detector assembly, the quadrupole beam pipe vacuum components, including the cooling channels (but minus the RFA vacuum cover,) were TIG-welded together.  A vacuum leak check was performed using a Viton gasket to seal off the RFA pocket to ensure the vacuum integrity for all the major vacuum welds.  Then to ensure that the temperature sensitive RFA parts (the flexible circuits, PEEK grid frames, etc.) were not subjected to heating from any heavy TIG-welding, water-cooled bars were added to the cover before welding. The RFA electrical properties (including capacitance measurements) were checked repeatedly at every step of the assembly.  The finished RFA beam pipe was leak checked and a 150$^\circ$C, 24-hr vacuum bakeout was carried out before back-filling with dry N$_2$ in preparation for installation.

Two RFA quadrupole beam pipe assemblies were constructed during the {\cesrta} Phase 1 program.  The first assembly was with a bare aluminum beam pipe. It was installed in the L3 experimental region (in the Q48W quadrupole magnet) in July~2009 and tested in {\cesrta} experimental runs between July~2009 and March~2010.  With the first quadrupole RFA beam pipe successfully assembled and tested in CESR, the second assembly was built with a TiN-coated aluminum beam pipe.  The TiN coating (of 150~nm to 200~nm in thickness) was applied via DC sputtering to the inner surfaces of the beam pipe before the assembly of the RFA.  The same RFA assembly procedure was followed as was employed for the first assembly.  The TiN-coated RFA beam pipe replaced the bare aluminum RFA beam pipe in April~2010, and has remained in the L3 experimental region until December of 2012.

\section{Summary}
%% Text of Summary

The modification of the storage ring CESR to support the creation of {\cesrta}, a test accelerator configured to study accelerator beam physics issues for a wide range of accelerator effects and development of instrumentation related to present light sources and future lepton damping rings, required the installation of a significant number of vacuum chambers with their associated diagnostics.  This paper has presented an overview plus many of the details for these vacuum system changes and it is a useful reference for any future analysis of measurements undertaken as part of the {\cesrta} program.  With careful planning, UHV-compatible designs and construction, and stringent vacuum QA/QC, the vacuum system conversion successfully supported the {\cesrta} program without any negative impact to the scheduled CHESS operations.
Not included in this paper are the details of the RFA detectors created for use within one of the superconducting wiggler magnets in CESR; it is described in a companion paper.  When operating for the {\cesrta} program, CESR's vacuum system and instrumentation has been optimized for the study of low emittance tuning methods, electron cloud effects, intra-beam scattering, fast ion instabilities as well as the development and improvement of beam diagnostics.

\label{sec:summary}

\section{Acknowledgements}

The authors would like to acknowledge the many contributions that have helped make the {\cesrta} research program a success. It would not have occurred without the support of the International Linear Collider Global Design Effort led by Barry Barish. Furthermore, our colleagues in the electron cloud research community have provided countless hours of useful discussion and have been uniformly supportive of our research goals.

We would also like to thank the technical and research staff at Cornell{Õ}s Laboratory for Accelerator ScienceS and Education (CLASSE) for their efforts in maintaining and upgrading CESR for Test Accelerator operations. 

Finally, the authors would like to acknowledge the funding agencies that helped support the program. The U.S. National Science Foundation and Department of Energy implemented a joint agreement to fund the {\cesrta} effort under contracts PHY-0724867 and DE-FC02-08ER41538, respectively. Further program support was provided by the Japan/US Cooperation Program. Finally, the beam dynamics simulations utilized the resources off the National Energy Research Scientific Computing Center (NERSC) which is supported by the Office of Science in the U.S. Department of Energy under contract DE-AC02-05CH11231.

\label{Acknowledgements}



%% References


\bibliographystyle{amsplain}
\cleardoublepage
\bibliography{Bibliography/CesrTA}






% \end{thebibliography}

\end{document}